\documentclass[twocolumn,superscriptaddress,aps,floatfix]{revtex4-2}

\usepackage{braket}
\usepackage{graphicx}
\usepackage{amsmath}
\usepackage{mathrsfs}
\usepackage[dvipsnames]{xcolor}
\usepackage{longtable}
\usepackage{amsfonts}
\usepackage{tikz}
\usepackage[colorlinks,linkcolor=blue,citecolor=blue,anchorcolor=blue]{hyperref}
\usepackage{bm}

	\newcommand{\un}[1]{\mathrm{\:#1}}
	\newcommand{\Yb}{$\mathrm{^{171}Yb}$~}
	\newcommand{\x}{x_c}
	\newcommand{\y}{x_a}
	\newcommand{\FSR}{\mathrm{FSR}} %Free spectral range
    
    \newcommand{\aver}[1]{\ensuremath{\langle {#1} \rangle}}

\begin{document}	
	\title{Collective Spin-Light and Light-Mediated Spin-Spin Interactions in an Optical Cavity
	}
	\date{\today}
	\author{Zeyang Li}
	\thanks{These authors contributed equally}
	\affiliation{Department of Physics, MIT-Harvard Center for Ultracold Atoms and Research Laboratory of Electronics, Massachusetts Institute of Technology, Cambridge, Massachusetts 02139, USA}
	\author{Boris Braverman}
    \altaffiliation{Current address: Department of Physics and Max Planck Centre for Extreme and Quantum Photonics, University of Ottawa, 25 Templeton Street, Ottawa, Ontario K1N 6N5, Canada}
	\affiliation{Department of Physics, MIT-Harvard Center for Ultracold Atoms and Research Laboratory of Electronics, Massachusetts Institute of Technology, Cambridge, Massachusetts 02139, USA}
	\author{\!\!$^{,*}$ Simone Colombo}
	\affiliation{Department of Physics, MIT-Harvard Center for Ultracold Atoms and Research Laboratory of Electronics, Massachusetts Institute of Technology, Cambridge, Massachusetts 02139, USA}
	\author{Chi Shu}
	\affiliation{Department of Physics, MIT-Harvard Center for Ultracold Atoms and Research Laboratory of Electronics, Massachusetts Institute of Technology, Cambridge, Massachusetts 02139, USA}
	\affiliation{Department of Physics, Harvard University, Cambridge, Massachusetts 02138, USA}
	\author{Akio Kawasaki}
    \affiliation{Department of Physics, MIT-Harvard Center for Ultracold Atoms and Research Laboratory of Electronics, Massachusetts Institute of Technology, Cambridge, Massachusetts 02139, USA}\affiliation{National Metrology Institute of Japan (NMIJ), National Institute of Advanced Industrial Science and Technology (AIST), 1-1-1 Umezono, Tsukuba, Ibaraki 305-8563, Japan}
	\author{Albert F. Adiyatullin}
	\altaffiliation{Current address: Universit\'{e} Lille, CNRS, UMR 8523-PhLAM-Physique des Lasers Atomes et Mol\'{e}cules, F-59000 Lille, France}
	\affiliation{Department of Physics, MIT-Harvard Center for Ultracold Atoms and Research Laboratory of Electronics, Massachusetts Institute of Technology, Cambridge, Massachusetts 02139, USA}
	\author{Edwin Pedrozo-Pe\~{n}afiel}
	\affiliation{Department of Physics, MIT-Harvard Center for Ultracold Atoms and Research Laboratory of Electronics, Massachusetts Institute of Technology, Cambridge, Massachusetts 02139, USA}
	\author{Enrique Mendez}
	\affiliation{Department of Physics, MIT-Harvard Center for Ultracold Atoms and Research Laboratory of Electronics, Massachusetts Institute of Technology, Cambridge, Massachusetts 02139, USA}
	\author{Vladan Vuleti\'{c}}
	\affiliation{Department of Physics, MIT-Harvard Center for Ultracold Atoms and Research Laboratory of Electronics, Massachusetts Institute of Technology, Cambridge, Massachusetts 02139, USA}

	\begin{abstract}
		The interaction between an atomic ensemble and a light mode in a high-finesse optical cavity can easily reach the strong-coupling regime, where quantum effects dominate. In this regime, the interaction can be used to generate both atom-light and atom-atom entanglement. We analyze the dominant effects on the collective atomic state and the light field, and derive a unified approach that can account for atomic entanglement induced both by measurements on the light field, and by ignoring the state of the light field altogether. We present analytical expressions for the entanglement induced by the interaction, and determine the conditions that maximize the entanglement-induced gain over the standard quantum limit in quantum sensors and atomic clocks.
	\end{abstract}

\pacs{}

\maketitle

\section{Introduction}

Atoms and atom-like systems placed in an optical resonator provide a versatile experimental platform to study many-body quantum physics, entanglement, quantum simulation and computation, quantum measurement, dissipative quantum systems, quantum networks, and a variety of other topics \cite{Raimond2001,Ritsch2013,Georgescu2014,Reiserer2015,Wendin2017,Georgakopoulos2018,Ozawa2019,Muniz2020,Davis2020}. 
In these cavity quantum electrodynamics (cQED) systems, the optical cavity allows a photon to interact multiple times with all atoms on successive round trips, which substantially enhances the atom-light interaction as compared to free space. Often, the atom-light interaction is dominated by the coupling of the atoms to a single mode of the cavity. For a single atom, such a system is a nearly ideal experimental realization of the Jaynes-Cummings Hamiltonian \cite{Jaynes1963}, while for an atomic ensemble, the system approximates the Tavis-Cummings Hamiltonian \cite{Tavis1968}. The interaction of atoms with multimode optical cavities has also been investigated \cite{Kollar2017,Georgakopoulos2018}, and this type of system can produce even more complex dynamics both in the atomic and photonic subspaces \cite{Vuletic2001,Domokos2002,Guo2019a,Guo2019b,Clark2020,Kollar2017}. Typically, two or more atomic energy levels are relevant to the problem, so that the atomic state can be described by a spin degree of freedom, while each cavity mode constitutes a harmonic oscillator.

To the lowest order, the atom-cavity interaction generates quantum correlations (entanglement) between the collective atomic spin and the light mode. Higher-order terms can lead to nonclassical states of the light (effective photon-photon interactions) \cite{Sayrin2011,Vlastakis2013,Hacker2016}, as well as nonclassical collective spin states (effective atom-atom interactions) \cite{Leroux2010,Schleier-Smith2010a,Bohnet2014a,Cox2016a,Hosten2016,Hosten2016a,Braverman2019,Zhao2021}, or even hybrid spin-light states \cite{Wilk2007a,Hacker2019}. Such entangled states are useful as building blocks for quantum metrology \cite{Ma2011,Pezze2018,Beckey2020}, quantum networks \cite{Kimble2008}, quantum information processing \cite{Weedbrook2012}, and as testbeds for the foundations of quantum mechanics \cite{Hensen2015}. 

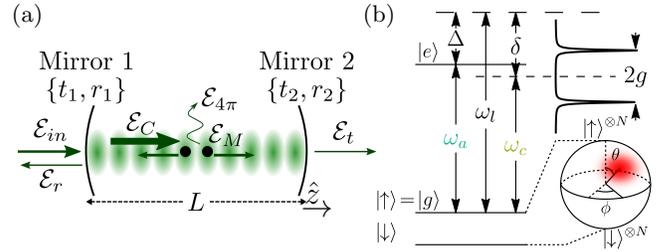
\begin{figure}
    \centering
    \setlength{\unitlength}{\columnwidth}
    \begin{picture}(1,0.38)
    \put(0.02,0.06){\includegraphics[width=0.55\columnwidth]{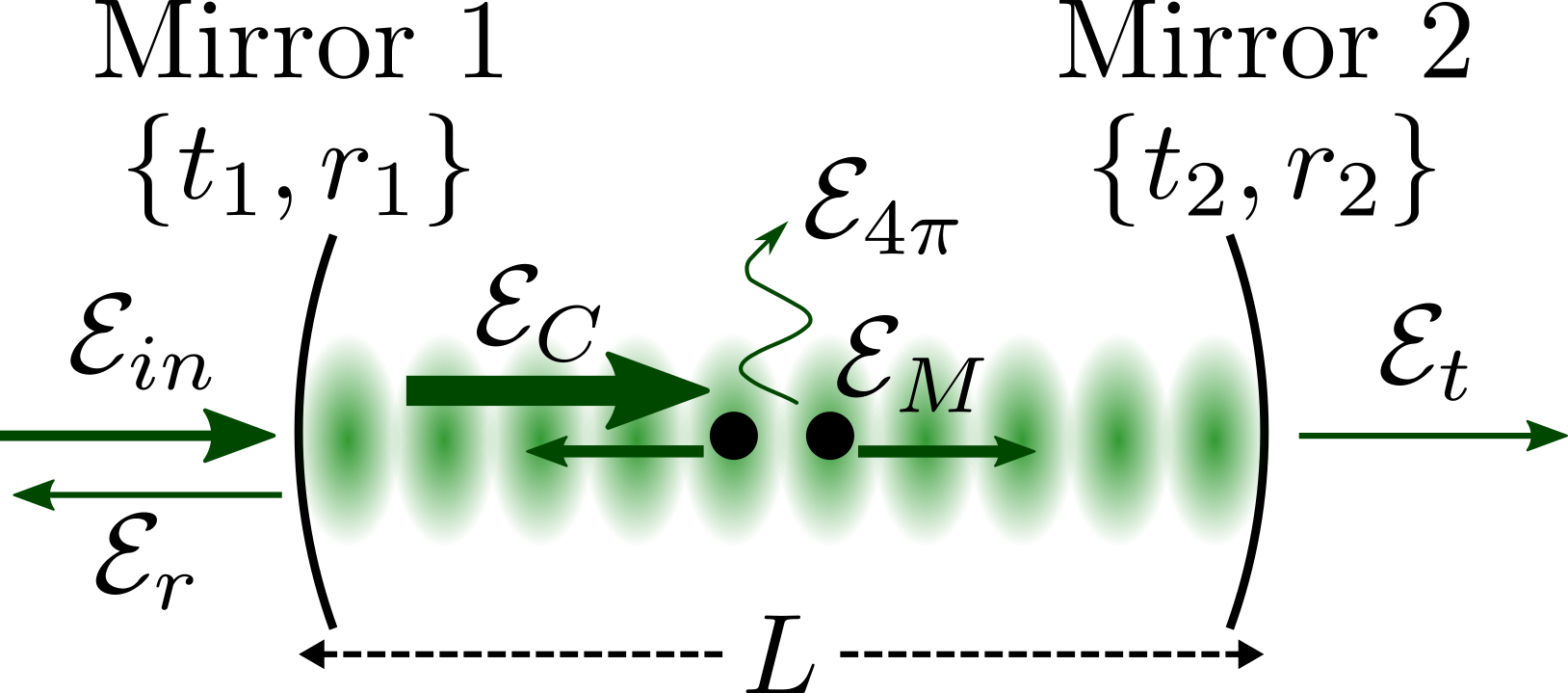}}
    \put(0.57,0){\includegraphics[width=0.43\columnwidth]{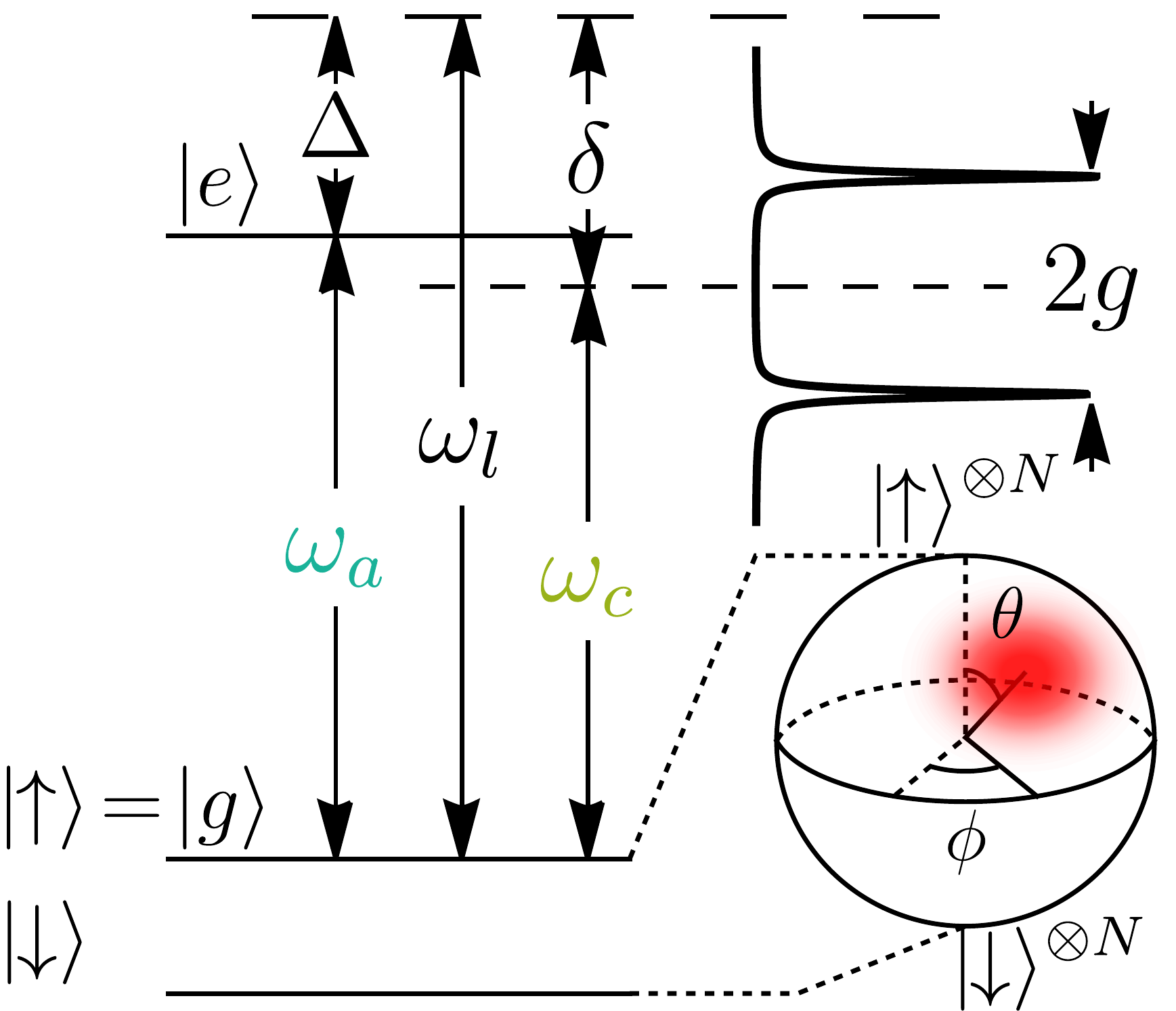}}
    \put(0.455,0.05){\large$\to$}
    \put(0.46,0.07){\large$\hat{z}$}
    \put(0.01,0.35){(a)}
    \put(0.55,0.35){(b)}
    \end{picture}
    \caption{(a) Schematic diagram of an ensemble of atoms interacting with a single-mode cavity of length $L$. The incident field $\mathcal{E}_\mathrm{in}$ builds up in the cavity to the value $\mathcal{E}_c$. The cavity field is coherently scattered by the atoms into the cavity with amplitude $\mathcal{E}_M$, and coherently radiated through the cavity mirrors to give the transmitted and reflected fields $\mathcal{E}_t$ and $\mathcal{E}_r$. The cavity field scatters from the atoms into free space with amplitude $\mathcal{E}_{4\pi}$, and from the cavity mirrors with amplitude $\mathcal{E}_{sc}$, which is assumed to be zero in a lossless cavity. 
    (b) Level diagram of the atomic three-level system. $\ket{\uparrow}=\ket{g}$ and $\ket{e}$ are the two states interacting with the cavity light. $\omega_a, \omega_c$ and $\omega_l$ are the atomic, cavity resonance frequencies and the probing light frequency, respectively. $\Delta=\omega_l-\omega_a$ and $\delta=\omega_l-\omega_c$ are the detunings of probing light from atomic and cavity resonance, respectively. The curve on the right represents the cavity transmission spectrum in the presence of atoms, and $2g$ is the vacuum Rabi splitting due to the light-atom interaction when $\omega_a=\omega_c$. The sphere in the bottom right represents a Bloch sphere formed by the $N$ atoms of spin $\frac{1}{2}$, and the shaded distribution represents a coherent spin state (CSS) pointing along polar and azimuthal angles ($\theta,\phi$) in spherical coordinates where each atom is in a superposition state $\cos(\theta/2)\ket{\uparrow}+\sin(\theta/2)\mathrm{e}^{i\phi}\ket{\downarrow}$.}
    \label{fig:AtomCavitySchematic}
\end{figure}

Previous theoretical analyses of the generation of entangled atomic states in cQED systems have adopted one of two viewpoints: the first focuses on the measurement aspect, whereby the atom-light interaction transfers information about the atomic spin state onto the light field, and therefore a measurement of the light field can conditionally project the state of the atoms onto a collective entangled state ~\cite{Hagley1997,Kuzmich1998,Vernac2000,Sorensen2002,Saffman2009,Inoue2013,Chen2014,McConnell2015,Barontini2015}. 
Alternatively, one can view the system from a quantum feedback perspective \cite{Wiseman1994,Lloyd2000,Appel2009,Zhang2012,Pawlowski2016}, where the light-atom interaction maps the spin quantum noise onto the light field \cite{Appel2009,Hammerer2004,Hammerer2010}, and then the light field acts back onto the spin \cite{Inoue2013}, creating atomic entanglement without the necessity to perform any measurements (cavity feedback squeezing)~\cite{Schleier-Smith2010,Liu2011,Lee2014a,Davis2016}.
While the first approach to generating atomic entanglement builds upon the use of information contained in the light field, the second approach neglects that information altogether \cite{Takeuchi2005}. Hence, these two effects have so far been treated separately. Nonetheless, the two effects originate from the same Hamiltonian so that they are related, and both need to be considered in cQED experiments.

In this article, we present a unified description of both the measurement-induced and quantum-feedback-induced generation of entangled many-body spin states in cQED, considering both of these aspects on an equal footing. Our analysis remains valid in the regime where both atom-light and effective atom-atom interactions play a significant role in the system dynamics, and thus neither can be neglected. We assume that the atomic transition is not saturated by the cavity field, which allows us to treat the optical field semi-classically, and derive analytical expressions for the intracavity, transmitted, reflected, and scattered fields, as well as the collective spin state. We also limit the majority of this article to analyzing optical cQED problems, where the coupling strength is far from reaching the ultrastrong coupling regime, see Ref.~\cite{FornDiaz2019} for details. In principle the framework developed here can also be used for the superconducting circuit QED problems like \cite{Fink2009,Barends2014} of circuit QED systems; however, the majority of circuit QED systems contain much larger phase noise than the atom-number relaxation, i.e., $T_2\ll 2T_1$~\cite{Kjaergaard2020,Blais2021} and therefore cannot be described by the coherent dipolar coupling that we are considering here.

The analysis presented here reveals universal features of the cQED system that remain valid for arbitrary detuning of the probe light frequency from the cavity and atomic resonances. We show that the atom-light entanglement strength (i.e., the measurement strength \cite{Hammerer2004,Hammerer2010,Julsgaard2001,Saffman2009} when the light field is used as a `meter' for the atomic quantum state) at any light detuning or coupling strength is given by the product of three terms: the single-atom cooperativity, the normalized cavity transmission, and the the total number of photons scattered by the atoms into free space. Thus, when the probe laser frequency is close to the cavity resonance (whether bare or dressed by atom-cavity interactions) and transmission is high, the atom-photon entanglement is the dominant effect. Consequently, the generation of interatomic entanglement in a `heralded' fashion by performing a measurement on the light field~\cite{McConnell2015,Chen2015,Frowis2017,Duan2019} is most efficiently accomplished by tuning the light field close to the (dressed) cavity resonance. 

On the other hand, when the detuning of the probe laser frequency from the cavity resonance far exceeds the cavity linewidth, and transmission through the atom-cavity system is low, the atom-light entanglement remains relatively weak, and higher-order terms take on the leading role \cite{Zhang2015}, generating spin-spin correlations \cite{Schleier-Smith2010a,Leroux2010,Braverman2019}. 
In this regime, first discovered in Ref.~\cite{Zhang2015} using numerical simulations, the light acts as a {\it quantum catalyst}: the light field mediates the interaction between distant atomic spins, resulting in entangled many-body atomic states, while becoming only weakly entangled with the atoms. Thus at the end of the interaction process, the state of the atom-light systems approximately factorizes into a fixed coherent state of the light field and a collective atomic spin state displaying interatomic entanglement. Under appropriate conditions, this entanglement can take the form of a squeezing process where the quantum projection noise of the collective atomic spin is redistributed along specific directions, resulting in a squeezed spin state (SSS) \cite{Kitagawa1993,Wineland1992,Wineland1993}. 
Zhang \textit{et al.}~\cite{Zhang2015} have previously pointed out that it is possible to reduce the impact of atom-light entanglement on cavity-induced spin squeezing by optimizing the the detuning of the probe beam from cavity resonance. However, their analysis is only valid in the limit of a large detuning between the atomic and the cavity resonances, while a number of experiments use smaller detunings to maximize the light-atom interaction strength, and reduce the sensitivity to technical noise \cite{Cox2016a, Braverman2019}. 

Our results predict the attainable levels of atom-light entanglement which can be used for measurement-based squeezing \cite{Chen2014,Bohnet2014a}, as well as to what extent it is possible to engineer effective spin-spin interactions. The latter can be used for a variety of purposes, including metrology with squeezed spin states \cite{Schleier-Smith2010,Hosten2016a,Vincenzo2020}, quantum simulation with atomic motional degrees of freedom \cite{Ritsch2013,Leonard2017a,Leonard2017b,Vaidya2018,Guo2019a,Guo2019b,Morales2019} or internal spin degrees of freedom \cite{Hung2016,Davis2019,Bentsen2019a,Bentsen2019b}, and the study of other interesting quantum many-body phenomena such as the out-of-time-order correlations in many-body system \cite{Garttner2017}, quantum phase transition \cite{Evrard2019,Makhalov2019}, quantum non-equilibrium behaviors \cite{Sierant2019,Pizzi2021,Muniz2020} and quantum gravity \cite{Periwal2021}.
In particular, it is possible to change the sign of the effective spin Hamiltonian via the detuning of the light from the cavity resonance, and generate system evolution `backwards in time' for measuring, e.g., out-of-time-order correlation functions \cite{Swingle2016,Garttner2017,Li2017,Ken2019}, or for improving quantum measurements \cite{Davis2016}. 
The treatment presented here provides a complete analytical description that applies for arbitrary detunings between the incident light field, the cavity mode, and the atomic resonance, and for both weak and strong atom-light coupling, both in terms of the single-atom cooperativity and the collective cooperativity. Out analysis provides a unified framework that describes both effects of measurement and quantum feedback, and can thus be used for analyzing and comparing, over the full range of parameters, cQED experiments that aim to generate entangled atomic states.

Our manuscript is organized as follows: We first develop a general unified theory. In Sec.~\ref{sec:ALI} we describe the general setup of the cQED system and the assumptions underlying our approach. 
We also define the collective spin state and obtain 
an effective spin Hamiltonian with the semi-classical coherent light field in the cavity. 
In Sec.~\ref{sec:Non-unitary}
we discuss the 
generated light-atom entanglement, especially the concept of 
quantum Fisher information (QFI) of the light in the context of the cQED system. 
In Sec.~\ref{sec:Chirp} we use the concept of QFI to discuss measurement-based spin squeezing as it is used in experiments. In Sec.~\ref{sec:FLS} we generalize the discussion in
Sec.~\ref{sec:ALI} and Sec.~\ref{sec:Fbroadening} 
to a richer atomic model with four internal states. In Sec.~\ref{sec:Application} we apply the complete four-level model to \Yb atoms to analyze the experimental results of Refs. \cite{Braverman2019,Pedrozo2020} that generate spin squeezing in an optical-transition atomic clock. We conclude with a discussion in Sec.~\ref{sec:Conclusion}.

\section{Interaction between an atomic ensemble and a cavity mode}\label{sec:ALI}

An ensemble of two-level atoms inside a cavity with non-degenerate modes is well-described by the Tavis-Cummings Hamiltonian \cite{Jaynes1963,Tavis1968}. This Hamiltonian is valid if the interaction with one electromagnetic mode dominates. The excitation of the cavity mode by an incident light field and dissipation by decay from the cavity or by atomic scattering into free space are not part of the Tavis-Cummings Hamiltonian, but can be included phenomenologically \cite{Gardiner1985,Scala2007}, or from first principles \cite{Alsing1992,Sames2014}. This approximation of a single cavity mode with external driving and dissipation is in excellent agreement with the majority of cQED experiments where the interaction with other transverse or longitudinal cavity modes can be ignored \cite{Bohnet2014a, Spethmann2016, Moller2017, Schine2019, Hacker2019}. For many applications it is desirable to have an analytical and more intuitive description of the atom-light interaction that can be directly applied to existing experiments, and used to guide future applications. Such experiments often deal with nearly classical light beams containing many photons, where a quantization of the electromagnetic field (beyond the effects of photon shot noise) is not needed. In this case it is possible to derive simpler and analytical expressions for the light-atom and the light-mediated spin-spin interaction, as we discuss below.

When the optical cavity containing an ensemble of atoms is illuminated by a continuously applied laser beam or a long laser pulse (coherent state of light), one can often approximate the intracavity, transmitted and reflected light fields as remaining coherent states, even when the interaction with the atoms is included (for a discussion of the limits of this approximation, see Sec.~\ref{sec:AtomLightInteractionTwoLevelSystemApproximation}). In this limit, the photon annihilation and creation operators can be substituted by their expectation values plus a small correction, reducing the problem to the interaction of atoms with a classical electromagnetic field plus photon shot noise. This approach is possible as long as the tuning of the cavity by the atomic spin variation results in small changes of the intracavity field, i.e. when the atom-cavity interaction can be linearized. Based on the corresponding effective Hamiltonian describing the interaction of an ensemble of atomic spins with a classical electromagnetic field, we can then find analytical solutions for the evolution of the coupled atom-cavity system, and use these expressions to gain physical insight into the entanglement process. 

\subsection{Atomic states and operators}\label{sec:HilbertSpace}

We consider an atomic level structure where each individual atom in the ensemble has ground-state levels $\ket{\downarrow}$ and $\ket{\uparrow}$, constituting a spin-$\frac{1}{2}$ system, as well as an electronically excited state $\ket{e}$ that couples only to the state $\ket{\uparrow}$. (For an extension to a more complex level structure, see Section \ref{sec:FLS}.) The collective spin operators are defined as the sum of the spin operators of the $N$ atoms,
\begin{align}
    \hat{S}_{x,y,z}=\frac{1}{2} \sum_{j=1}^N \hat{\sigma}_{x,y,z}^j
\end{align}
where the $\hat{\sigma}_{x,y,z}^j$ are the atoms' Pauli matrices. Among all possible spin states in the $2^N$-dimensional Hilbert space, there is a subspace of symmetric states which satisfy 
\begin{align}\label{eq:maxspin}
    \langle\hat{\bm{S}^2}\rangle=\left\langle\hat{S}_x^2+\hat{S}_y^2+\hat{S}_z^2\right\rangle=S(S+1),
\end{align}
where $S=N/2$ is the largest possible value for the total spin. As long as the system's Hamiltonian is symmetric in all spins $\bm{\sigma}^j$ and the initial state is in this subspace, the system will remain within this $(N+1)$-dimensional subspace that is much smaller than the full Hilbert space of dimension $2^N$. 

We can represent those collective atomic states on a Bloch sphere with radius $\sqrt{S(S+1)}$, e.g., in terms of their
Wigner quasiprobability distribution \cite{Schmied2011}. 
In quantum metrology and other applications, the atomic ensemble is initially prepared in a coherent spin state (CSS) \cite{Arecchi1972}, defined as the product state with all individual spins pointing along the same direction. The CSS along polar and azimuthal angles $(\theta, \phi)$, as illustrated in Fig.~\ref{fig:AtomCavitySchematic}~(b), has the following expansion in terms of Dicke states $\ket{S,S_z}$, the joint eigenstates of $\hat{\bm{S}}^2$ and $\hat{S}_z$ in the symmetric subspace \cite{Dicke1954,Arecchi1972,Schleich1994}:
\begin{align}\begin{split}
    \ket{\theta,\phi}=&\sum_{S_z=-S}^{S}\left(\begin{matrix}
    2S\\S+S_z
    \end{matrix}\right)^{1/2}\left(\sin\frac{\theta}{2}\right)^{S+S_z}\left(\cos\frac{\theta}{2}\right)^{S-S_z}\\
    &\times e^{-i(S+S_z)\phi}\ket{S,S_z}
\end{split}\end{align}

For a CSS pointing along the $x$-axis as in 
Fig.~\ref{fig:AtomCavitySchematic}(b), the uncertainties satisfy and saturate the Heisenberg uncertainty principle $\Delta S_y \Delta S_z=|\aver{[\hat{S}_y,\hat{S}_z]}|/2=S/2$. Due to rotational symmetry, we have $\Delta S_y = \Delta S_z=\sqrt{S/2}$.  In general, a CSS has a standard deviation $\sqrt{S/2}$ along any direction perpendicular to its mean spin $\aver{\bm{S}}$.

\subsection{Classical light field inside a cavity containing an atomic ensemble}
\label{sec:AtomLightInteractionTwoLevelSystemApproximation}

We consider the interaction of the atoms with a single standing-wave cavity mode, schematically shown in Fig.~\ref{fig:AtomCavitySchematic}. We assume that all $N$ atoms are fixed at the cavity antinodes, which provides the maximum atom-light coupling, which can be achieved by double wavelength trapping~\cite{Lee2014b}, site-dependent selection ~\cite{wu2021sitedependent}. The more general case of distributed atom positions can be mapped onto the case of uniform coupling by means of correction factors (of order unity) to the atom number and interaction strength \cite{Tanji-Suzuki2011,Hu2015}, where the attainable entanglement is suppressed if the temperature is too high. Apart from photon shot noise, we assume the light field to be classical, i.e. we use coherent states of light in the description. 
Then the two-level atom can be approximated as a harmonic oscillator. In the opposite limit, the ensuing saturation of the atomic state would prevent the atoms from absorbing further photons. This would effectively induce photon-photon interactions, and in this limit the intracavity field can no longer be described by a coherent state, as has been shown both in the optical domain cQED \cite{Schuster2008} and in circuit QED systems \cite{Bishop2009}. In this work, we always consider a linear atomic response far below saturation. Note that a linear response also applies to the intermediate probing regime where the average photon number can exceed one but is still much smaller than the number of atoms, as discussed in Appendix~\ref{sec:AppendixMultiPhoton}. 

The power transmission and reflection coefficients of the mirrors $i=1,2$ are denoted by $T_i=|t_i|^2$ and $R_i=|r_i|^2$, respectively. In general, the mirrors have loss and therefore $R_i+T_i<1$. In the following discussion, we consider lossless mirrors ($R_i+T_i=1$),  while in Appendix~\ref{sec:SimulatingLossyCavity} we establish a mapping of a cavity composed of realistic mirrors with loss onto one with lossless mirrors by means of additional optical components.

In the following part, we introduce several quantities and relations that characterize the coupling between the atomic ensemble and the cavity. These relations are derived in Appendix~\ref{sec:CavityFieldAppendix} following the treatment of Ref.~\cite{Tanji-Suzuki2011}.

The cavity-enhanced single-atom cooperativity at an antinode is defined as \cite[Eq.~(30)]{Tanji-Suzuki2011}
\begin{equation}\label{eq:SingleCooperativity}
    \eta=\frac{24\mathcal{F}}{\pi k^2w^2}, 
\end{equation}
where $k = 2\pi/\lambda$ is the wave number of the light, $w$ is the mode waist ($1/e^2$ intensity radius) of the TEM$_{00}$ cavity mode, and $\mathcal{F} \approx 2\pi/(2-R_1-R_2)$
is the cavity finesse for a lossless cavity with mirror power reflectivities $R_1, R_2 \approx 1$. 

To simplify the following expressions, we define a position-independent mode amplitude $\mathcal{E}$ via the relation \cite{Siegman86}
\begin{equation}\label{eq:ModeAmplitudeDefinition}
    \mathcal{E} = E(0,0,z) \sqrt{\frac{\epsilon_0 c \pi w^2(z)}{2}}.
\end{equation}
This definition applies to a Gaussian beam propagating in the $\hat{z}$ direction with an $1/e^2$ intensity radius $w(z)$ at a distance $z$ from the mode waist which is located at $z=0$. $E(0,0,z)$ is the electric field on the mode axis at an antinode at $z$. The thus defined mode amplitudes of the incident, cavity, transmitted and reflected fields are denoted by $\mathcal{E}_\mathrm{in}$, $\mathcal{E}_{c}$, $\mathcal{E}_{t}$, and $\mathcal{E}_{r}$, respectively.
 
The atomic parameter that determines the coupling of the atoms to the cavity is the atomic polarizability $\tilde{\alpha}$. For a single two-level atom with transition frequency $\omega_a$ and natural linewidth $\Gamma$, the polarizability is given by \cite[Eq.~(1)]{Tanji-Suzuki2011}
\begin{equation}\label{eq:AtomicPolarizability}
\tilde{\alpha} = 6 \pi \epsilon_0 c^3 \frac{\Gamma / {\omega_a}^2}{{\omega_a}^2 - {\omega_l}^2 - i (\omega_l^3 / {\omega_a}^2) \Gamma}.
\end{equation}
In extreme cases where the total solid angle subtended by the cavity modes are significant as in \cite{Heinzen1987}, the atomic linewidth $\Gamma$ in the denominator should be replaced by an effective linewidth \cite{Heinzen1987}; however, for most optical cavities the differences are negligible ($<\%1$). 
In the limit of moderate detuning from atomic resonance $\Delta= \omega_l - \omega_a\ll\omega_a$ and strong but not ultrastrong coupling between atoms and light, corresponding to the condition of the rotating-wave approximation, the atomic polarization can be approximated by a Lorentzian lineshape:
\begin{equation}\label{eq:AtomicPolarizabilityRWA}
\tilde{\alpha} \approx 6 \pi \epsilon_0 \frac{c^3}{\omega_a^3} \frac{i}{1 - \frac{i \Delta}{\Gamma/2}}=6 \pi \epsilon_0 \frac{c^3}{\omega_a^3}\left(\mathcal{L}_d(\y)+i\mathcal{L}_a(\y)\right).
\end{equation}
Here $\y{\equiv}\frac{\Delta}{\Gamma/2}$ is the normalized detuning of the probe laser ($\omega_l$) from the atomic resonance ($\omega_a$), while $\mathcal{L}_d(\y)=-\frac{\y}{1+\y^2}$ and $\mathcal{L}_a(\y)=\frac{1}{1+\y^2}$
are the Lorentzian dispersive and absorptive lineshapes, respectively \cite{Tanji-Suzuki2011}.

The amplitude of the intracavity field can then be expressed as
\begin{equation}\label{eq:CavityFieldFinalResultRWA}
\mathcal{E}_c = \frac{\mathcal{F}}{\pi} \frac{i t_1}
{1+N\eta\mathcal{L}_a(\y)-i(\x+N\eta\mathcal{L}_d(\y))}
\mathcal{E}_\mathrm{in},
\end{equation}
where $\x\equiv\frac{\delta}{\kappa/2}$ with $\delta=\omega_l - \omega_c$ is the normalized light-cavity detuning, $\kappa$ is the full-width half-maximum cavity linewidth, and $t_1$ is the amplitude transmission coefficient of the input mirror. The real part of the denominator of Eq.~\eqref{eq:CavityFieldFinalResultRWA} corresponds to the absorptive part of the atomic susceptibility, while the imaginary part combines the dispersive part of the atomic response with the phase shift that occurs when probing an empty cavity away from resonance. 

The transmitted and reflected field amplitudes $\mathcal{E}_t,\mathcal{E}_r$ are similarly given by
\begin{equation}\label{eq:TransmittedFieldFinalResult}
\mathcal{E}_t = 
-  \frac{\mathcal{F}}{\pi} \frac{t_1 t_2 e^{i k L}}{1+N\eta\mathcal{L}_a(\y)-i(\x+N\eta\mathcal{L}_d(\y))} \mathcal{E}_\mathrm{in} 
\end{equation}
and
\begin{equation}\label{eq:ReflectedFieldFinalResult}
\mathcal{E}_r = 
\left[ r_1-  \frac{\mathcal{F}}{\pi} \frac{ t_1^2 r_2 e^{2 i k L} }{1+N\eta\mathcal{L}_a(\y)-i(\x+N\eta\mathcal{L}_d(\y))} \right] \mathcal{E}_\mathrm{in}. 
\end{equation}
Here $L$ is the length of the standing-wave cavity, and $t_2$ the amplitude transmission coefficient of the output mirror. The transmitted field is simply proportional to the intracavity field, while the reflected field is a superposition of the light directly reflected off the input mirror (term with $r_1$) and the field leaking out of the cavity (second term).

From Eq.~\eqref{eq:TransmittedFieldFinalResult} it follows that the power transmission of a symmetric and lossless cavity is given by
\begin{equation}\label{eq:SymmetricTransmission}
    \mathcal{T}_0=\frac{\left|\mathcal{E}_t\right|^2}{\left|\mathcal{E}_r\right|^2}=\frac{1}{(1+N\eta\mathcal{L}_a(\y))^2+(\x+N\eta\mathcal{L}_d(\y))^2}.
\end{equation}
The atomic absorption (first term in the denominator) is determined by the atom number $N$, the single-atom cooperativity $\eta$, and the absorptive Lorentzian lineshape $\mathcal{L}_a$, while the dispersion of the atom-cavity system (second term in the denominator) is given by $N$, $\eta$, and the dispersive Lorentzian $\mathcal{L}_d$.
The transmission of an asymmetric cavity with mirror power transmissions $T_{1,2}=|t_{1,2}|^2$ is given by
\begin{equation}\label{eq:AsymmetricTransmission}
\mathcal{T}=\frac{4T_1T_2}{(T_1+T_2)^2}\mathcal{T}_0.
\end{equation}
The ratio of the photon scattering rate into free space $\mathcal{S}$ to the photon transmission rate $\mathcal{T}$ can be simply expressed as \cite{Tanji-Suzuki2011}:
\begin{equation}\label{eq:AtomScatteringTransmissionResultRWA}
\frac{\mathcal{S}}{\mathcal{T}}= \frac{2\pi}{T_2 \mathcal{F}} N\eta \mathcal{L}_a(\y)
\end{equation}
The scattering rate into free space is proportional to the rate of photon transmission, the collective cooperativity $N \eta$, and the absorptive atomic lineshape $\mathcal{L}_a$. 

\subsection{Effective light-induced Hamiltonian for collective atomic state}

The light shift experienced by a polarizable object, in this case an atomic ensemble, subject to an electric field $\hat{E} = 2\hat{E}_c$ at an antinode of the cavity is given by the Hamiltonian \cite{Grimm2000}
\begin{align}
    \hat{H}_\mathrm{dip}
    =-\frac{1}{2}\big|\hat{E}\big|^2\mathrm{Re}(\alpha)=-2\big|\hat{E}_c\big|^2\mathrm{Re}(\alpha), 
\end{align}
where $\alpha$ is the atomic polarizability. For $\hat{N}_\uparrow = \hat{S}_z+S$ atoms interacting with the intracavity field, this Hamiltonian using Eqs. \ref{eq:AtomicPolarizabilityRWA} and \ref{eq:CavityFieldFinalResultRWA} can be expressed as
\begin{align}\label{eq:CavityExpansion}
\begin{split}
    \hat{H}_{\textrm{dip}}&=-\left(\hat{S}_z+S\right)\eta\frac{|\hat{\mathcal{E}}_c|^2}{\omega}\frac{\pi}{\mathcal{F}}\mathcal{L}_d(\y) \\
    &= - \hbar \Omega \hat{n}_c \left(\hat{S}_z+S\right). 
\end{split}
\end{align}
Here $\Omega = \pi\eta \mathcal{L}_d(\y)\kappa /\mathcal{F} $ is the light shift per intracavity photon and $\hat{n}_c=|\hat{\mathcal{E}}_c|^2/(2\hbar\omega\kappa)$ is the intracavity photon number. The operator $\hat{\mathcal{E}}_c$, as in \eqref{eq:CavityFieldFinalResultRWA}, is a function of the number of atoms coupling to the cavity, which equals $\hat{N}_\uparrow$ because only atoms in $\ket{\uparrow}$ state couple to the cavity. Since $\hat{N}_\uparrow=\hat{S}_z+S$, the intracavity photon number $\hat{n}_c$ becomes a function of $\hat{S}_z$.
The above Hamiltonian corresponds to an interaction term of the form $\hbar \Omega \hat{c}^\dagger \hat{c} S_z$ in the second quantization formalism of Ref.~\cite[Eq.~(2)]{Leroux2012}, with $c$ being the photon annihilation operator for the cavity mode, plus an $S_z$-independent average shift of the cavity mode frequency $\omega_c \to \tilde{\omega}_c=\omega_c - \Omega S$. This additional, $S_z$-independent term can take a different form when both spin levels are coupled to excited states as in Ref.~\cite{Leroux2012}, instead of just one of the levels, as in the present analysis.
Subtracting the average frequency shift, i.e. measuring the cavity detuning relative to $\tilde{\omega}_c$, we arrive at the effective atomic Hamiltonian
\begin{equation}\label{eq:atomicHamiltonian}
\hat{H} = - \hbar \Omega \hat{n}_c \hat{S}_z
\end{equation}
This Hamiltonian depends on atomic parameters only through the collective (symmetric) spin operator $\hat{S}_z$, and therefore the atomic state remains in the symmetric subspace of maximal spin $S$ defined in \eqref{eq:maxspin} under the action of this Hamiltonian. The Hamiltonian depends on the light field only through the intracavity photon number $\hat{n}_c$ or intensity $|\hat{\mathcal{E}}_c|^2$. The Hamiltonian $H$, as given by Eq.~\eqref{eq:atomicHamiltonian}, can also be obtained by generalizing the Tavis-Cummings Hamiltonian \cite{Tavis1968}, or through the procedure described in the Supplemental Information of Ref.~\cite{Davis2016}. 

\subsection{Light-induced coherent effects on the collective atomic spin} 
\label{sec:CavityFeedbackSqueezing}

For a CSS at the equator with $\langle \hat{S}_z\rangle=0$, it is natural to expand the $\hat{S}_z$-dependent intracavity photon number $\hat{n}_c(\hat{S}_z)$ in the Hamiltonian \eqref{eq:atomicHamiltonian} into a Taylor series in terms of $\hat{S}_z$:
\begin{equation}\label{eq:CavityAtomsStarkShiftTwoLevelAtomTaylorSeries}
\hat{H} = -\hbar \Omega \hat{S}_z \sum_{j=0}^\infty \frac{S_z^j}{j!}\left(\frac{\partial^j \hat{n}_c}{\partial S_z^j}\right)_{S_z=0} .
\end{equation}
The leading term of this expansion is the  phase-shift $\hat{H}_0= -\hbar \Omega\langle{\hat{n}_c}\rangle \hat{S}_z$ with $\aver{\hat{n}_c}=\hat{n}_c(S_z=0)$, which generates a rotation of the collective atomic spin about the $z$-axis of the Bloch sphere (Fig.~\ref{fig:QFeffectonstate}(d)). This simply represents the light shift on the atoms associated with the average intracavity photon number $\aver{\hat{n}_c}=n_c(S_z=0)$. Using Eqs. \eqref{eq:CavityFieldFinalResultRWA} and \eqref{eq:AtomScatteringTransmissionResultRWA}, the atomic phase shift after a time $\tau$ is
\begin{equation}\label{eq:CavityAtomsStarkShiftTwoLevelAtomPhaseShiftVsTransmission}
\Delta \phi = {\Omega\aver{n_c} \tau} =  \eta n_\gamma^\mathrm{t} \frac{T_1+T_2}{2 T_2}\mathcal{L}_d(\y) = 
-\frac{n_\gamma^\mathrm{sc}}{2\aver{N_\uparrow}}\y,  
\end{equation}
which has been expressed in terms of the average transmitted photon number $n_\gamma^\mathrm{t}$ or the average scattered photon number $n_\gamma^\mathrm{sc}$. Here $\hat{N}_\uparrow =\hat{S}_z+S$ is the atom number operator in the state $\ket{\uparrow}$, and we can use $\aver{\hat{N}_\uparrow}=N/2$ in the vicinity of the equator. The ratio of atomic phase shift $\Delta \phi$ to the transmitted photon number is independent of the atom number, but is proportional to the single-atom cooperativity $\eta$, which makes the phase shift measurement a convenient way to measure $\eta$ without requiring an accurate calibration of the atom number.

The first-order term in Eq.~\eqref{eq:CavityAtomsStarkShiftTwoLevelAtomTaylorSeries} is the non-linear light-mediated spin-spin interaction term
\begin{equation}\label{eq:TwistingHamiltonian}
\hat{H}_1 = -\hbar \chi \hat{S}_z^2. 
\end{equation}
This term represents an $\hat{S}_z$-dependent rotation of the atomic spin about the $z$-axis. $H_1$ is also known as the one-axis twisting Hamiltonian \cite{Kitagawa1993, Sorensen2002}, as depicted in Fig.~\ref{fig:QFeffectonstate}(e). Here we see how the light can act as a quantum catalyst that causes an effective spin-spin interaction $S_z^2$ between distant atoms. This type of interaction can, e.g., produce SSSs, as first proposed by Kitagawa and Ueda \cite{Kitagawa1993}. The shearing parameter $\chi$ in $\hat{H}_1$ is given by
\begin{equation}\label{eq:CavityAtomsSz2}
\chi = - \eta \mathcal{L}_d(\y) \left(1-\x\y+\langle \hat{N}_\uparrow\rangle\eta\right)\mathcal{T}_0 \frac{\mathcal{S} }{\aver{\hat{N}_\uparrow}}.
\end{equation}
We see that the light-mediated one-axis twisting $\hat{S}_z^2$ is proportional to the scattered photon number: without photon scattering, there is no shearing of the atomic spin. On the other hand, the shearing strength is proportional to the cooperativity, and therefore a better cavity (or a larger number of atoms) allows one to reduce the decoherence associated with the scattering of photons into free space for a given desired shearing \cite{Manzoni2018}.

The shearing acting over a time $\tau$ produces a shearing strength $Q$, which is defined as the differential phase displacement along $S_y$ per unit $S_z$, as in Fig.~\ref{fig:QFeffectonstate}(e)~\cite{Schleier-Smith2010a,Leroux2010},
\begin{align}
    Q=S\frac{\mathrm{d}^2\arg\left(e^{iS_z^2\chi \tau}\right)}{\mathrm{d}S_z^2}= 2S \chi \tau = N \chi \tau. 
\end{align}
In the vicinity of the equator of the Bloch sphere where $S_z=0$, the shearing strength $Q$ is given by 
\begin{equation}\label{eq:CavityAtomsQ}
Q = -2 \eta\mathcal{L}_d(\y)
(1 - \x \y + S \eta)\mathcal{T}_0 n_\gamma^\mathrm{sc}, 
\end{equation} where $\langle \hat{N}_\uparrow\rangle=N/2=S$, such that $\langle\hat{S}_z\rangle=0$, has been used. $Q$ is proportional to the number of photons scattered by the ensemble, and can be used as a parameter that characterizes the spin squeezing \cite{Schleier-Smith2010a,Leroux2010,Braverman2019}, as explained later in Sec.~\ref{sec:WinelandParameter}. 

\setlength{\unitlength}{\columnwidth}
\begin{figure}[t]\centering
\begin{picture}(1,1)
    \put(0.1,0.525){\includegraphics[width=0.85\columnwidth]{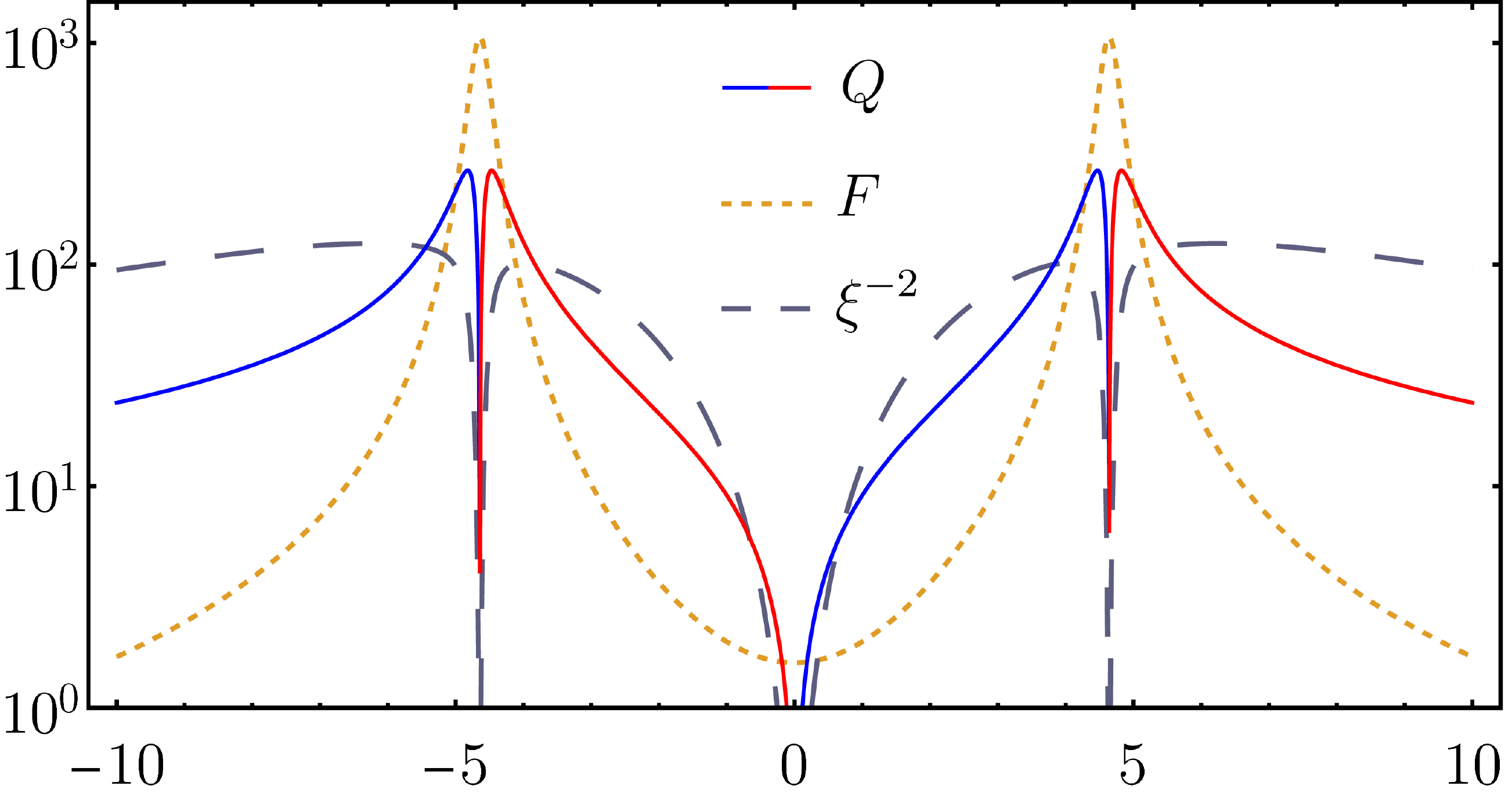}}
    \put(0.,0.775){{\makebox(0,0)[l]{\rotatebox{90}{$|Q|, F, \xi^{-2}$}}}}
    \put(0.55,0.505){{\makebox(0,0)[c]{Normalized Detuning $2\Delta/\Gamma$}}}
    \put(0.1,0.02){\includegraphics[width=0.85\columnwidth]{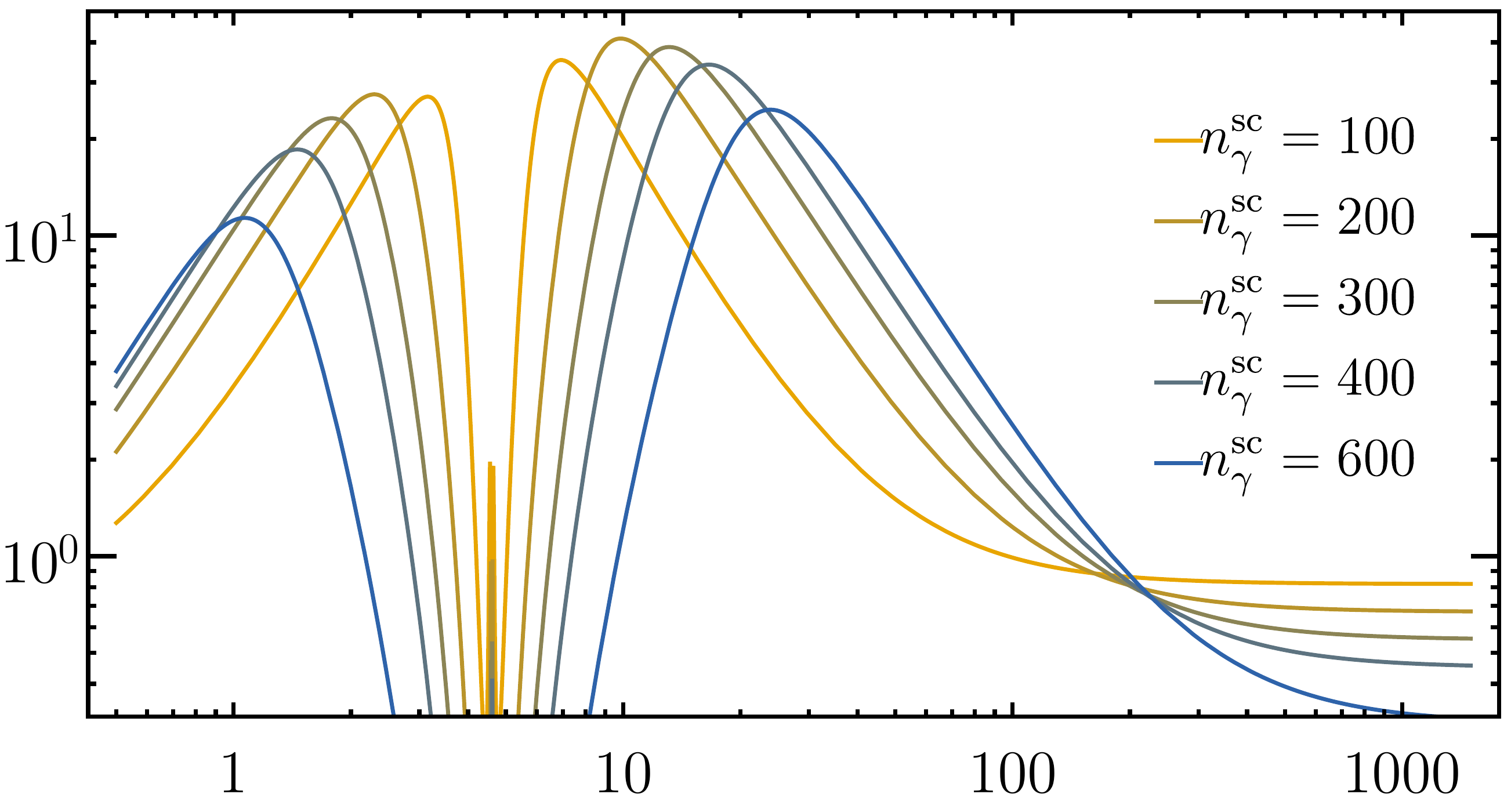}}
    \put(0.025,0.26){{\makebox(0,0)[c]{\rotatebox{90}{$\xi^{-2}$ including Bloch}}}} 
    \put(0.075,0.265){{\makebox(0,0)[c]{\rotatebox{90}{sphere curvature}}}}
    \put(0.55,0.){{\makebox(0,0)[c]{Normalized Detuning $2\Delta/\Gamma$}}}
    \put(0.19,0.92){(a)}
    \put(0.19,0.42){(b)}
\end{picture}
\caption{Results of cavity feedback squeezing for a cavity that is resonant with the atomic transition, $\omega_c=\omega_a$. (a) Squeezing strength $Q$, quantum Fisher information $F$ and inverse Wineland parameter $\xi^{-2}$ at fixed scattered photon number $n_\gamma^\mathrm{sc}=400$ and collective cooperativity $N_\uparrow\eta=900$. (These parameters correspond to the conditions of the experiment described in Ref. \cite{Braverman2019}). For $Q$ plot, red branches stand for positive $Q$ and blue branches for negative $Q$. In this calculation, the curvature of the Bloch sphere has been ignored, and the results do not depend on the single-atom cooperativity $\eta$. (b) Inverse Wineland parameter including the effect of the Bloch sphere curvature for different scattered photon number $n_\gamma^\mathrm{sc}$ calculated for the parameters $N=1000$, $\aver{S_z}=0$, and $\eta=1.8$. Both (a) and (b) assume $\kappa/\Gamma=2.8$ as in the experiment of Refs. \cite{Braverman2019,Pedrozo2020}. }
\label{fig:WinelandFixN}
\end{figure}
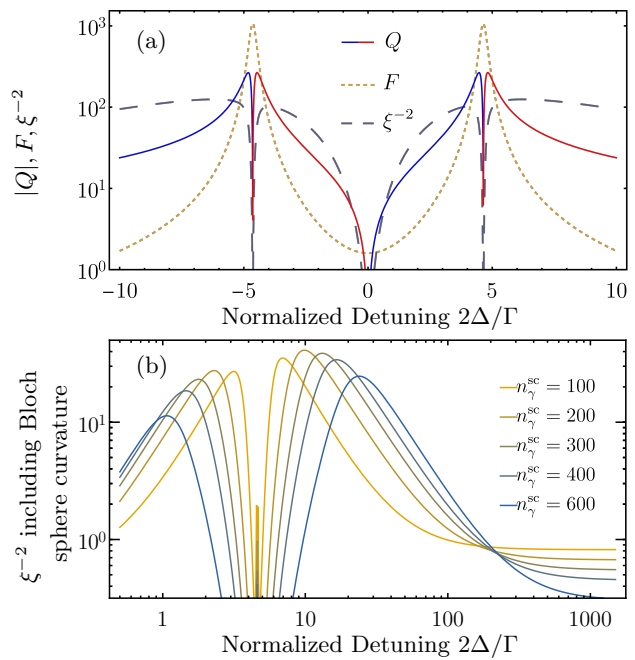

\section{Light-atom entanglement and non-unitary evolution of the atomic state}\label{sec:Non-unitary}

When deriving the above results for the evolution of the collective atomic spin state and spin squeezing, we have treated the light field as classical, and described the atomic evolution as a unitary evolution governed by the effective Hamiltonian $\hat{H}$ given in Eq.~\eqref{eq:atomicHamiltonian}. In doing so, we have so far ignored the effect of photon shot noise fluctuations on the atomic spin, or equivalently, the light-atom entanglement that builds up during this process \cite{Leroux2012}. If the information contained in the light field about the atomic ensemble remains unused, or equivalently, if the quantum state of the reflected, scattered and transmitted light fields is traced over, this results in a decoherence in the atomic subspace, and a mixed, rather than pure, quantum state of the atoms. In the following, we first characterize the information about the atomic state contained in a light field after the interaction, as quantified by the Fisher information of the light (Section \ref{sec:FI}). We then discuss the back action on the atomic state if this information remains unused (Section \ref{sec:Fbroadening}), and finally quantify the contrast loss due to scattering of photons into free space (Section \ref{sec:Contrast}).

\subsection{Quantum Fisher information available in a coherent state of light}\label{sec:FI}

\begin{figure}[!htbp]
    \centering
    \includegraphics[width=.65\columnwidth]{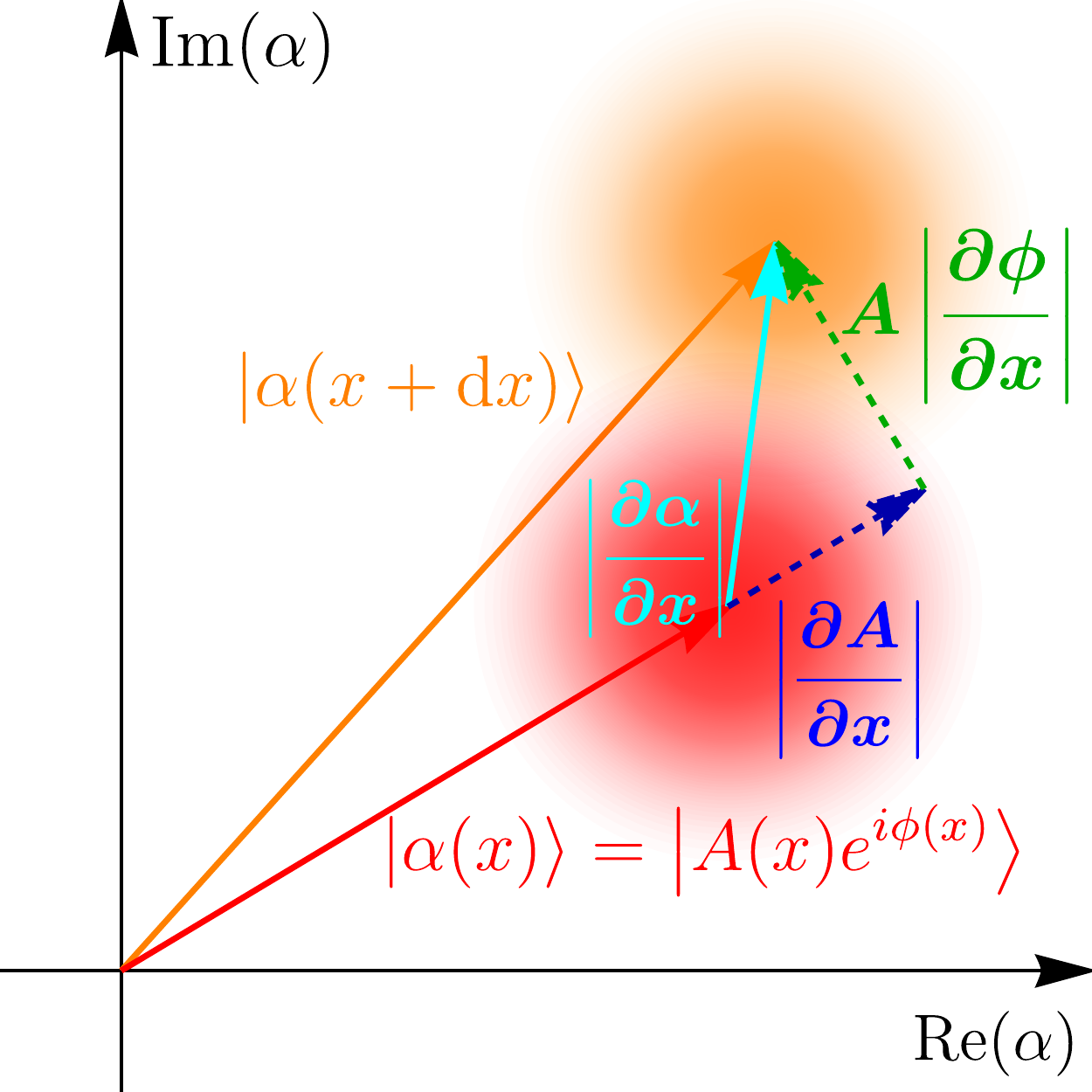}
    \caption{Precision of estimation of an unknown parameter $x$ by measurement of 
    a coherent state of light $\ket{\alpha}$. A change $dx$ in some parameter $x$ modifies the amplitude (blue) and phase (green) of $\ket{\alpha(x)}$. The optimum amount of information about $dx$ can be extracted by measuring the state quadrature along the direction of change $\partial \alpha/\partial x$ of the coherent state, which in general is a combination of phase and amplitude measurements. The shot noise of the coherent state of light (color-shaded area) limits the precision of the measurement.
    }
    \label{fig:NewFigureDisplacedCoherentState}
\end{figure}

In Sec.~\ref{sec:AtomLightInteractionTwoLevelSystemApproximation} we derived the amplitude and phase of the coherent state of light inside the cavity in the presence of the atoms, Eq.~\eqref{eq:CavityFieldFinalResultRWA}. This coherent state $\ket{\alpha}$ will depend on the number of atoms and their spin state. The light field leaking out of the cavity, as given by Eqs. \eqref{eq:TransmittedFieldFinalResult} and \eqref{eq:ReflectedFieldFinalResult}, can then be used to deduce information about the atomic ensemble. In general, if a coherent state of light $\ket{\alpha}$ depends on some physical parameter $x$, i.e. $\ket{\alpha} = \ket{\alpha(x)}$, then the quantum Cram\'er-Rao bound \cite{Helstrom1969,Braunstein1994,Ma2011,Pezze2018} for the variance associated with estimating the parameter $x$ is given by
\begin{equation}\label{eq:ErrorInXEstimateWithLight}
\left(\Delta x \right)^2 = \left( 2 \left|\frac{\partial \alpha}{\partial x} \right| \right)^{-2}
\end{equation}
This relation is derived in Appendix.~\ref{sec:FIderivation}. We recast this expression in terms of the QFI contained in the light field,
\begin{equation}\label{eq:FisherInformationInXEstimateWithLight}
\tilde{F}(x) = 4 \left|\frac{\partial \alpha}{\partial x} \right|^{2},
\end{equation}
where the QFI is just the inverse of the Cram\'er-Rao bound.

Often measurements are performed on either the power or the phase of the light field emanating from the cavity. In this case, it is useful to consider the amplitude $A$ and phase $\phi$ of the coherent state, $\alpha = A e^{i \phi}$, in order to separate the QFI into its amplitude component $\tilde{F}_A$ and phase component $\tilde{F}_{\phi}$,
\begin{equation}\label{eq:FisherInformationInXEstimateWithLightTransmissionPhase}
\tilde{F}(x) = \tilde{F}_A + \tilde{F}_{\phi} = 4 \left|\frac{\partial A}{\partial x} \right|^{2} 
+
4 A^2
\left|\frac{\partial \phi}{\partial x} \right|^{2}.
\end{equation}
In general, neither amplitude nor phase alone contain the full information about the atomic system available in the light field. Nevertheless, the full QFI can always be extracted by performing a homodyne measurement of the coherent state $\ket{\alpha}$ with optimized phase offset and amplitude (see Fig. \ref{fig:NewFigureDisplacedCoherentState}). If we only measure the transmitted and reflected power instead of performing the optimum state detection, the variance in the estimation of parameter $x$ will equal $\tilde{F}_A^{-1}$ with the best analysis. 

In the context of the cQED system displayed in Fig.~\ref{fig:AtomCavitySchematic}, the expression for the QFI, Eq.~\eqref{eq:FisherInformationInXEstimateWithLight}, leads to two consequences. If the parameter $x$ is chosen to be the collective atomic spin $S_z$ (see Section \ref{sec:HilbertSpace}), then Eq.~\eqref{eq:FisherInformationInXEstimateWithLight} quantifies the amount of entanglement between the collective atomic state and light, or equivalently, the information imprinted onto the state of the light by the atoms. So the QFI of the light with respect to the atomic state, normalized to the atomic state shot noise, tells us how much measurement-based spin squeezing can be attained by an optimal probing scheme. Second, even if the light is not used for measurement, as in cavity feedback squeezing \cite{Leroux2010,Schleier-Smith2010a}, the QFI still describes the amount of excess quantum backaction on the atomic state by the input light: discarding the information contained in the light field, or equivalently, tracing over the state of the light field, results in excess variance of the atomic spin state above the minimum imposed by the Heisenberg uncertainty relations for angular momentum.

\setlength{\unitlength}{\textwidth}

\begin{figure*}
\begin{picture}(1,0.4)
\put(0,0.2){\includegraphics[width=.5\columnwidth]{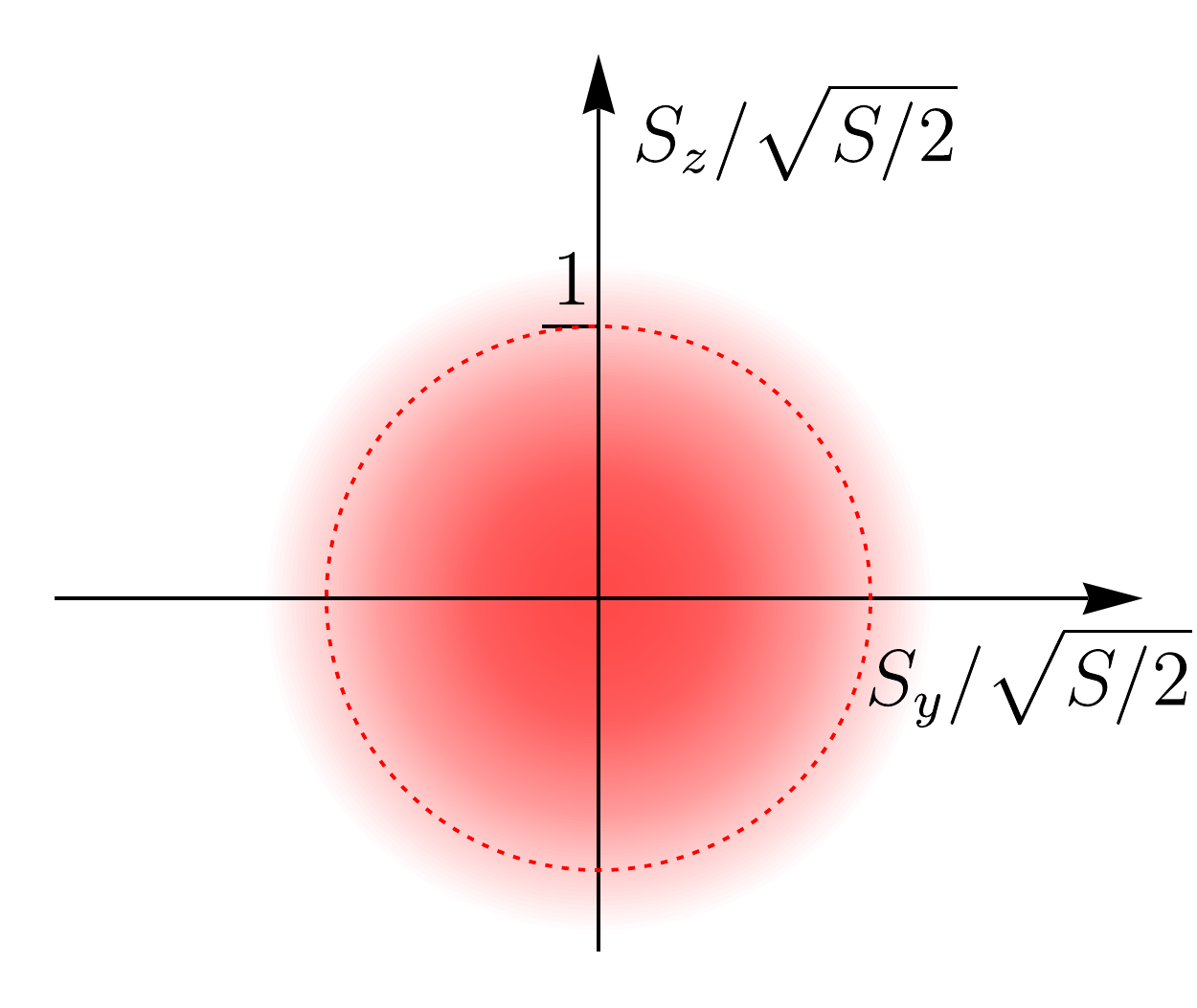}}
\put(0.34,0.2){\includegraphics[width=.5\columnwidth]{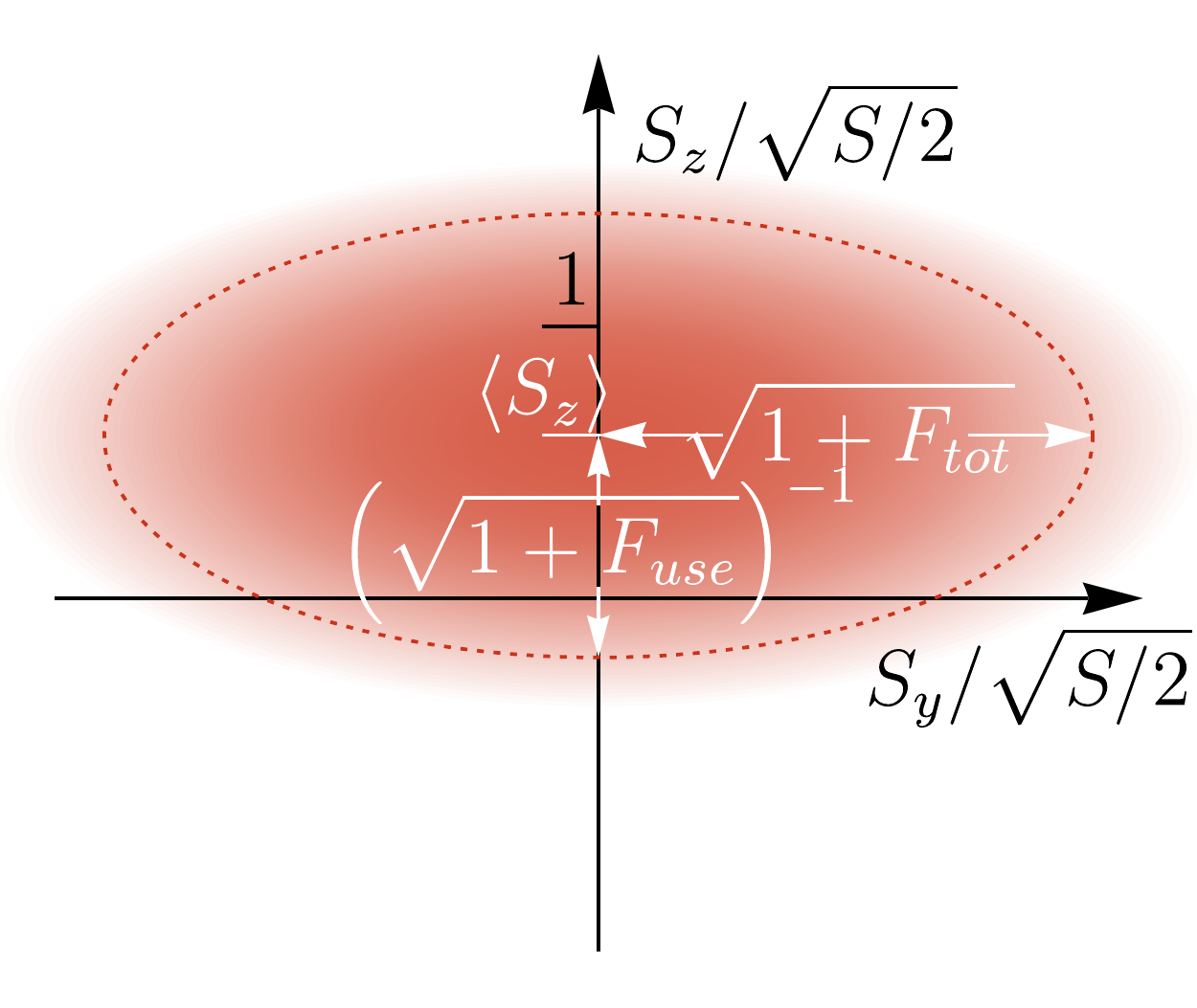}}
\put(0.7,0.2){\includegraphics[width=.5\columnwidth]{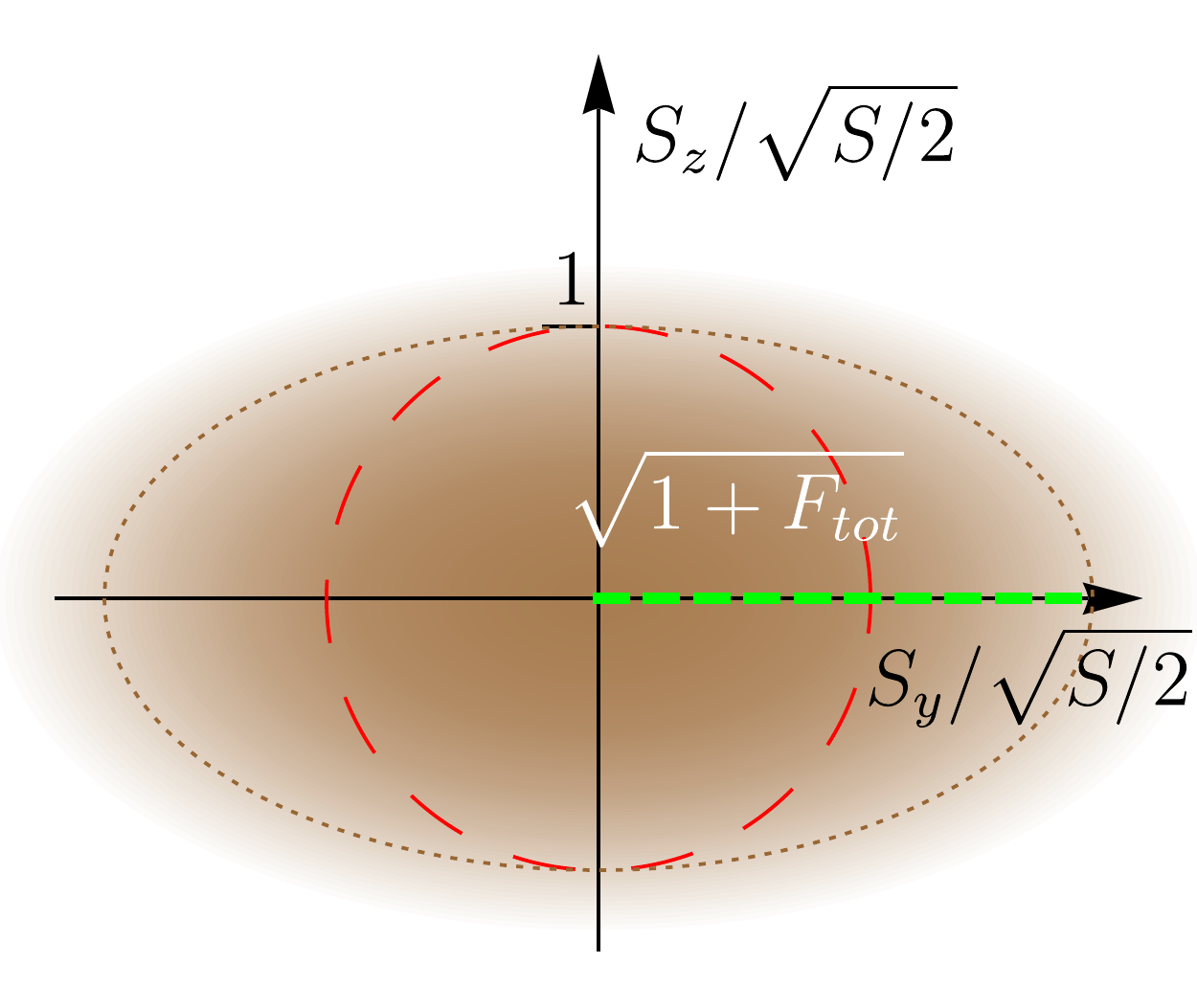}}
\put(0,0){\includegraphics[width=.5\columnwidth]{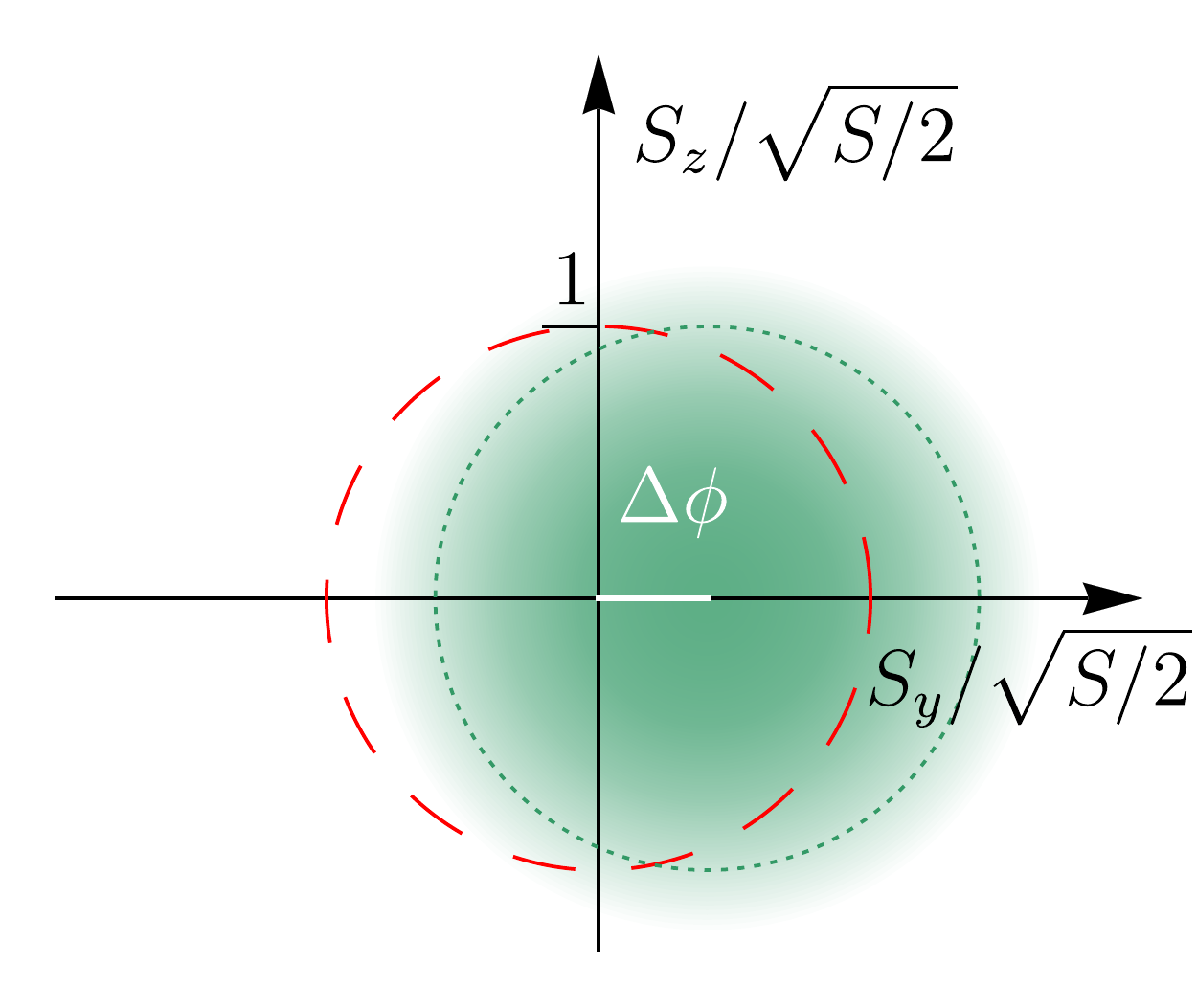}}
\put(0.34,0){\includegraphics[width=.5\columnwidth]{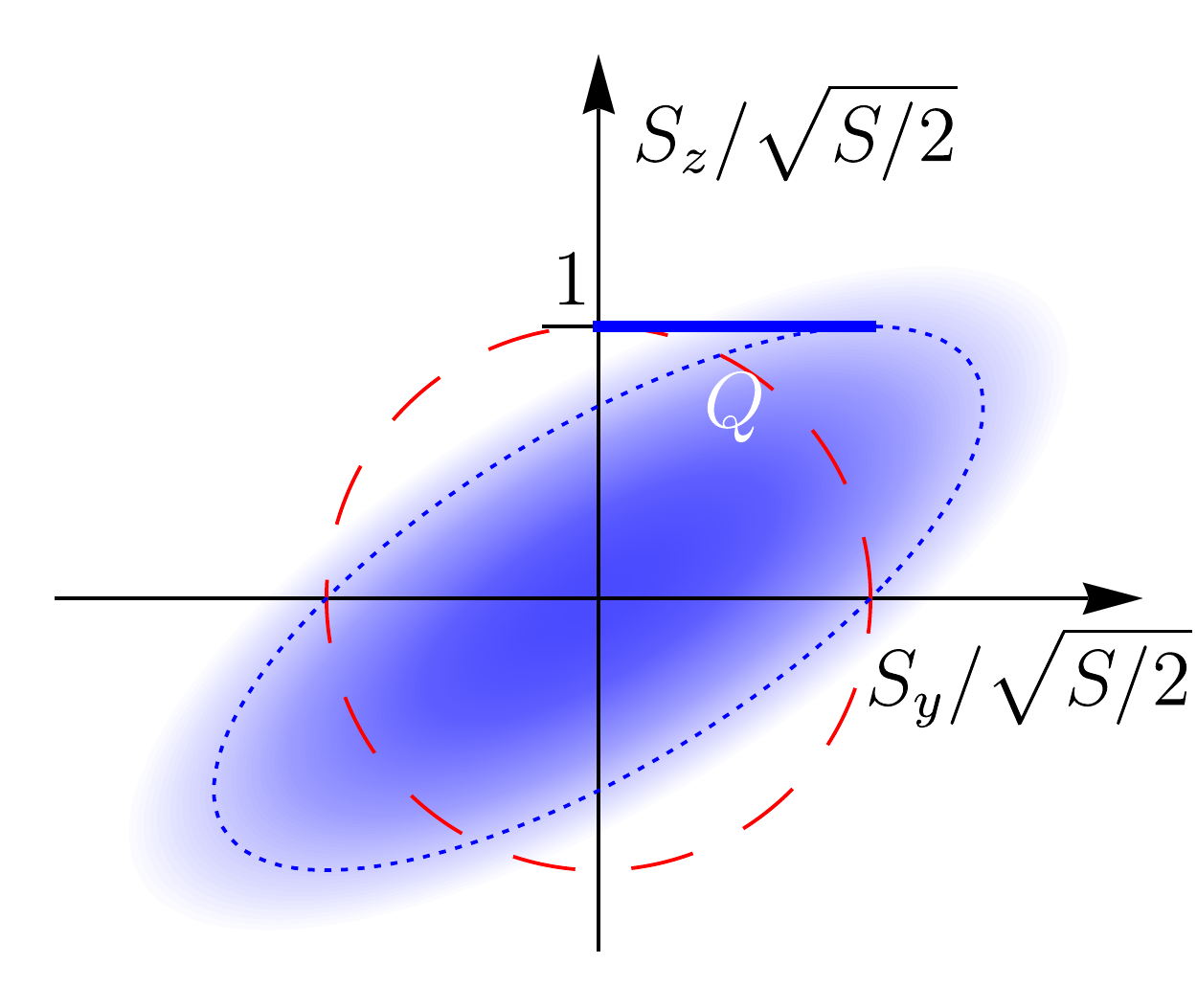}}
\put(0.7,0){\includegraphics[width=.591\columnwidth]{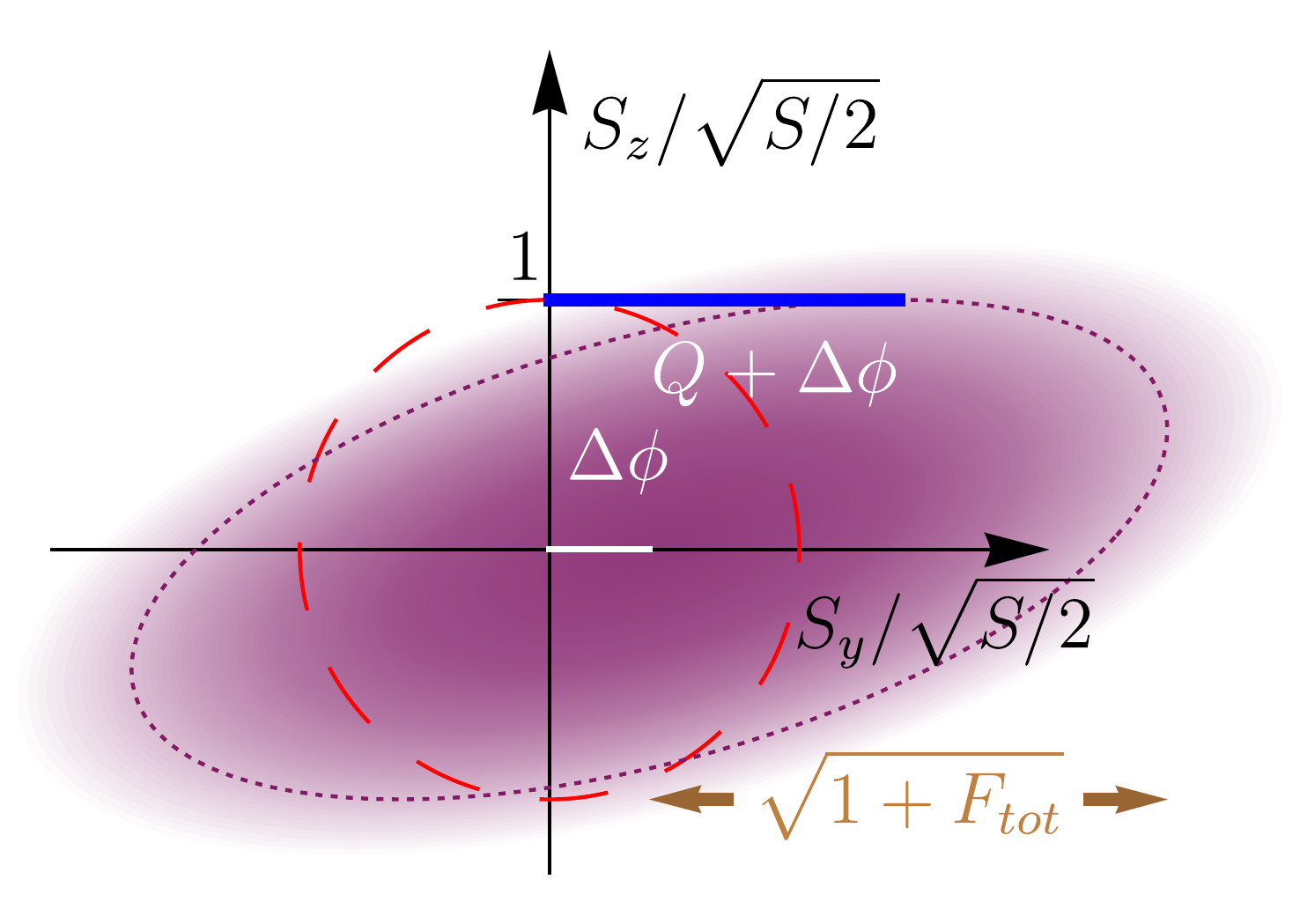}}
\put(0.236,0.31){{Projecting}}
\put(0.26,0.29){\includegraphics[width=.1\columnwidth]{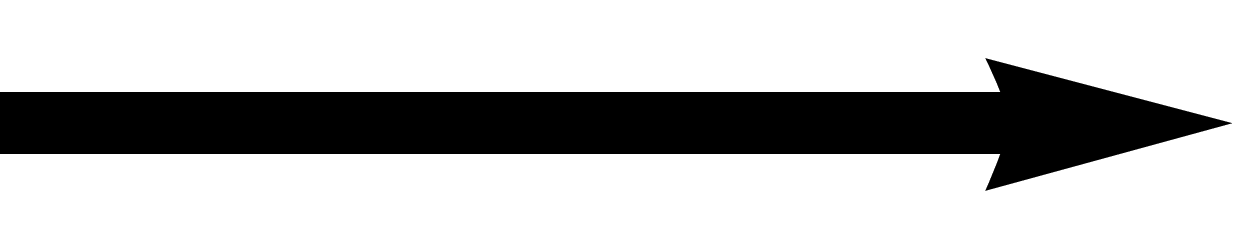}}
\put(0.61,0.33){{Envelope}}
\put(0.58,0.31){{(Information Loss)}}
\put(0.62,0.29){\includegraphics[width=.1\columnwidth]{figure/arrow.pdf}}
\put(0.087,0.2){{\makebox(0,0)[c]{\rotatebox{-90}{Phase Shift}}}}
\put(0.105,0.2){{\makebox(0,0)[c]{\rotatebox{-90}{First Order}}}}
\put(0.07,0.2){{\makebox(0,0)[c]{\rotatebox{-90}{\includegraphics[width=.1\columnwidth]{figure/arrow.pdf}}}}}
\put(0.28,0.213){{\makebox(0,0)[c]{\rotatebox{-45}{Squeezing}}}}
\put(0.275,0.2){{\makebox(0,0)[c]{\rotatebox{-45}{\includegraphics[width=.1\columnwidth]{figure/arrow.pdf}}}}}
\put(0.64,0.14){{Together}}
\put(0.62,0.1){\includegraphics[width=.25\columnwidth]{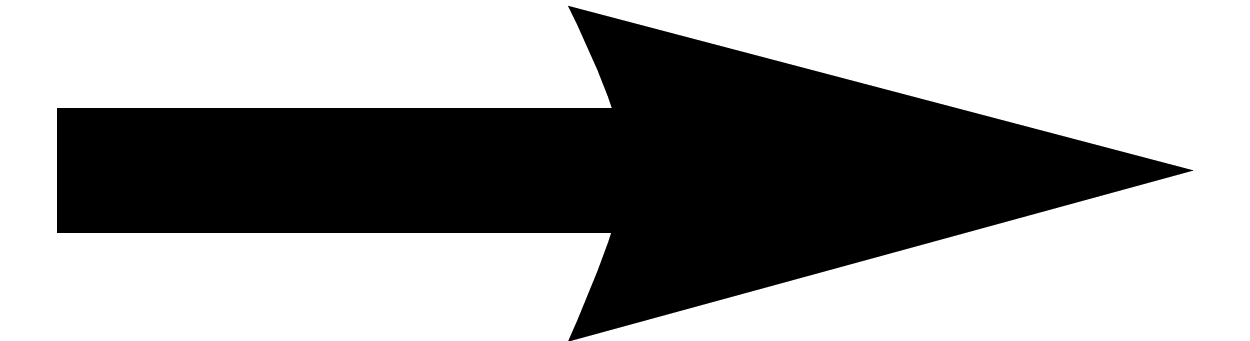}}
\put(0.03,0.38){(a)}
\put(0.35,0.38){(b)}
\put(0.68,0.38){(c)}
\put(0.03,0.17){(d)}
\put(0.35,0.17){(e)}
\put(0.68,0.17){(f)}
\put(0.3,0.35){{$\displaystyle\sum_{\langle S_z\rangle}P(\langle S_z\rangle)\times$}}
\put(0.13,0.22){\textcolor{red}{$(\cos\theta, \sin\theta)$}}
\put(0.35,0.25){\textcolor[rgb]{0.8,0.2,0.1}{$(\sqrt{1+F_\mathrm{tot}}\cos\theta, $}}
\put(0.35,0.23){\textcolor[rgb]{0.8,0.2,0.1}{$\langle S_z\rangle+(\sqrt{1+F_\mathrm{use}})^{-1}\sin\theta)$}}
\put(0.35,0.02){\textcolor{blue}{$(\cos\theta+Q\sin\theta, \sin\theta)$}}
\put(0.68,0.22){\textcolor{brown}{$(\sqrt{1+F_\mathrm{tot}}\cos\theta, \sin\theta)$}}
\put(0.03,0.02){\color[rgb]{0.2,0.6,0.4}{$(\cos\theta+\Delta\phi, \sin\theta)$}}
\put(0.68,0.025){\color[rgb]{0.5,0.1,0.4}{$(\sqrt{1+F}\cos\theta$}}
\put(0.68,0.005){\color[rgb]{0.5,0.1,0.4}{$+Q\sin\theta+\Delta\phi, \sin\theta)$}}
\begin{tikzpicture}
\draw[thick] (0,0) -- (11,0) -- (11,3.5) -- (17.85,3.5) -- (17.85,7.2) -- (0,7.2) -- (0,0);
\end{tikzpicture}
\end{picture}
    \caption{Illustration of the effects of the atom-light interaction on a CSS. (a) Original CSS on the equator of the Bloch sphere, $\aver{S_z}=0$. (b) The effect of a single-shot optical measurement (see text) with usable QFI $F_\mathrm{use}$ and total QFI $F_\mathrm{tot}$, which displaces the state along $S_z$ based on the measurement outcome and yields a resolution beyond the SQL (measurement-based spin squeezing), while broadening it along $S_y$ due to the measurement backaction. (c) The envelope arising from averaging over many repeated measurements or equivalently, tracing over the light field, showing broadening along $S_y$. (d) Phase shift $\Delta\phi$ due to the atom-light interaction, Eq.~\eqref{eq:CavityAtomsStarkShiftTwoLevelAtomPhaseShiftVsTransmission}. (e) The effect of single-axis twisting with shearing strength $Q$. (f) The complete final state for cavity spin squeezing including all the discussed effects: First-order phase shift, shearing, and state broadening due to unused information $F$ contained in the light field. The equations below each plot give parametric expressions $(\tilde{S}_y(\theta),\tilde{S}_z(\theta))$ with $\theta\in[0,2\pi)$ for the uncertainty ellipses of the atomic states which are indicated by dotted lines in the same colors as the states. In (c-f) the dashed red circle represents the uncertainty circle for the original CSS for comparison.}
    \label{fig:QFeffectonstate}
\end{figure*}

\subsection{Measurement Backaction}\label{sec:Fbroadening}

In quantum mechanics, any process that yields information about some variables is associated with backaction onto any non-commuting variables to maintain the uncertainty relation \cite{Clerk2010,Elliott2015,Yang2019, Minev2019}. If that information is not used, the uncertainty relation becomes a true inequality, and the initially pure quantum state turns into a mixed state, exhibiting excess broadening. In the case of spin squeezing, the effective Hamiltonian, Eq.~\eqref{eq:atomicHamiltonian}, entangles the $S_z$ component of the atomic spin with the intracavity photon number, such that the light field leaving the cavity serves as a meter of $S_z$. Hence in this case one can think of the unused information contained in the light field as causing additional broadening in the $S_z$ direction compared to the information about $S_z$ that would be available to a perfect observer. From an alternative viewpoint, the original $S_z$ distribution remains unchanged, but the photon shot noise of the light causes an additional broadening of $S_y$ compared to a (noiseless) classical light field due to the uncertainty in the light shift on the atoms \cite{Schleier-Smith2010a,Leroux2012}. These processes result in a mixed state of the atomic spin with $\Delta S_z \Delta S_y > S/2$.

To quantify those effects, consider a CSS pointing along the $x$-axis as in Fig.~\ref{fig:QFeffectonstate}(a). From Eq.~\eqref{eq:atomicHamiltonian} we know that photons only couple to the $S_z$ quadrature, and therefore the backaction-induced antisqueezing must manifest itself in the $S_y$ quadrature. 
The coherent light probes the physical quantity $S_z$ and therefore generates a back action quantified by the QFI in the light field $\tilde{F}$ (see Eq.~\eqref{eq:FisherInformationForCoherentStates}). We define the normalized QFI of the light,
\begin{equation}\label{eq:NormalizedQFI}
    F=\frac{2}{S} \tilde{F}.
\end{equation}
The quantity $1+F$ represents the factor by which a perfect measurement of the light field can reduce the spin variance below the SQL variance of the CSS, 
\begin{align}
    \left( \Delta S_z \right)^2_\mathrm{cond}=\frac{S}{2 \left(1+ F \right)},
\end{align}
and the back action by the probing light thus causes broadening in the $S_y$ direction, resulting in a total variance
\begin{align}\label{eq:ExcessiveBroadening}
    \left( \Delta S_y \right)^2=\frac{S}{2} \left(1+ F \right),
\end{align}
Here the first term $S/2$ accounts for the variance of the original CSS, while the second term $FS/2$ is due to the back action.

The complete information about the spin state of the ensemble is contained in the field amplitudes and phases of the transmitted light, reflected light, and free-space scattering. 
We note that the electric field amplitude $\mathcal{E}$ is related to the coherent-state index $\alpha$ in Sec.~\ref{sec:FI} via $\mathcal{E}=\alpha\sqrt{2\hbar\omega/\tau}$, where $\tau$ is the probing time. Therefore, we can use Eqs. \eqref{eq:TransmittedFieldFinalResult}, \eqref{eq:ReflectedFieldFinalResult} and \eqref{eq:FisherInformationInXEstimateWithLight} to evaluate the total QFI about the atomic state that is available in the light field:
\begin{align}\label{eq:TotalCavityFisherInformationCalc1}
\begin{split}
F =
\frac{\langle \hat{N}_\uparrow\rangle\tau}{\hbar\omega} \left(T_2 +  T_1 R_2 +  \frac{\pi}{\mathcal{F}} \langle \hat{N}_\uparrow\rangle\eta
\mathcal{L}_a(\y) \right)
\left|\frac{\partial \mathcal{E}_c}{\partial N_\uparrow} \right|^{2}.
\end{split}
\end{align}

Using the expression for the intracavity field, Eq.~\eqref{eq:CavityFieldFinalResultRWA}, we can simplify \eqref{eq:TotalCavityFisherInformationCalc1} to obtain
\begin{align}\label{eq:QFIforBroadening}\begin{split}
    F &= 2\,n_\gamma^\mathrm{sc}\,\eta\,\mathcal{L}_a(\y)
\left(1+\y^2+\frac{S \eta}{2}\right)\mathcal{T}_0\\
 &={2\,n_\gamma^\mathrm{sc}\eta\mathcal{L}_a(\y)\left(1+\y^2+\frac{\nu}{4}\right)\mathcal{T}_0, }
\end{split}
\end{align}
where $\nu=N\eta$ and $\aver{N_\uparrow}=N/2=S$ has been substituted. 
For fixed $\langle N_\uparrow\rangle\eta$, the spectrum of the coupled atom-cavity system \eqref{eq:CavityFieldFinalResultRWA}-\eqref{eq:ReflectedFieldFinalResult} remains the same, while according to \eqref{eq:QFIforBroadening} the QFI carried by photons scales linearly with the cooperativity $\eta$. 
This equation also demonstrates an important principle of optical measurements: The information available about the system is inevitably proportional to the average number of photons that have been scattered into free space $n_\gamma^\mathrm{sc}$. The higher the cavity cooperativity $\eta$, the more information about the atomic system can be extracted per scattered photon. Note that in the limit of small photon number it is possible to gain more information per scattered photon by the use of postselected measurements, see, e.g., Refs.~\cite{McConnell2015,Haas2014}. However, averaged over many realizations of the measurement, the available information gain is proportional to the scattered photon number, as expressed by Eq.~\eqref{eq:QFIforBroadening}.
When this information $F$ is extracted from the light field, it can serve to conditionally squeeze the atomic spin noise in the $S_z$ direction by a factor $(1+F)^{-1}$, i.e. $\left(\Delta S_z \right)^2_{cond}=\frac{S/2}{1+F}$. At the same time, $\Delta S_y$ is increased by the back action to $\left(\Delta S_y \right)^2 = \frac{S}{2}\left(1+F \right)$, thus maintaining the Heisenberg uncertainty for angular momentum, $\Delta S_y \Delta S_z \geq \frac{S}{2}$. If the information in the light field remains unused, then the increased $\Delta S_y$ describes the broadening of the state along $S_y$ that at the microscopic level is due to the photon shot noise in the incident light.

The complete QFI expressed by Eq.~\eqref{eq:QFIforBroadening} can further be separated into two parts. The first two terms inside the parenthesis correspond to the information contained in the transmitted and reflected light fields that do not provide access to the state of any individual atom, but only to the collective spin, i.e. the symmetric state of the atoms. On the other hand, the last term which is proportional to the collective cooperativity $S \eta$, originates from free-space scattering, and reveals local spin fluctuations that take the system out of the symmetric spin space.

With the help of Eq.~\eqref{eq:CavityAtomsQ}, we can write a relation between the Fisher information $F$ and the squeezing parameter $Q$:
\begin{align}\label{eq:FQRelation}
    \frac{Q}{F} = \y \frac{1 - \x \y + S \eta}{1 + \y^2 + \frac{S \eta}{2}}.
\end{align}

\subsection{Gaussian State Representation}\label{sec:GaussianRep}

In this section we develop a simple formalism to describe the transformation of Gaussian states in the limit where the state occupies a small region on the Bloch sphere. Collective spin states near the CSS are good examples of Gaussian states that have significant roles in quantum information science \cite{Weedbrook2012}. 
The CSS state can be described by a symmetric Gaussian distribution when projected along any spin quadrature perpendicular to the spin's mean direction. When the shearing parameter $Q$ and the QFI $F$ are small, the local Bloch sphere can be approximated by a 2D Cartesian frame, and the state can be represented by a Gaussian distribution. This also corresponds to the Holstein-Primakoff approximation~\cite{Holstein1940}. 

The CSS along the $x$-axis has a Wigner quasi-probability distribution $P(S_y,S_z)\propto \mathrm{e}^{-2(S_y^2+S_z^2)/N}$, or, 
\begin{align}\label{eq:CovarianceSQL}
    P(S_y,S_z)\propto\mathrm{exp}\left[-\frac{1}{2}
    \begin{matrix}
    ( S_y & S_z )\\
    &
    \end{matrix}
    \Sigma^{-1}
    \left(\begin{matrix}
    S_y\\
    S_z
    \end{matrix}\right)\right],
\end{align}
where the inverse covariance matrix $\Sigma^{-1}$ is an identity matrix with a prefactor of the standard quantum limit (SQL) unit of variance: $\Sigma^{-1}=\left(2/S\right)\mathbb{I}_2$. Other Gaussian states that are centered along the $x$-axis can all be described similarly but with a different inverse covariance matrix.

\subsection{The Effect of the Probing Light on the Representation of the Atomic State in the Gaussian Approximation}\label{sec:Ellipses}

Starting from a CSS along the $x$ axis, entangled atomic states can be generated via a Hamiltonian unitary evolution as discussed in Sec.~\ref{sec:ALI}, or by a measurement as described in Sec.~\ref{sec:Fbroadening}. 

We first consider the one-axis twisting, $\hat{H}_2=-\hbar\chi\hat{S}_z^2$. In an infinitesimal time step $\mathrm{d}t$, 
the one-axis twisting transforms a point on the collective Bloch sphere according to
\[\left(\begin{matrix}
    S_y(t+\mathrm{d}t)\\ S_z(t+\mathrm{d}t)
    \end{matrix}\right) = \left(\begin{matrix}
    1 & N\chi\mathrm{d}t \\
    0 & 1
    \end{matrix}\right)\left(\begin{matrix}
    S_y(t)\\ S_z(t)
    \end{matrix}\right), \]
which generates a unitary evolution of the covariance matrix as 
\begin{align}\label{eq:covarianceUnitary}
    \Sigma(t+\mathrm{d}t)= U\Sigma(t)U^{\dagger},
\end{align}
where
\[U = \left(\begin{matrix}
    1 & N\chi\mathrm{d}t \\
    0 & 1
    \end{matrix}\right).  \]

When we include the non-unitary effect of the QFI, the covariance matrix transforms such that the product of eigenvalues is no longer normalized to the SQL. With constant strength of measurement, i.e. a QFI that increases linearly in time, the broadening along the $S_y$ axis is also linear in time, i.e., the $(1,1)$ element of the covariance matrix evolves as
\begin{align}\label{eq:covarianceNonUnitary}
    \Sigma(1,1)(t+\mathrm{d}t)=\Sigma(1,1)(t)+\dot{F}\mathrm{d}t
\end{align}
where $\dot{F}$ is the normalized QFI rate, and with a constant probing rate, $\dot{F}=F/t$ and $\dot{Q}=Q/t=N\chi$. It turns out that the effects described by Eqs. \eqref{eq:covarianceUnitary} and \eqref{eq:covarianceNonUnitary} commute with each other, giving 
\begin{align}\label{eq:CovarianceWithFandQ}
    \Sigma=\frac{N}{4}\left(\begin{matrix}1+F+Q^2 & Q\\
    Q & 1
    \end{matrix}\right). 
\end{align}

We can apply the above discussion of the covariance matrix to the parametric description of the $1/e$ boundary of the quasi-probability distribution on the Bloch sphere. For example, the squeezing transforms the curve $(\cos\theta, \sin\theta)^\mathrm{T}$ to $U\cdot(\cos\theta, \sin\theta)^\mathrm{T} = (\cos\theta+N\chi\mathrm{d}t\sin\theta, \sin\theta)^\mathrm{T}$, as shown in Fig.~\ref{fig:QFeffectonstate}~(e). For the effect of QFI induced broadening, using \eqref{eq:CovarianceSQL} we have that the new boundary is given by
\[\tilde{S}_y^2/(1+\dot{F}\mathrm{d}t)+\tilde{S}_z^2=1, \]
which corresponds to a parametric description of $(\tilde{S}_y,\tilde{S}_z)=(\sqrt{1+\dot{F}\mathrm{d}t}\cos\theta,\sin\theta)$ where $\theta\in[0,2\pi)$. Note that for individual measurements, the state is actually projected to a random state centered at $(\langle S_z\rangle,0)$ with a probability $P(\langle S_z\rangle)\propto\exp(-2S_z^2/N)$ and 
a variance along the $z$-direction that is reduced by a factor of $1+F_\mathrm{use}$. Here $F_\mathrm{use}$ is the usable fraction of the total QFI, 
but broadened along the $y$-direction as quantified by the same $F_\mathrm{tot}$ we discussed above. This individual projection looks like the one in Fig.~\ref{fig:QFeffectonstate}~(b); however, if we only care about the deterministic effect
where the light state is traced over, the effect is depicted as in Fig.~\ref{fig:QFeffectonstate}~(c). 

In addition, in this description the atomic state can deviate from the $x$-axis by the first order phase shift effect $\hat{H}_0=-\hbar\Omega\langle\hat{n}_c\rangle\hat{S}_z$, which translates $(\cos\theta,\sin\theta)$ to $(\cos\theta+\Delta\tilde{\phi}, \sin\theta)$, as in Fig.~\ref{fig:QFeffectonstate}~(d). Here $\Delta\tilde{\phi}$ is the normalized phase shift and relates to the phase shift $\Delta\phi$ in \eqref{eq:CavityAtomsStarkShiftTwoLevelAtomPhaseShiftVsTransmission} by $\Delta\tilde{\phi} = \sqrt{N}\Delta\phi$.

For a more general discussion of the Gaussian representation, please refer to the Appendix~\ref{sec:Gaussian}. 

\subsection{Contrast loss induced by photon scattering and Bloch sphere curvature}\label{sec:Contrast}

When one photon is scattered by the atomic ensemble into free space, the particular atom that scattered this photon gains a random phase (or equivalently, is projected into a particular internal state) and therefore no longer retains coherence with the other atoms. 
Given $n_\gamma^\mathrm{sc}$ scattered photons and an average number of scattered photon per atom $p=n_\gamma^\mathrm{sc}/N$, the Poissonian probability for any particular atom to not have scattered any photons is given by
\begin{align}\label{eq:ContrastLossSC}
    C_\mathrm{sc}=\frac{p^0e^{-p}}{0!}=e^{-n_\gamma^\mathrm{sc}/N}. 
\end{align}
This is also the contrast limit of a phase sensitive atomic signal, as measured, e.g., in a Ramsey sequence, due to photon scattering. As we have seen above (Eq.~\eqref{eq:CavityAtomsQ}), light-induced spin squeezing is impossible without scattering photons, and thus some contrast loss is inevitable. Therefore, in practical applications, one must carefully balance the potential metrological gain provided by spin squeezing against the detrimental effects of decoherence. 

Besides the scattering-induced contrast loss, there exists another type of contrast loss that is due to the antisqueezing of the state around the Bloch sphere, perpendicular to the squeezed direction, which shortens the average collective spin vector $|\aver{\mathbf{S}}|=C_\mathrm{Bloch} S$. (This process occurs even in the absence of photon scattering simply due to the deformation of the state.) As described in Appendix \ref{sec:AppendixContrastLoss}, this contrast loss is well approximated by $C_\mathrm{Bloch} = \exp\left(-\frac{Q^2}{2N}\right)$. Combining those two contrast loss mechanisms, we obtain the total contrast as
\begin{align}\label{eq:ContrastLoss}
    C=C_\mathrm{sc}C_\mathrm{Bloch}=\exp\left(-\frac{n_\gamma^\mathrm{sc}+Q^2/2}{N}\right).
\end{align}
This fundamental contrast reduction for phase-sensitive measurements affects the performance of spin squeezed atomic clocks \cite{Hosten2016, Cox2016a, Pedrozo2020,huang2020self} and other interference-based atomic sensors \cite{Appel2009,sewell2012magnetic,Braverman2019,bao2020spin}.

\subsection{Wineland parameter}\label{sec:WinelandParameter}

The combined effects of light-induced measurement with corresponding backaction, atom-atom interaction, and contrast loss on atomic states can produce a variety of final atomic states. We quantify the utility of these SSSs for metrological applications using the Wineland parameter \cite{Wineland1992, Wineland1994}
\begin{align}\label{eq:WinelandDefinition}
    \xi^{2}=\frac{\xi^2_\mathrm{KU}}{C^2} = \frac{1}{C^2}\frac{\Delta\theta^2}{\Delta\theta_\mathrm{SQL}^2}, 
\end{align}
where $\Delta\theta^2$ is the variance of the phase estimation by the quantum system, $\Delta\theta_\mathrm{SQL}^2=(2S)^{-1}$ is the same quantity for uncorrelated particles in the absence of contrast loss, which also corresponds to the SQL in units of phase, $C$ is the contrast, and $\xi^2_\mathrm{KU} = \Delta\theta^2/ \Delta\theta_\mathrm{SQL}^2$ is the Kitagawa-Ueda parameter \cite{Kitagawa1993} which quantifies the metrological gain without contrast loss. 

In the approach discussed above in Section \ref{sec:Ellipses}, the reduction of collective spin projection noise 
\cite{Wineland1993} of the final state depends both on $Q$ and the normalized QFI, $F$. The Kitagawa-Ueda parameter is given by the smaller eigenvalue of the covariance matrix \eqref{eq:CovarianceWithFandQ}: 
\begin{equation}\label{eq:UsableSqueezingFromQandI}
\xi^2_\mathrm{KU} = \frac{2+F+Q^2-\sqrt{4Q^2+(F+Q^2)^2}}{2},
\end{equation}
and in the limit of $Q\gg F\sim1$,
\[\begin{split}
    \xi^2_\mathrm{KU}
    &\sim\frac{1+F}{Q^2}. 
\end{split}\]
\setlength{\unitlength}{\columnwidth}
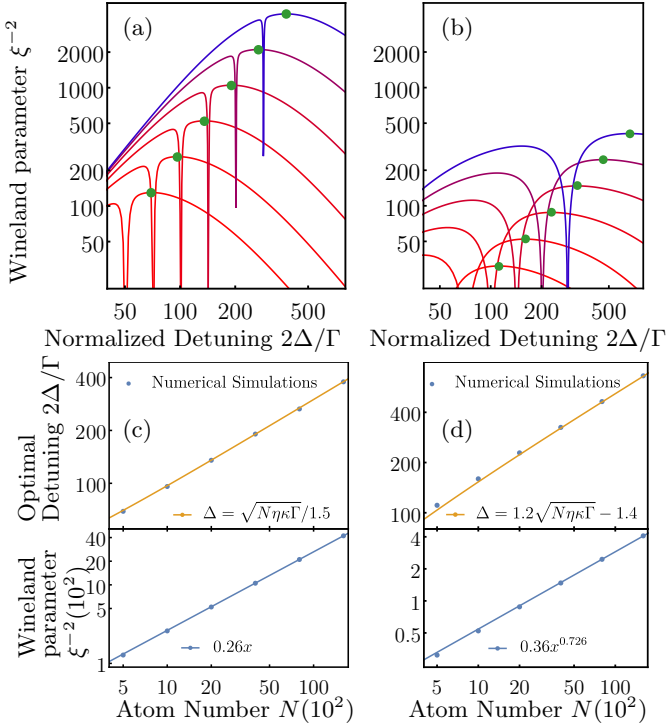
\begin{figure}
\begin{picture}(1,1.1)
\put(0.06,0.6){\includegraphics[height=.5\columnwidth]{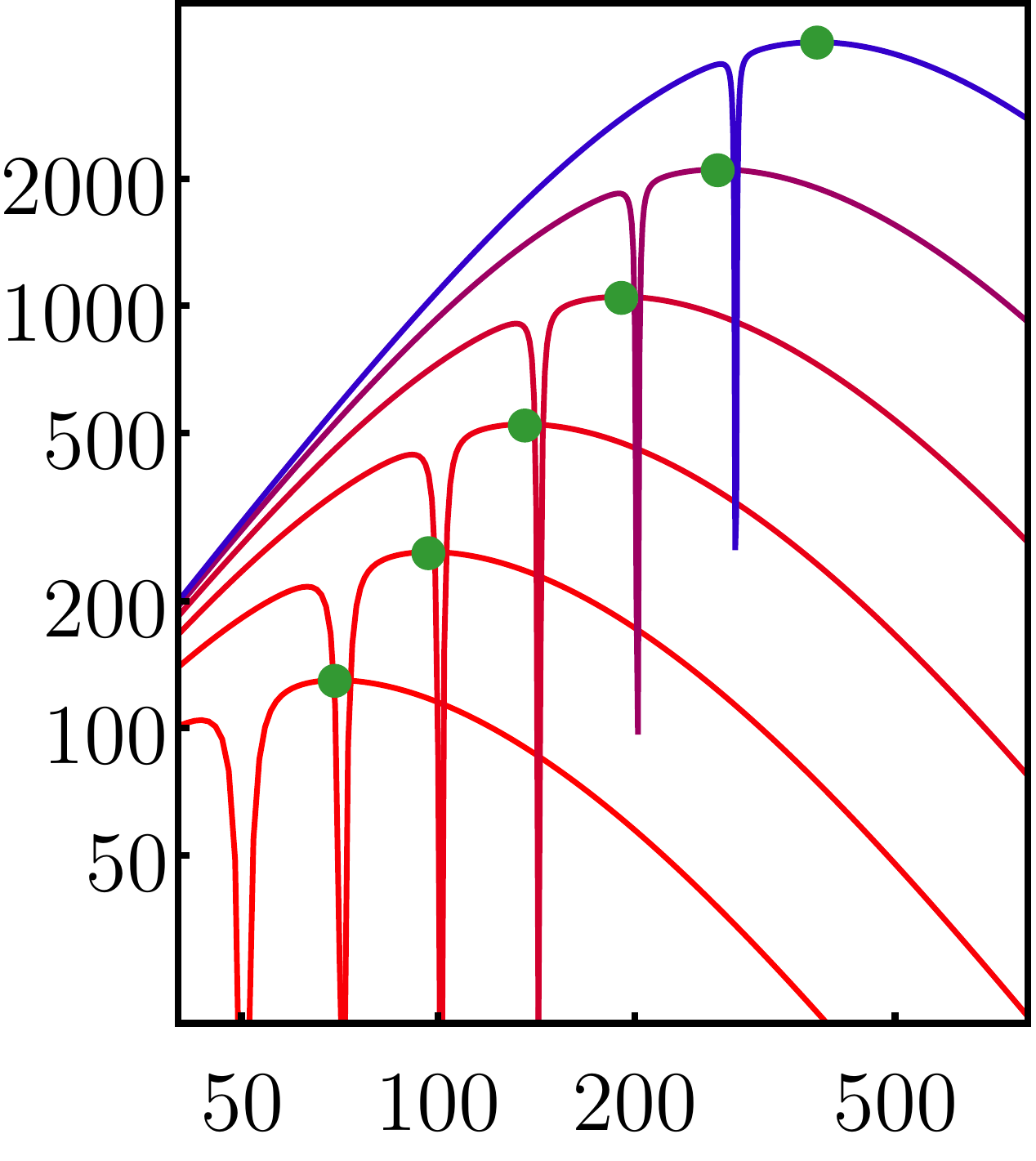}}
\put(0.545,0.6){\includegraphics[height=.5\columnwidth]{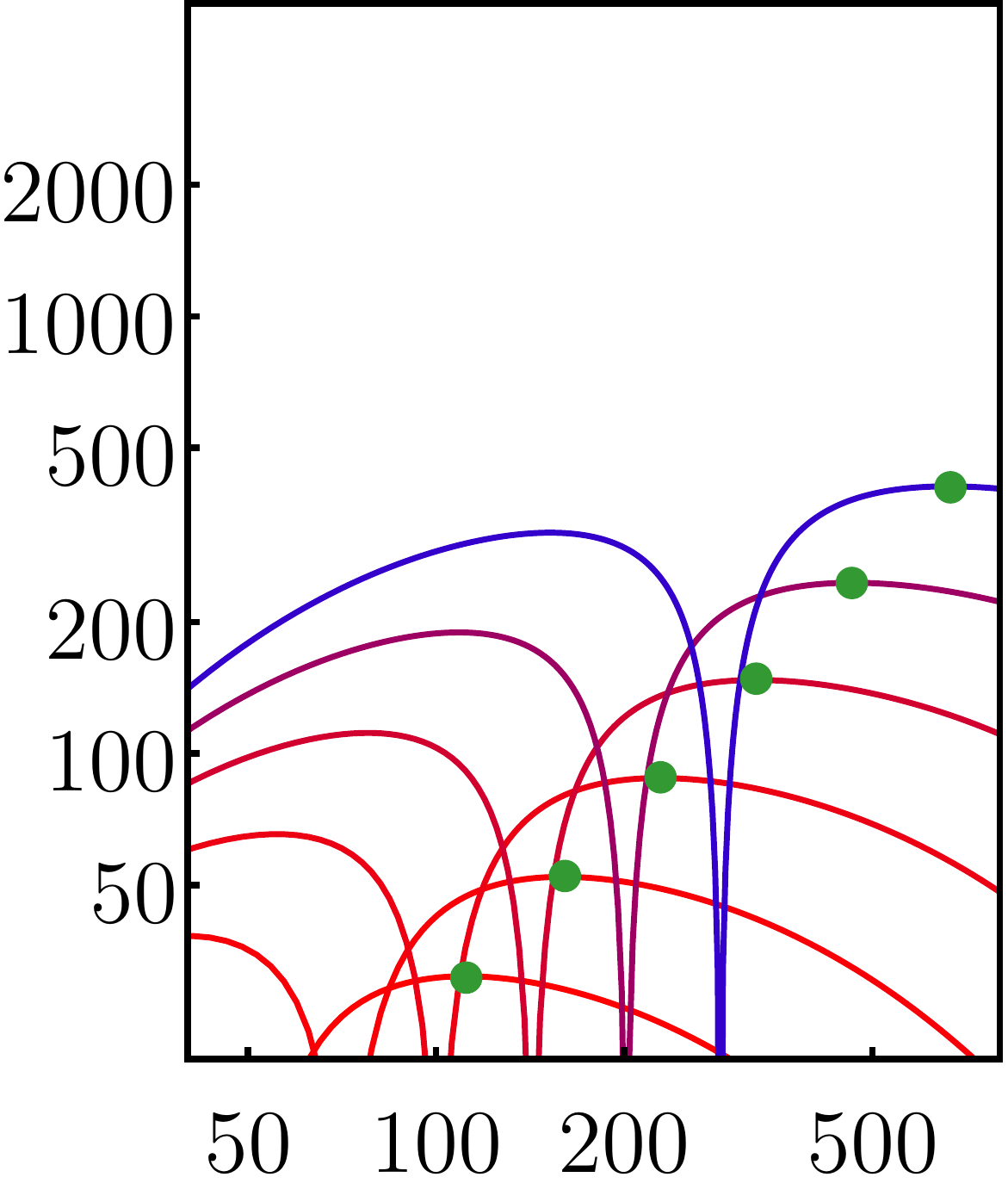}}
\put(0.16,1.05){(a)}
\put(0.65,1.05){(b)}
\put(0.16,0.43){(c)}
\put(0.65,0.43){(d)}
\put(0.04,0.57){Normalized Detuning $2\Delta/\Gamma$}
\put(0.54,0.57){Normalized Detuning $2\Delta/\Gamma$}
\put(0.,0.87){{\makebox(0,0)[c]{\rotatebox{90}{Wineland parameter $\xi^{-2}$}}}}
\put(0.081,0.032){\includegraphics[height=.514\columnwidth]{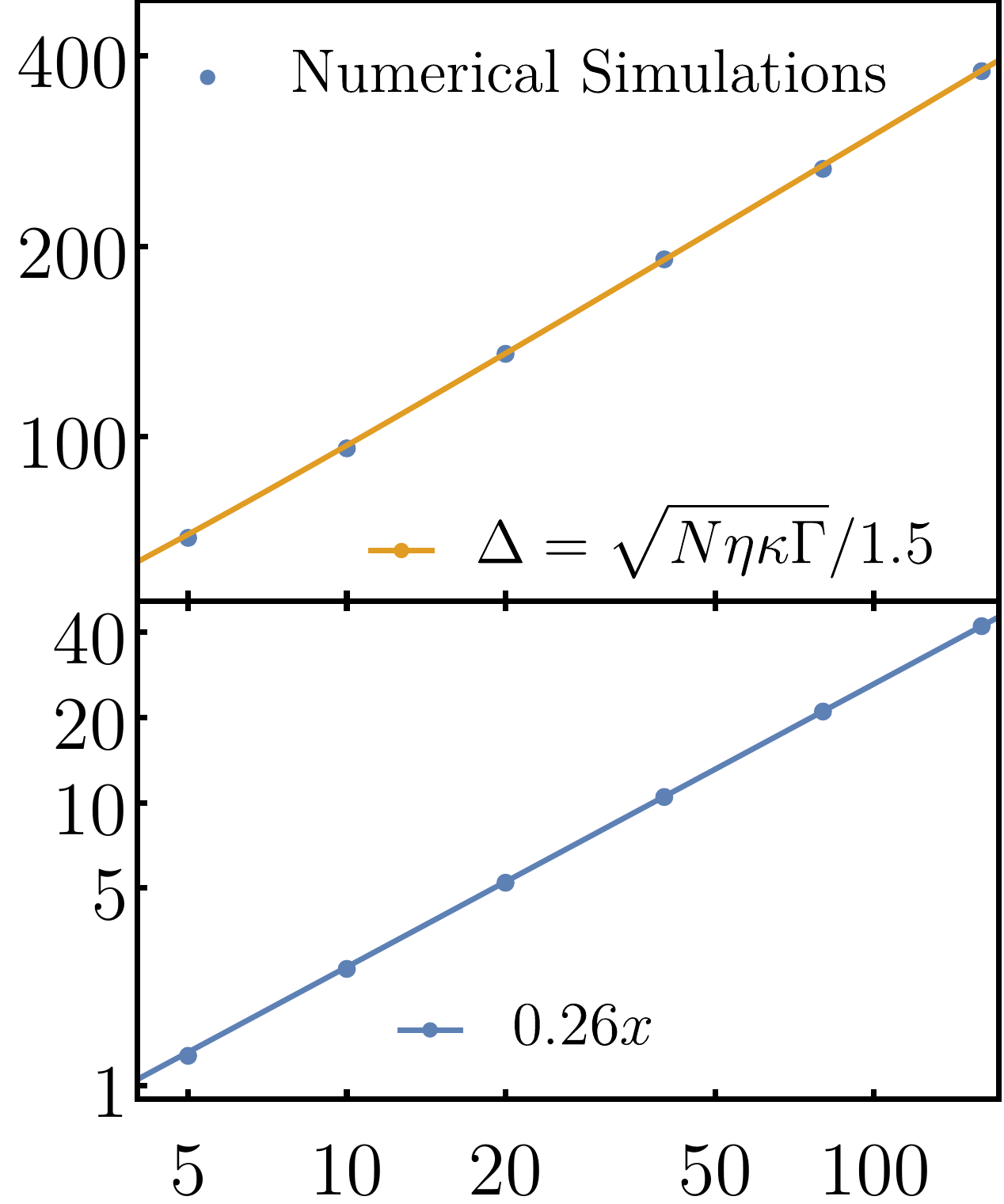}}
\put(0.565,0.032){\includegraphics[height=.514\columnwidth]{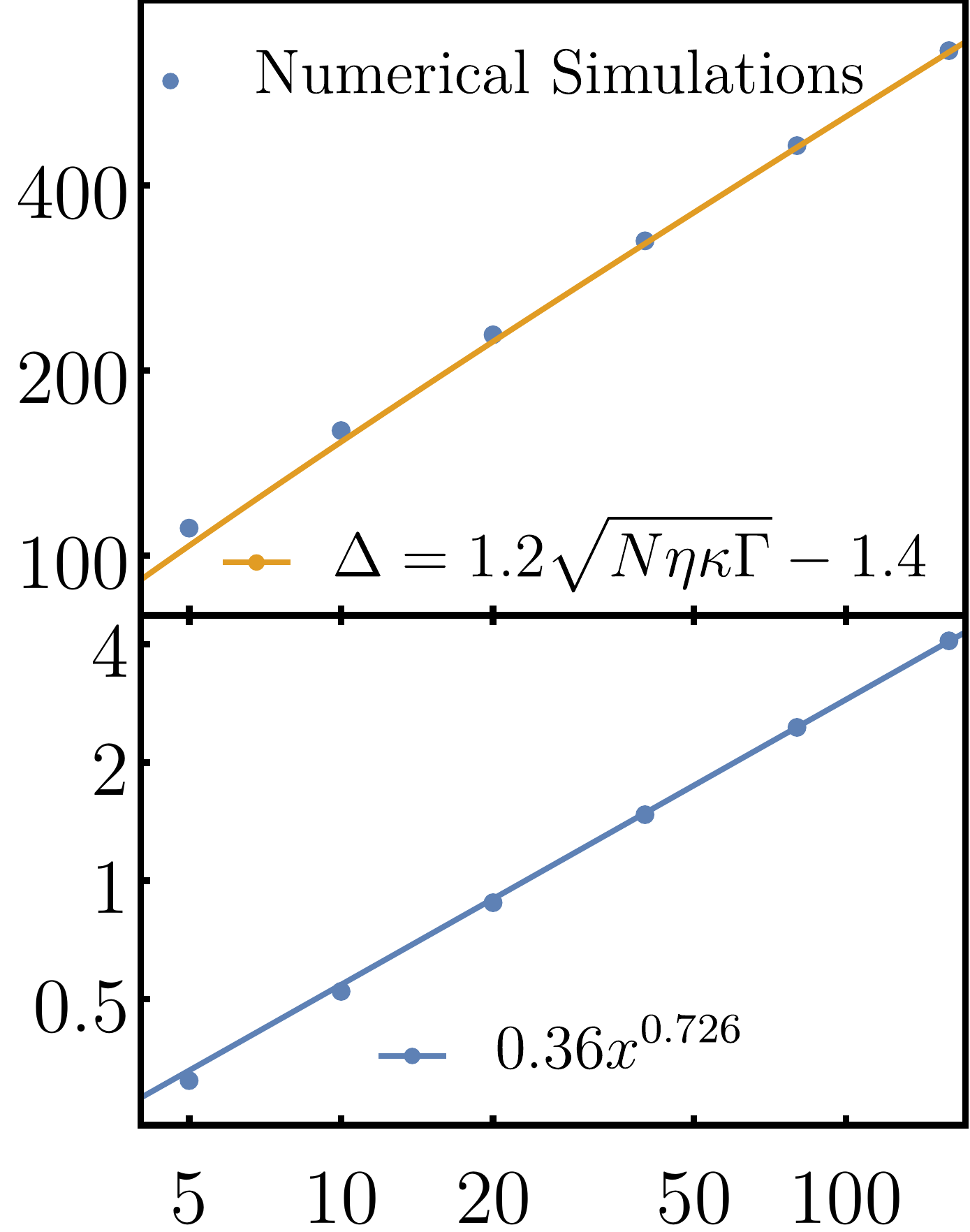}}
\put(0.0,0.3){{\makebox(0,0)[lb]{\rotatebox{90}{Optimal}}}}
\put(0.035,0.3){{\makebox(0,0)[lb]{\rotatebox{90}{Detuning $2\Delta/\Gamma$}}}}
\put(0.0,0.09){{\makebox(0,0)[lb]{\rotatebox{90}{Wineland}}}}
\put(0.035,0.09){{\makebox(0,0)[lb]{\rotatebox{90}{parameter}}}}
\put(0.07,0.09){{\makebox(0,0)[lb]{\rotatebox{90}{$\xi^{-2}(10^2)$}}}}
\put(0.15,0.0){Atom Number $N(10^2)$}
\put(0.6,0.0){Atom Number $N(10^2)$}
\end{picture}
\caption{(a, b) the optimal Wineland parameter with respect to the optimized photon number at given $N$ and detuning, without (a) and with (b) Bloch sphere curvature effect, where from bottom to top the atom numbers are $N=\{1,2,4,8,16\}\times500$. The single-atom cooperativity is $\eta=1.8$. The green dots indicate the maximum Wineland parameter for a given atom number. (c, d) show the optimum detuning and maximum Wineland parameter as a function of atom number, corresponding to the parameters of the green dots in subfigures (a, b); the data points correspond to numerical simulations and the solid lines represent fits that hold for large atom number $N \gg 1$. }
\label{fig:WinelandParameter}
\end{figure}

We see from Eqs. \eqref{eq:CavityAtomsQ} and \eqref{eq:QFIforBroadening} that for fixed probing parameters $\delta$, $\Delta$, and $N$, both the squeezing parameter $Q$ and the QFI $F$ scale linearly with the number of photons scattered into free space. This means that as long as the QFI is small, i.e., $F<1$, the amount of squeezing $\xi^2$ will grow quadratically with the number of scattered photons, but when $F\gtrsim1$, the scaling will become linear. However, for sizeable OAT squeezing, $Q\sim\mathcal{O}(\sqrt{S})$, the curvature of the Bloch sphere induces an additional broadening of the distribution \cite{Kitagawa1993,Schleier-Smith2010}: 
\begin{align}\label{eq:WinelandWithBloch}
    \xi_\mathrm{KU}^2\to\xi_\mathrm{KU}^2+\frac{Q^4}{24S^2}. 
\end{align}
Thus the variance of the squeezed direction reaches a minimum for an optimal number of scattered photons, as displayed in Fig.~\ref{fig:WinelandFixN}~(b). The increasing widths of the dips around the atom-cavity resonance at $\y=50$ with increasing scattered photon number indicate the effect of the Bloch sphere curvature. 

By fixing the atom number $N$ and centering the atomic state at the equator, $\aver{S_z}=0$, i.e., $\langle \hat{N}_\uparrow\rangle=N/2$, one can obtain an optimal frequency that yields the largest inverse Wineland parameter $\xi^{-2}$, as shown in Fig.~\ref{fig:WinelandParameter}. When varying the atom number, we can see that the optimum detuning is proportional to the vacuum Rabi splitting, as displayed in Fig.~\ref{fig:WinelandParameter}(c). Therefore, no matter how many atoms are loaded, the $Q, F$ parameters per scattered photon (denoted as $\hat{Q}\equiv Q/n_\gamma^\mathrm{sc}, \hat{F}\equiv F/n_\gamma^\mathrm{sc}$) for the optimal detuning remain the same. The optimal Wineland parameter as a function of the scattered photon number $n_\gamma^\mathrm{sc}$ for large $N$ therefore simplifies to
\begin{align}
    \xi^{2}=\frac{1+\hat{F}n_\gamma^\mathrm{sc}}{(\hat{Q}n_\gamma^\mathrm{sc})^2}e^{2 n_\gamma^\mathrm{sc}/N}\sim
    \frac{\hat{F}}{\hat{Q}^2N} p^{-1} e^{2 p}
\end{align}
where $p=n_\gamma^\mathrm{sc}/N$ is the average photon number scattered per atom. The minimum Wineland parameter is obtained for $p=1/2$, and the metrological gain scales as total atom number $N$, i.e., exhibits Heisenberg scaling \cite{Colombo2021}. 
If we consider the curvature effect of the Bloch sphere, Eq.~\eqref{eq:WinelandWithBloch}, then the metrological gain exhibits a scaling with atom number as $N^{0.73}$, as shown in Fig.~\ref{fig:WinelandParameter} (d) based on numerical results. For the same collective cooperativity $N\eta$, the higher the single-atom cooperativity $\eta$, i.e. the smaller the atom number, the more severely the finite size of the Bloch sphere limits the Wineland parameter. We note, however, that there are approaches beyond spin squeezing that are not limited by the curvature of the Bloch sphere
\cite{Davis2016,Schulte2020,Gilmore2021,Colombo2021}.

\subsection{Brief Summary}
In the preceding subsections~\ref{sec:FI} to \ref{sec:WinelandParameter}, we have derived analytical results for the cavity-induced effects on the collective atomic state. Those effects are summarized in Fig.~\ref{fig:QFeffectonstate}. 
The first effect is induced by the QFI carried by the photons leaving the system. The photons that are transmitted, reflected, or scattered into free space all carry QFI about the atomic state. The transmitted and reflected photons only carry information about the collective spin state of the atoms, and a measurement of those fields projects the atomic system into a Gaussian state with a narrowed $S_z$ distribution. This atom-light entanglement is the foundation of measurement-based squeezing \cite{Kuzmich1998,Vasilakis2015}; however, the final obtained state is not deterministic, but depends on the measurement result. The distribution along $S_y$ is broadened not only by the backaction associated with the QFI of the transmitted and reflected light, but also by the QFI encoded in the photons scattered into the free space. The latter carry information about the spatial structure of the atomic spin, and not only the collective spin. As such, they take the system out of the Hilbert space associated with the maximum spin $|\bm{S}|=\sqrt{S(S+1)}$. 

If one performs measurement-based squeezing, the transmitted and/or reflected photons carry QFI that can be used to infer the atomic $S_z$ value. Conditioned on the measurement result, the collective atomic state is projected into a state with some mean value of $\aver{S_z}$ sampled from the original CSS distribution of $S_z$ and reduced state variance $(\Delta S_z)^2$. Note that in this scenario, the photons scattered into free space take the system out of the maximally symmetric subspace; nevertheless, these photons still contribute to the total QFI $F_\mathrm{tot}$ which causes excessive broadening in the antisqueezing direction $S_y$, just as in the feedback-squeezing case. 

For the cavity feedback squeezing, which aims at deterministic state preparation, the information contained in the light field is typically not collected or used. In other words, the light pulses project the states into $S_z$ distributions of reduced width with a $\aver{S_z}$ value that is different on each trial, and unknown to the experimenter. The amount of unused QFI is then given by Eq.~\eqref{eq:QFIforBroadening}:
\[F = 2\,n_\gamma^\mathrm{sc}\,\eta\,\mathcal{L}_a(\y)
\left(1+\y^2+\frac{S \eta}{2}\right)\mathcal{T}_0\]
which causes excess broadening in the $S_y$ direction (Eq.~\eqref{eq:ExcessiveBroadening})
\[(\Delta S_y)^2=\frac{S}{2}(1+F).  \]

On the other hand, the light pulse also generates unitary evolution of the collective atomic states, i.e. the zeroth-order collective phase shift and first-order spin squeezing as in \eqref{eq:CavityAtomsStarkShiftTwoLevelAtomPhaseShiftVsTransmission} and \eqref{eq:CavityAtomsQ}:
\[\Delta\phi = -\frac{n_\gamma^\mathrm{sc}}{N}\y, \]
\[Q = -{2n_\gamma^\mathrm{sc}}
\eta\mathcal{L}_d(\y)
(1-\x\y+N_\uparrow\eta)\mathcal{T}_0. \]
By means of a spin-echo sequence~\cite{Hahn1950,Braverman2019}, i.e., by adding the $\pi$ pulse between two squeezing pulses, the terms that are odd in $\hat{S}_z$, such as the unwanted phase shift $\Delta \phi$, cancel out, while the even terms, such as the squeezing term, are retained.  

The Wineland parameter was defined as \eqref{eq:WinelandDefinition}
\[\xi^{2} = \frac{1}{C^2}\frac{\Delta\theta^2}{\Delta\theta_\mathrm{SQL}^2}= \frac{2S}{C^2} \Delta\theta^2 ,\]
which describes the metrological gain of the SSS. Without finite-curvature effect, the squeezing produces a Heisenberg scaling of the Wineland parameter, $\xi^2 \propto N^{-1}$, while the finite curvature of the Bloch sphere reduces the scaling to $\xi^2 \propto N^{-0.73}$. 

\section{Atomic-State Detection}\label{sec:Chirp}

The above discussions are based on the assumption that we can measure the collective atomic state sufficiently precisely to fully resolve the spin variance of the squeezed state. In this section, we derive the spin resolution that can be obtained with an optical measurement probing the cavity. A more specific discussion is provided in Ref.~\cite{Measurement2021}. Related approaches to the non-destructive detection of the atom number using a high-finesse cavity have been investigated in Ref.~\cite{Chen2014,Hobson2019}. 

Our goal here is to derive the measurable amount of Fisher information $F_\mathrm{meas}(S_z)$ about the collective atomic spin operator $S_z=N_\uparrow-N_\downarrow$ that is available in the light field. This information is contained in the reflected and transmitted fields $\mathcal{E}_r$ and $\mathcal{E}_t$, but not in the scattered light, so it is not the same as total Fisher information, Eq.~\eqref{eq:TotalCavityFisherInformationCalc1}. The measurement resolution of $S_z$ is thus characterized by the Fisher information contained in the reflected and transmitted fields,
\begin{equation}\label{eq:MeasurableCavityFisherInformation}
F_\mathrm{meas}(S_z) =  4 \tau T_\mathrm{tot}
\left|\frac{\partial \mathcal{E}_c}{\partial S_z} \right|^{2},
\end{equation}
where $T_\mathrm{tot}=T_2$ in the case where only the transmitted photons are measured, and $T_\mathrm{tot}=T_1+R_2T_1\approx T_1+T_2$ when both the transmitted and reflected photons are detected. 
Evaluating Eq.~\eqref{eq:MeasurableCavityFisherInformation} using Eq.~\eqref{eq:CavityFieldFinalResultRWA}, we find
\begin{align}
    F_\mathrm{meas}(S_z) =  4 \tau T\frac{\mathcal{F}^2}{\pi^2}T_1|\mathcal{E}_\mathrm{in}|^2\eta^2\mathcal{L}_a(\y)\mathcal{T}_0^2, 
\end{align}

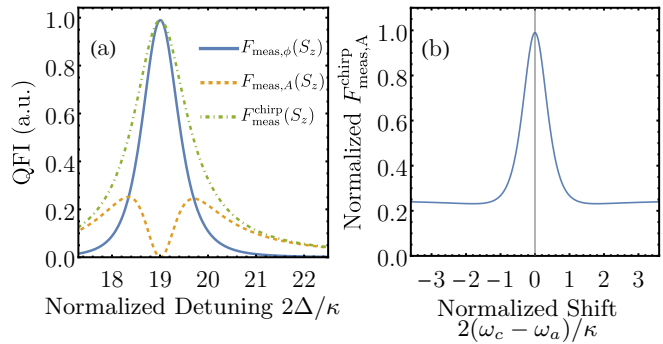
\begin{figure}
    \centering
    \begin{picture}(1,0.495)
        \put(0,0){\includegraphics[width=.495\columnwidth]{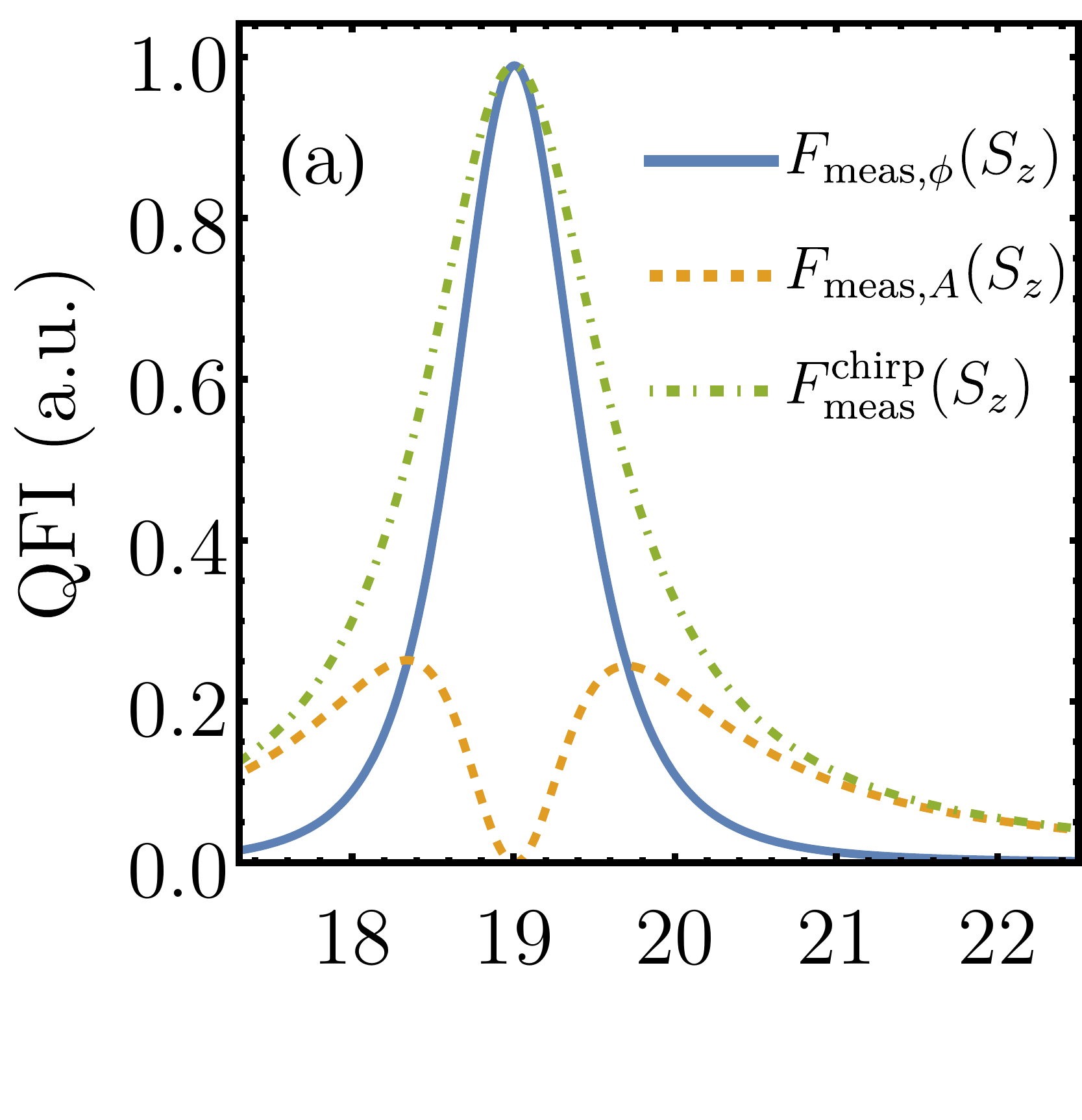}}
        \put(0.505,0){\includegraphics[width=.5\columnwidth]{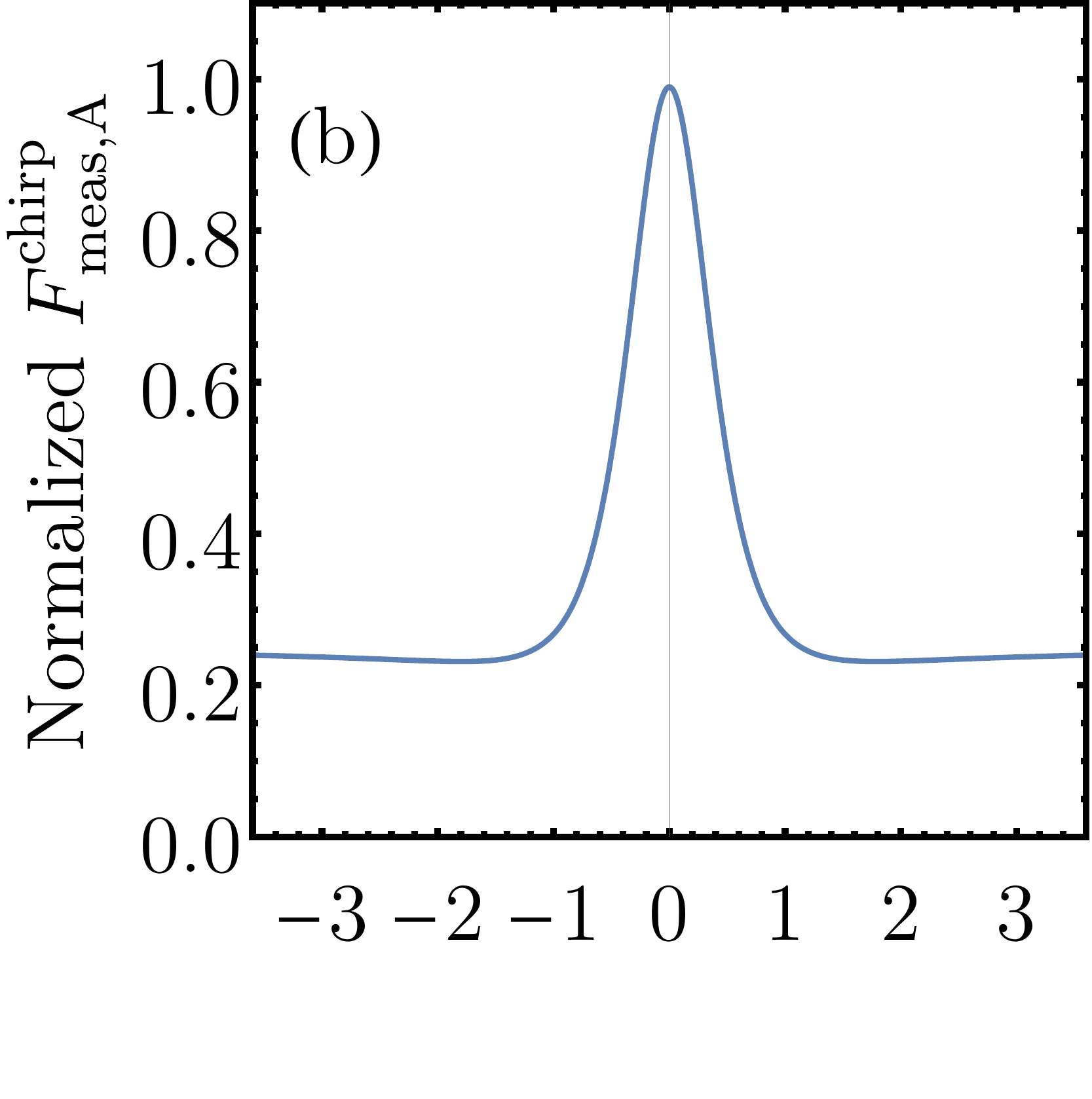}}
        \put(0.285,0.04){{\makebox(0,0)[c]{Normalized Detuning $2\Delta/\kappa$}}}
        \put(0.8,0.04){{\makebox(0,0)[c]{Normalized Shift}}}
        \put(0.8,0.005){{\makebox(0,0)[c]{$2(\omega_c-\omega_a)/\kappa$}}}
    \end{picture}
    \caption{(a) QFI per scattered photon per atom for different types of measurements. The blue solid line and the yellow dashed line correspond to the amplitude and phase components of the total QFI $F_\mathrm{A}$ and $F_\mathrm{\phi}$, respectively. 
    The green dotted line represents the QFI $F_\mathrm{A}^\mathrm{chirp}$ for a chirped two-color measurement of the transmission amplitude through the cavity, which equals the total QFI $F_\mathrm{A}+F_\mathrm{\phi}$; (b) The normalized total QFI for a chirped two-color measurement as a function of the cavity-atom detuning. The QFI is quite sensitive to the symmetry of the transmission lineshape, see text.}
    \label{fig:FI_measure}
\end{figure}

Experimentally, the intensity of the light field can be measured directly by a photodetector, while a measurement of the phase requires interference with a signal that oscillates at a similar optical frequency. As shown in Fig.~\ref{fig:FI_measure}(a), the measurable Fisher information is contained primarily in the phase component of the light field. Usually a homodyne or heterodyne measurement is used to retrieve the information contained in the phase. 
For a situation where the cavity is on resonance with the atomic transition, and the coupled system exhibits a spectrum with Rabi splitting, it is possible to use the transmitted or reflected light in a two-color chirp measurement with two balanced sidebands \cite{Braverman2018a,Braverman2019}. In this measurement scheme, one generates two balanced sidebands with frequencies $\omega_\pm=\omega_c\pm\omega_m$, where $\omega_m$ is the modulation frequency that can be controlled. 
The transmitted field represents an interference of the two components with a beatnote at $2 \omega_m$, which contains the desired phase information. The QFI of the transmitted light in such a measurement is
\begin{align}
    F_\mathrm{t}^\mathrm{chirp}=4 \tau T
\left|\frac{\partial \mathcal{E}_c^\mathrm{chirp}}{\partial S_z} \right|^{2}, 
\end{align}
which equals the sum of of the QFI contained in the two sidebands. In addition to recovering all of the QFI from the transmitted field, this two-color measurement scheme suppresses the effects of laser and cavity noise by probing both cavity resonances at the same time. 
However, this approach requires a symmetric transmission lineshape, as in Fig.~\ref{fig:FI_measure}(b). The simulation is performed with a cavity linewidth 
$\kappa=2\pi\times520\un{kHz}$ and an atomic linewidth $\Gamma=2\pi\times184\un{kHz}$, corresponding to the parameters of the experiment in Refs. \cite{Braverman2019,Pedrozo2020}.
The simulation shows that for $N_\uparrow\eta=N_\downarrow\eta=900$, when the cavity is detuned by only $\kappa/10$ from atomic resonance, the total Fisher information in the two-color chirp measurement drops by a factor of 2. Therefore when using this scheme it is very important to keep the cavity on resonance with the atomic transition, or equivalently, to make the cavity transmission lineshape symmetric around the bare-cavity resonance. 

\section{Four-level atom}\label{sec:FLS}

In practice, the three-level system of Fig. \ref{fig:AtomCavitySchematic}(b) is usually only an approximation, and one or more other excited states come into play and couple to the other ground state $\ket{\downarrow}$. Here we consider a four-level system as typical of a $J=\frac{1}{2} \rightarrow J=\frac{3}{2}$ transition in the presence of circularly polarized light. 
In that system, the $\ket{\downarrow}$ level will not only act as a phase reference for the $\ket{\uparrow}$ state, but also be coupled to an excited state as depicted in Fig.~\ref{fig:energylevel} (a). We thus consider two excited states ($\ket{e\uparrow},\ket{e\downarrow}$) and two ground states ($\ket{\uparrow},\ket{\downarrow}$) in the interaction Hamiltonian. 

\setlength{\unitlength}{\columnwidth}
\begin{figure}[!tb]
    \centering
    \begin{picture}(1,0.5)
    \put(0.0,0){\includegraphics[height=.485\columnwidth]{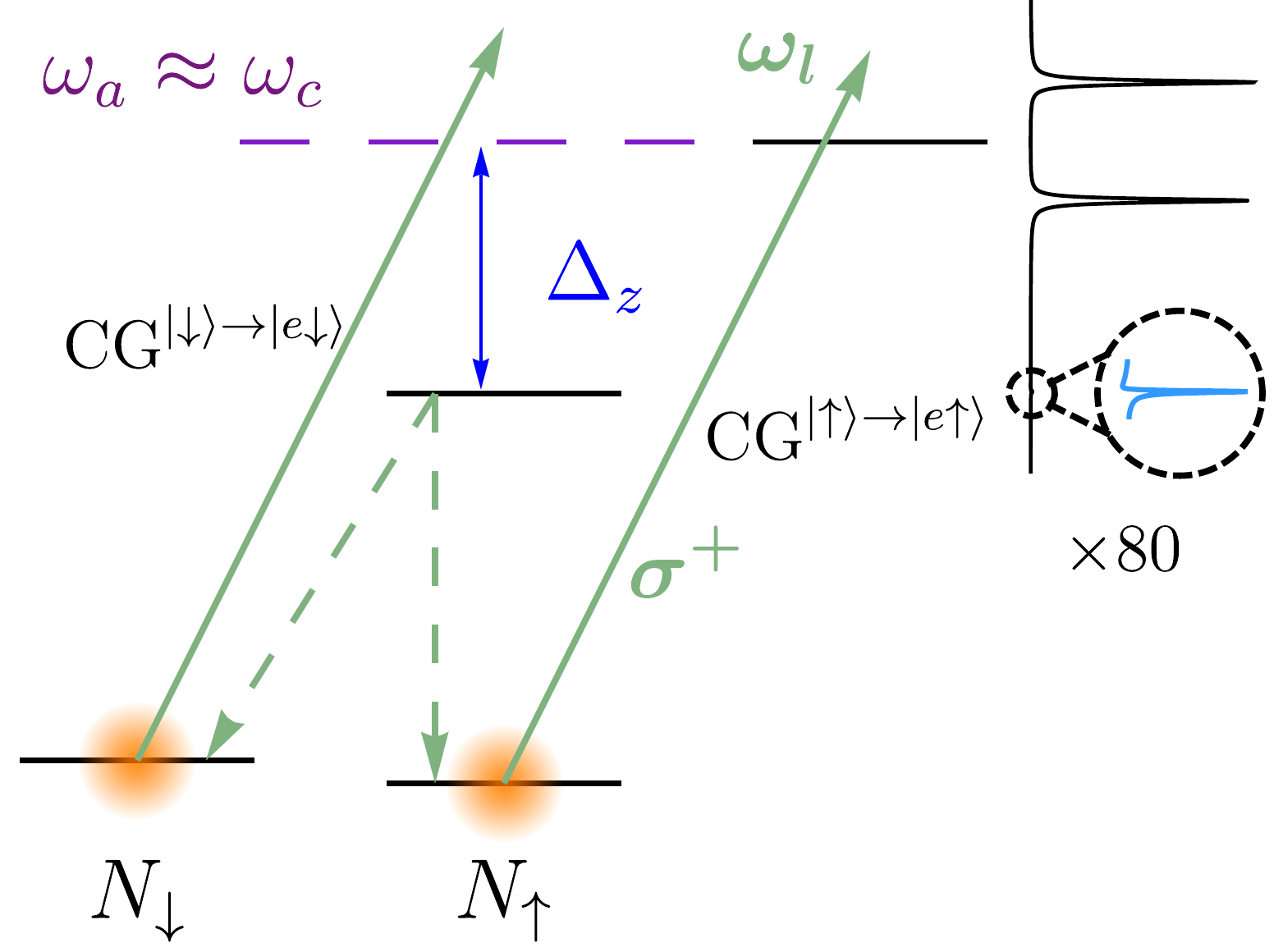}}
    \put(0.7,0){\includegraphics[height=.485\columnwidth]{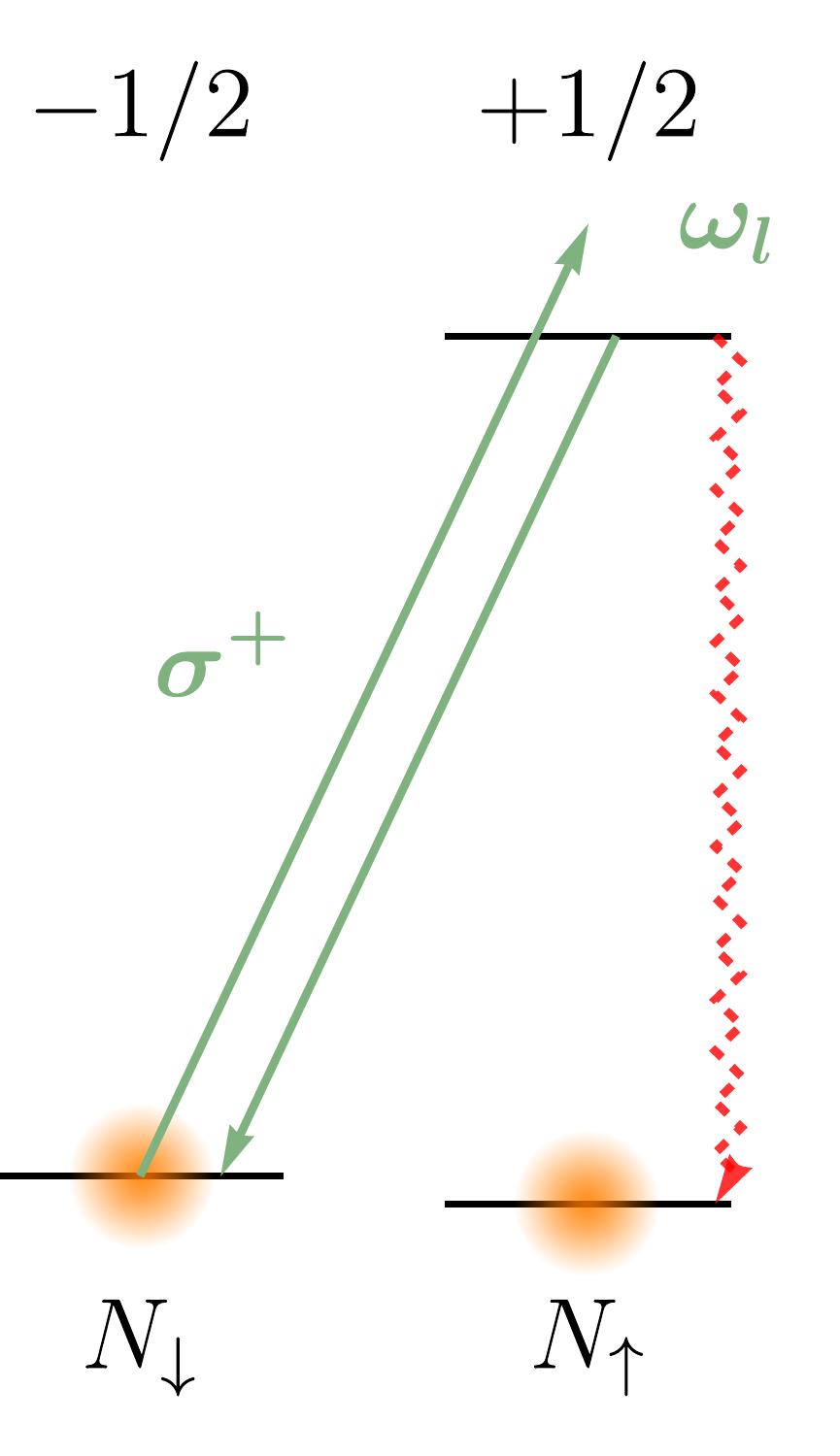}}
    \put(0.0,0.49){{(a)}}
    \put(0.7,0.49){{(b)}}
    \put(0.02,0.13){$\ket{\downarrow}$}
    \put(0.25,0.25){$\ket{e\downarrow}$}
    \put(0.32,0.11){$\ket{\uparrow}$}
    \put(0.42,0.37){$\ket{e\uparrow}$}
    \end{picture}
    \caption{(a) Energy level diagram for a four-level atom in a magnetic field. In this case we assume that the Zeeman splitting $\Delta_z\gg\kappa, \Gamma$, and $\Delta_z\gtrsim g$. The curve on the right in this subfigure is the cavity transmission spectrum, and the magnified blue line is the small transmission peak due to the $\ket{\downarrow}\leftrightarrow\ket{e\downarrow}$ transition. (b) Forward Rayleigh scattering $\ket{\downarrow} \rightarrow \ket{\downarrow}$ that is responsible for the collective atom-light interaction (green lines) and spontaneous Raman scattering $\ket{\downarrow} \rightarrow \ket{e\downarrow} \rightarrow \ket{\uparrow}$ (green and red lines) that causes additional spin noise.}
    \label{fig:energylevel}
\end{figure}
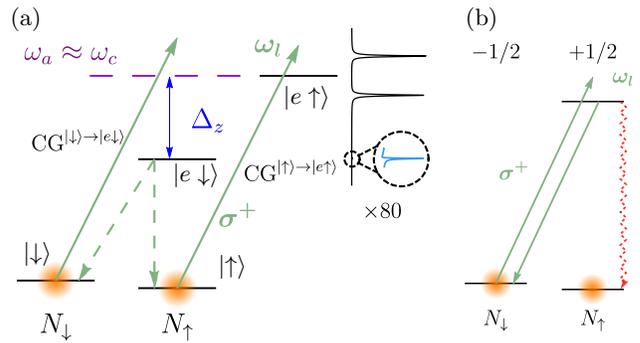

The probing light simultaneously couples to both transitions, so the light field amplitude and phase will have some residual sensitivity to $N_\downarrow$. Since circularly polarized ($\sigma^+$) cavity light only couples levels with specific angular momentum, the transition between $\ket{e\downarrow}$ and $\ket{\uparrow}$ is not coupled and there is no collective stimulated Raman scattering between the two ground states. However, the spontaneous emission can still cause the transition and contribute to a randomization of the atomic spin, as depicted in Fig.~\ref{fig:energylevel}(b). Besides this effect, we can treat the two transitions as entirely independent, and sum their individual contributions to the atomic polarization.

\subsection{Raman scattering}

For the four-level atomic structure depicted Fig.~\ref{fig:energylevel}(a), the attainable cavity feedback squeezing is limited by the effects of Raman scattering, in addition to the residual atom-light entanglement we derived for the 3-level case in Section~\ref{sec:Non-unitary}. An atom in the state $\ket{\downarrow}$ can be excited by a cavity photon to the state $\ket{e,\downarrow}$ and subsequently spontaneously decay into $\ket{\uparrow}$, as depicted in Fig. \ref{fig:energylevel} (b). This causes an undesired random spin flip and results in an additional variance
\begin{align}\label{eq:RamanNoise}
    (\Delta S_z)^2_\mathrm{R} = R_\mathrm{\downarrow \uparrow} \tau
\end{align}
where $R_\mathrm{\downarrow \uparrow}$ is the Raman scattering rate on the particular spin flipping transition $\ket{\downarrow}\to\ket{e\downarrow}\to\ket{\uparrow}$, and $\tau$ is the probing time.

\begin{figure}
    \centering
    \includegraphics[width=\columnwidth]{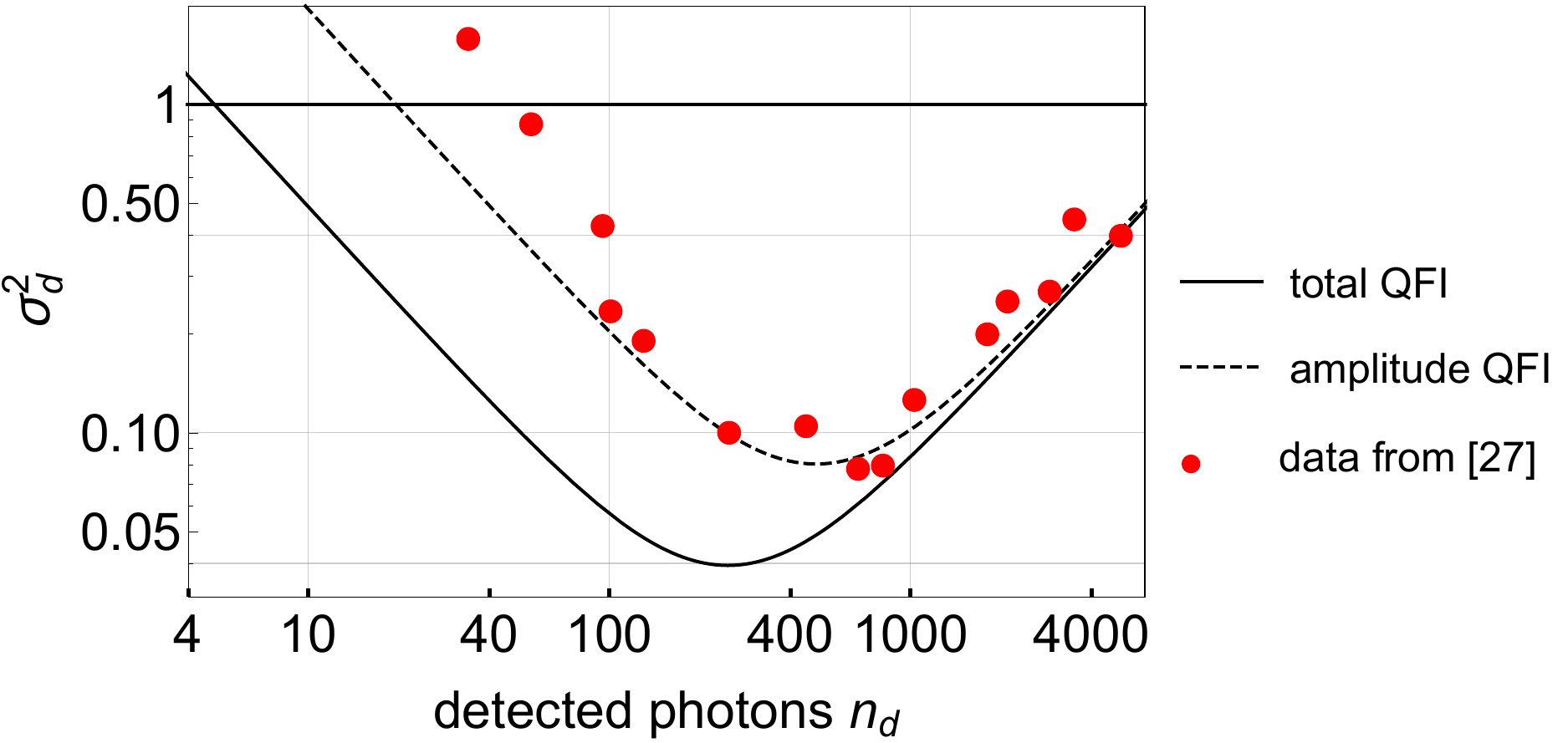}
    \caption{Measurement quality of the system $\sigma_d^2$ (variance of measurement spin noise normalized to SQL) for the chirped two-color measurement, with total atom number $N=1000$, single atom cooperativity $\eta=1.8$ and the total quantum efficiency for detecting photons is $q=0.15$. The solid black curve corresponds to the theoretical model including both the total QFI and Raman scattering noise, and the dashed black curve only includes the QFI contained in the transmission amplitude. The red points are experimental data from \cite{Braverman2019}. The detection limit $\sigma_d^2$ first improves with photon number $n_d$ as the fractional photon shot noise is reduced, and then increases with $n_d$ due to spontaneous Raman scattering. 
    }
    \label{fig:Raman}
\end{figure}

The Raman scattering effect sets the limit of the amount of spin squeezing as in \cite{Torre2013}, and the performance of the optical $S_z$ measurement discussed in Sec.~\ref{sec:Chirp}. The variance of the $S_z$ measurement is composed of two terms, one being the inverse attainable QFI contained in the light field, and the other the Raman scattering induced spin flip noise as in Eq.~\eqref{eq:RamanNoise}:
\begin{align}
    (\Delta S_z)^2_\mathrm{meas}=\frac{1}{1+F_\mathrm{meas}} + (\Delta S_z)^2_\mathrm{R}
\end{align}

The result from both contributions is summarized in Fig.~\ref{fig:Raman}. At small photon number $n_\gamma^\mathrm{meas}$, the detected variance is dominated by the low QFI contained in the light field, and therefore scales as $1/n_\gamma^\mathrm{meas}$. In the limit of large photon number, the variance is dominated by the Raman scattering and scales linearly with $n_\gamma^\mathrm{meas}$. For the parameters of Fig.~\ref{fig:Raman}, corresponding to the experiment in Ref.~\cite{Braverman2019}, the best measurement quality is obtained with $\sim 500$ detected photons, and corresponds to a resolution of 11~dB below the SQL.  
The measured resolution was 3~dB above the theoretical optimal resolution with the same parameters. 

\subsection{Cavity field for four-level atoms}

Atoms in the $\ket{\downarrow}$ state have a similar functional form for their contribution to the atomic polarizability as atoms in $\ket{\uparrow}$, but the corresponding transition will be detuned due to the Zeeman shift $\Delta_z$ if a magnetic field is applied. Writing $b=2\Delta_z/\Gamma$ for the normalized Zeeman shift, and taking into account that the differences in matrix elements for the two transitions give different cooperativities $\eta_{\uparrow/\downarrow}$,  we can modify Eq.~\eqref{eq:CavityFieldFinalResultRWA} for the three-level model to obtain the corresponding expression for the four-level model as
\begin{align}\label{eq:CavityFieldFinalResultRWA4level}
\begin{split}
    \mathcal{E}_c^\mathrm{FL}=&it_1\mathcal{E}_\mathrm{in}\frac{\mathcal{F}}{\pi}\Big(1+N_\uparrow\eta_\uparrow\mathcal{L}_a(\y)+N_\downarrow\eta_\downarrow\mathcal{L}_a(\y+b)\\
    &-i(\x+N_\uparrow\eta_\uparrow\mathcal{L}_d(\y)+N_\downarrow\eta_\downarrow\mathcal{L}_d(\y+b))\Big)^{-1}
\end{split}
\end{align}
where $\eta_{\uparrow,\downarrow}$ are the cavity enhanced single-atom cooperativities for the $\ket{\uparrow}\leftrightarrow\ket{e,\uparrow}$ and $\ket{\downarrow}\leftrightarrow\ket{e,\downarrow}$ transitions, respectively. The transmitted field is then
\begin{align}\begin{split}
    \mathcal{E}_t^\mathrm{FL}=&-t_1t_2\mathcal{E}_\mathrm{in}\frac{\mathcal{F}}{\pi}\Big(1+N_\uparrow\eta_\uparrow\mathcal{L}_a(\y)+N_\downarrow\eta_\downarrow\mathcal{L}_a(\y+b)\\
    &-i(\x+N_\uparrow\eta_\uparrow\mathcal{L}_d(\y)+N_\downarrow\eta_\downarrow\mathcal{L}_d(\y+b))\Big)^{-1},
\end{split}\end{align}
and the cavity transmission is therefore
\begin{align}\begin{split}
    \mathcal{T}^\mathrm{FL}=&\frac{4T_1T_2}{(T_1+T_2)^2}\Big((1+N_\uparrow\eta_\uparrow\mathcal{L}_a(\y)+N_\downarrow\eta_\downarrow\mathcal{L}_a(\y+b))^2\\
    &+(\x+N_\uparrow\eta_\uparrow\mathcal{L}_d(\y)+N_\downarrow\eta_\downarrow\mathcal{L}_d(\y+b))^2\Big)^{-1}. 
\end{split}\end{align}
For the symmetric case with $T_1=T_2$ we label the corresponding transmission $\mathcal{T}_0^\mathrm{FL}$. 

Substituting Eq.~\eqref{eq:CavityFieldFinalResultRWA4level} into the Hamiltonian, Eq.~\eqref{eq:atomicHamiltonian}, we can derive a Hamiltonian for the four-level atom
\begin{align}\label{eq:4levelHamiltonian}
    H_\mathrm{FL}
    =S_z\frac{|\mathcal{E}_c^\mathrm{FL}|^2}{\omega}\frac{\pi}{\mathcal{F}}\left(\eta_\uparrow\mathcal{L}_d(\y)-\eta_\downarrow\mathcal{L}_d(\y+b)\right)
\end{align}
As for Eq.~\eqref{eq:CavityAtomsStarkShiftTwoLevelAtomPhaseShiftVsTransmission}, this Hamiltonian gives a phase difference between each atom's $\ket{\uparrow}$ and $\ket{\downarrow}$ components that is given by
\begin{align}
    \Delta\phi=\frac{1}{2\epsilon} \frac{P_t\tau}{\hbar\omega}
    \left(\eta_\uparrow\mathcal{L}_d(\y)-\eta_\downarrow\mathcal{L}_d(\y+b)\right)
\end{align}
We can also expand the four-level Hamiltonian as in Eq.~\eqref{eq:CavityAtomsStarkShiftTwoLevelAtomTaylorSeries} to derive the shearing strength $Q$ as in Eq.~\eqref{eq:CavityAtomsQ},
\begin{align}\label{eq:4levelQ}
    Q = N\frac{\tau}{\hbar\omega}\frac{\partial|\mathcal{E}_c^\mathrm{FL}|^2}{\partial S_z}\left(\eta_\uparrow\mathcal{L}_d(\y)-\eta_\downarrow\mathcal{L}_d(\y+b)\right)
\end{align}
where at $N_\uparrow=N_\downarrow=N/2$, 
\begin{align}\begin{split}
    \frac{\partial|\mathcal{E}_c^\mathrm{FL}|^2}{\partial S_z}=|\mathcal{E}_c^\mathrm{FL}|^2\mathcal{T}_0^\mathrm{FL}\Bigg(&\frac{1-\x\y+N\eta_\uparrow/2}{1+\y^2}-\\
    &\frac{1-\x(\y+b)+N\eta_\downarrow/2}{1+(\y+b)^2}\Bigg)
\end{split}\end{align}

For the QFI, Eq.~\eqref{eq:TotalCavityFisherInformationCalc1}, we now find
\begin{align}\begin{split}
    F_\mathrm{tot}(S_z) = 4 \tau\left|\frac{\partial\mathcal{E}_c^\mathrm{FL}}{\partial S_z}\right|^{2} \Big(&T_1+T_2 R_1+\frac{\pi}{\mathcal{F}}\langle N_\uparrow\rangle\eta_\mathrm{\uparrow}\mathcal{L}_a(\y)\\
    &+\frac{\pi}{\mathcal{F}}\langle N_\downarrow\rangle\eta_\mathrm{\downarrow}\mathcal{L}_a(\y+b)\Big)
\end{split}\end{align}

\begin{figure}
\centering
\begin{picture}(1,0.61)
    \put(0.0,0.03){\includegraphics[width=.98\columnwidth]{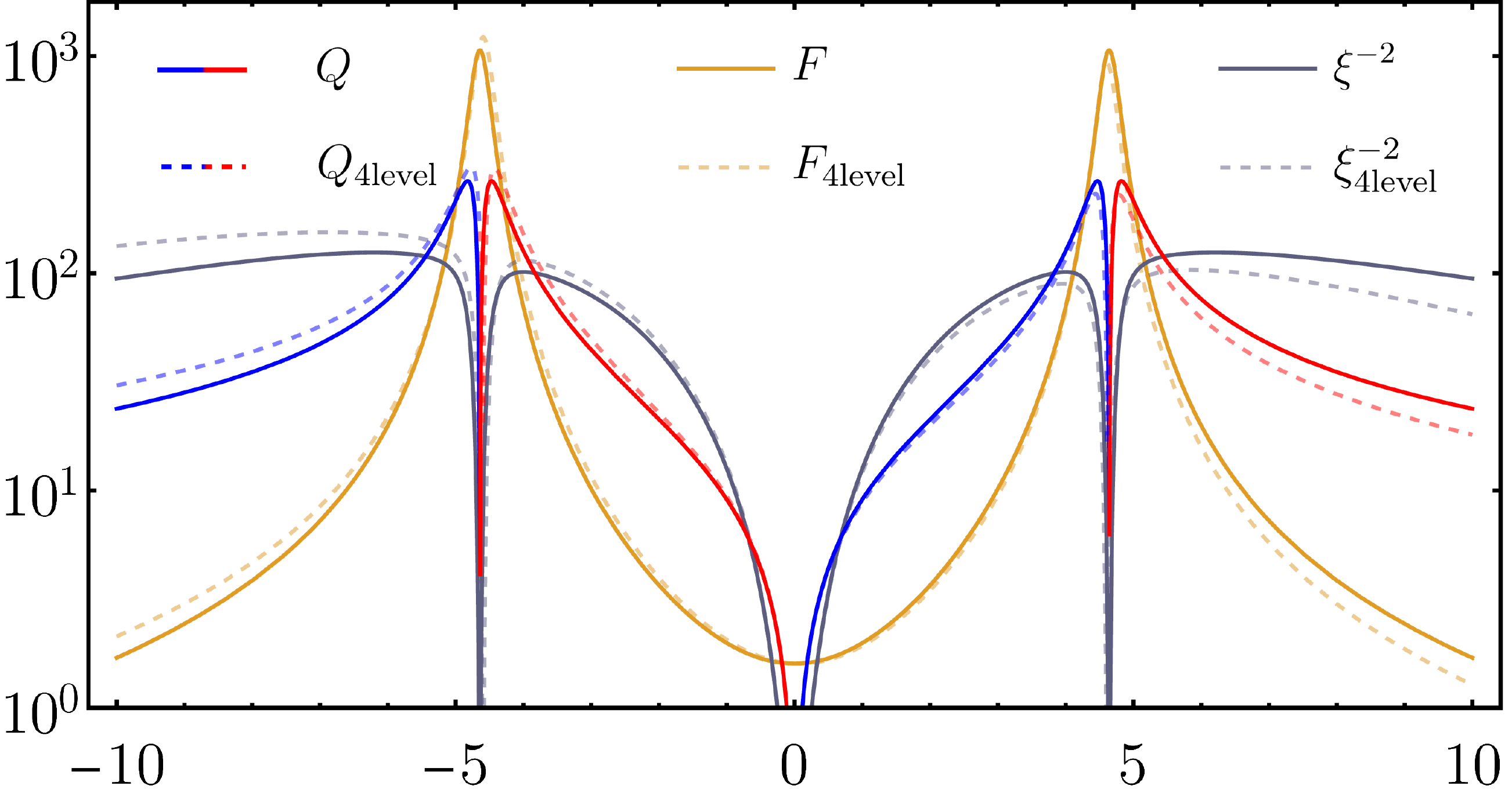}}
    \put(0.33,0){Detuning $\Delta/2\pi$(MHz)}
    \put(0.06,0.57){$|Q|, F, \xi^{-2}$ in condition of $n_\gamma^\mathrm{sc}=400, N_\uparrow\eta=N_\downarrow\eta=900$}
\end{picture}
\caption{Comparisons of the squeezing parameters $Q, F$ and $\xi^{-2}$ between the three-level and four-level models. The solid lines represent the results of the three-level model (see also Fig. \ref{fig:WinelandParameter}), and the dashed lines correspond to the four-level model. For $Q$ plots, red branches stand for positive $Q$ and blue branches for negative $Q$. The four-level model only provides a small correction to the three-level model. }
\label{fig:WinelandFixNCompare}
\end{figure}

In Fig. \ref{fig:WinelandFixNCompare} we compare these results with those of the three-level model (Sec.~\ref{sec:ALI} and Sec.~\ref{sec:Fbroadening}), assuming similar conditions where $N_\uparrow\eta=N_\downarrow\eta=900$ and $b=2\Delta_z/\Gamma=230$. The dashed lines correspond to the four-level model and the solid lines indicate the three-level results. 
The effects are small but important for experimental aspects, which will be discussed in Sec.~\ref{sec:Application}. 

\section{Application to Ytterbium-171} \label{sec:Application}

One particular atom of interest is \Yb which is one of the leading atomic species used in optical lattice clocks \cite{Hinkley2013,Kobayashi2020,Pedrozo2020}. 
Following Ref.~\cite{Braverman2019}, the two interacting states are chosen as the ground state $\ket{\uparrow}=\ket{^1S_0,F=\frac{1}{2}, m_F = \frac{1}{2}}$ 
and excited state $\ket{e\uparrow}=\ket{^3P_1, F=\frac{3}{2}, m_F = \frac{3}{2}}$, 
while the two detuning-suppressed weakly interacting states are $\ket{\downarrow}=\ket{1S_0, F=\frac{1}{2}, m_F = -\frac{1}{2}}$ 
and the excited state $\ket{e\downarrow} = \ket{^3P_1, F=\frac{3}{2}, m_F = \frac{1}{2}}$. 
Experimentally, we can reach this configuration by probing the system with $\sigma^+$ light on resonance with the transition $\ket{\uparrow} \rightarrow \ket{e \uparrow}$. 
The probing light generates effects described in Sec.~\ref{sec:ALI} and Sec.~\ref{sec:Fbroadening} such as spin squeezing between the Zeeman sublevels of the ground states 
($\ket{\uparrow}$ and $\ket{\downarrow}$).

\subsection{Cavity detuning for symmetric transmission}\label{sec:CavityDetuning}

The additional transition ($\ket{\downarrow}\leftrightarrow\ket{e\downarrow}$) in a four-level model leads to several additional coorections to three-level model. The first one affects the two-color measurement as discussed in Section \ref{sec:Chirp}. In the four-level model,  when the $\ket{\uparrow}\leftrightarrow\ket{e\uparrow}$ atomic transition and the cavity transition are on resonance, the cavity field will no longer have the balancing and fixed phase properties as in three-level model. However, since the weakly interacting transition is far detuned, its effect can be treated to lowest order as a cavity resonance shift. 
In the experiment described in Ref.~\cite{Braverman2019}, $\Delta_z$ and $\Gamma$ are $2\pi\times18.5\un{MHz}$ and $2\pi\times0.18\un{MHz}$, respectively, and therefore $b=2\Delta_z/\Gamma\approx 230$. 
The scattering rate by the weakly interacting atoms $N_\downarrow\eta\mathcal{L}_a(x_a+b)/3\to0$ 
but $N_\downarrow\eta\mathcal{L}_d(x_a+b)/3$  
remains at the same order of $\x$. Based on this and the experimental condition where $N_\downarrow\eta\sim900$ is routinely achieved \cite{Braverman2019}, the effective cavity field could be described as 
\begin{align}
    \mathcal{E}_c^\mathrm{FL,eff}\to\frac{\mathcal{F}}{\pi}\frac{it_1\mathcal{E}_\mathrm{in}}{1+N_\uparrow\eta\mathcal{L}_a(\y)-i(\x+N_\uparrow\eta\mathcal{L}_d(\y)-1.3)}
\end{align}
and by choosing a detuning between atomic and cavity resonances 
\[\delta_c\equiv\delta-\Delta=\frac{N_{\downarrow}\eta_{\downarrow}\Gamma\kappa}{4\Delta_{Z}}\]
so that $\x\to\x+1.3$,
the field is approximately compensated back to the symmetric dispersion of the three-level model when the cavity is on resonance with atomic transition. Specifically in the setting of Ref.~\cite{Braverman2019}, $\delta_c=2\pi\times0.34\un{MHz}$ and is exactly what is used in the experiments.

\subsection{Unitary squeezing by two-color pulses}

\begin{figure}[ht]
    \centering
    \includegraphics[width=\columnwidth]{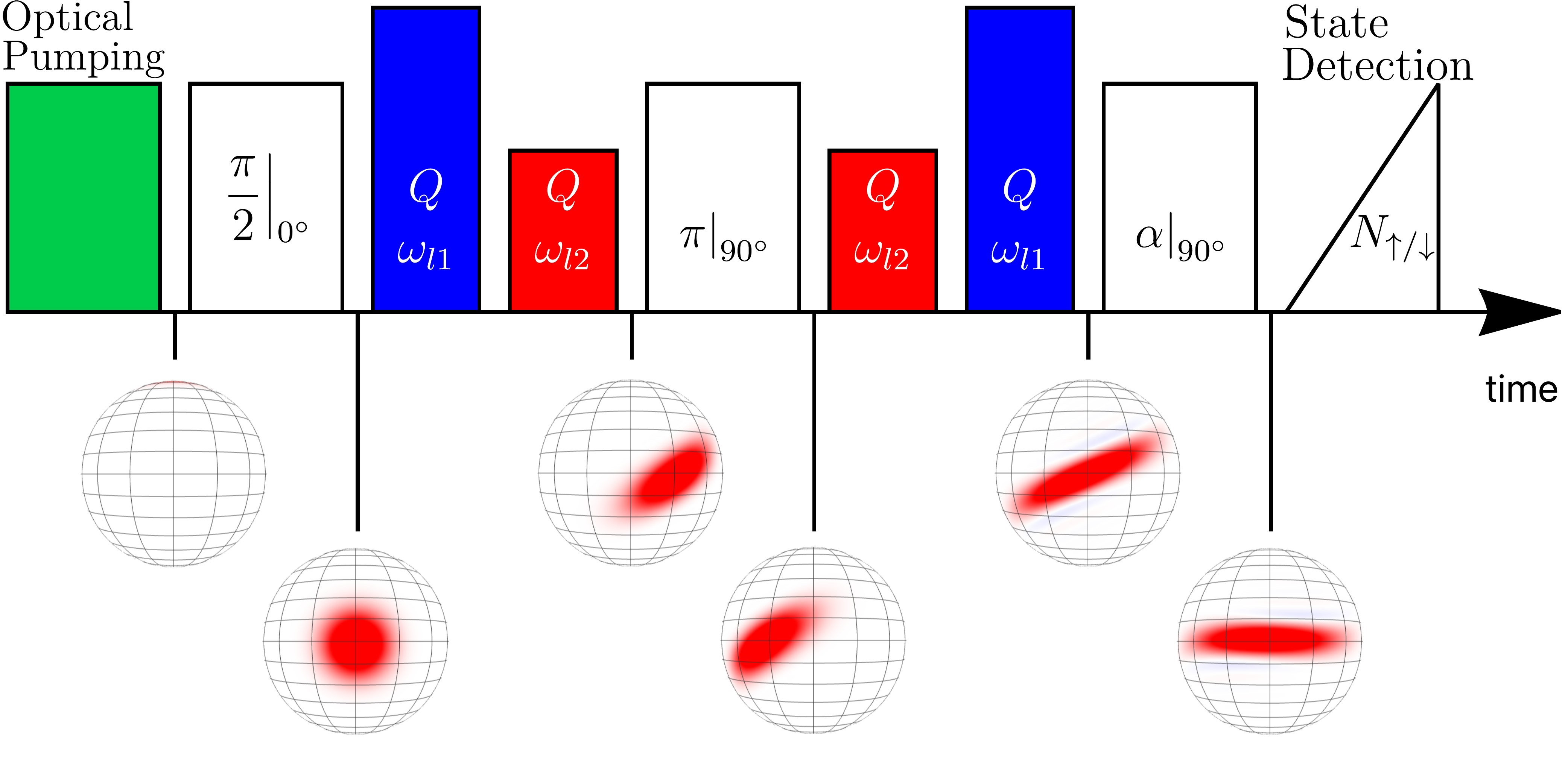}
    \caption{Full experimental sequence for cavity spin squeezing as used in Ref.~\cite{Braverman2019}. The subscript for the radiofrequency (RF) pulses indicates the axis of the rotation. The Bloch spheres below the sequence indicate the collective spin state at the corresponding time.}
    \label{fig:fullsequence}
\end{figure}

The degree of unitarity in the spin squeezing process has a significant impact for metrological purposes \cite{Braverman2018b, Braverman2018a}.
When using cavity feedback squeezing for metrology \cite{Braverman2019}, the major contributions to non-unitarity arise from:
(1) The dependence of the $Q$ factor and phase shift $\Delta\phi$ on the total atom number $N$, resulting in a non-unitary evolution as the total atom number fluctuates, and (2) the QFI of the squeezing pulse which is not used for atomic state detection and therefore causes a broadening of the $S_z$ distribution. The first of these effects can be minimized using a two-color squeezing scheme \cite{Braverman2019} with a sequence as illustrated in Fig. \ref{fig:fullsequence}. 

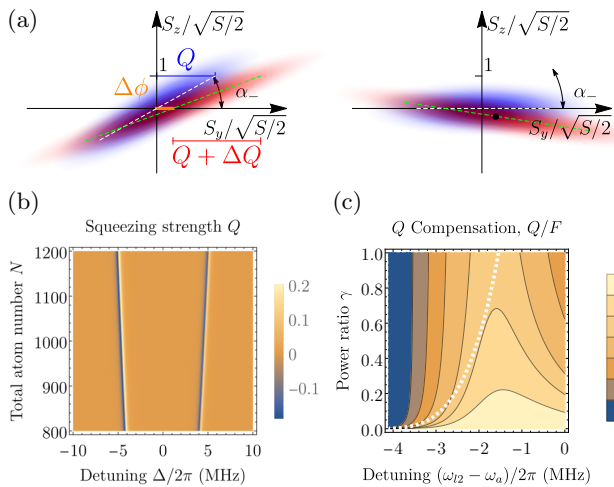
\begin{figure}[ht]
\begin{picture}(1,0.73)
    \put(0,0.71){{(a)}}
    \put(0.,0.46){\includegraphics[width=.5\columnwidth]{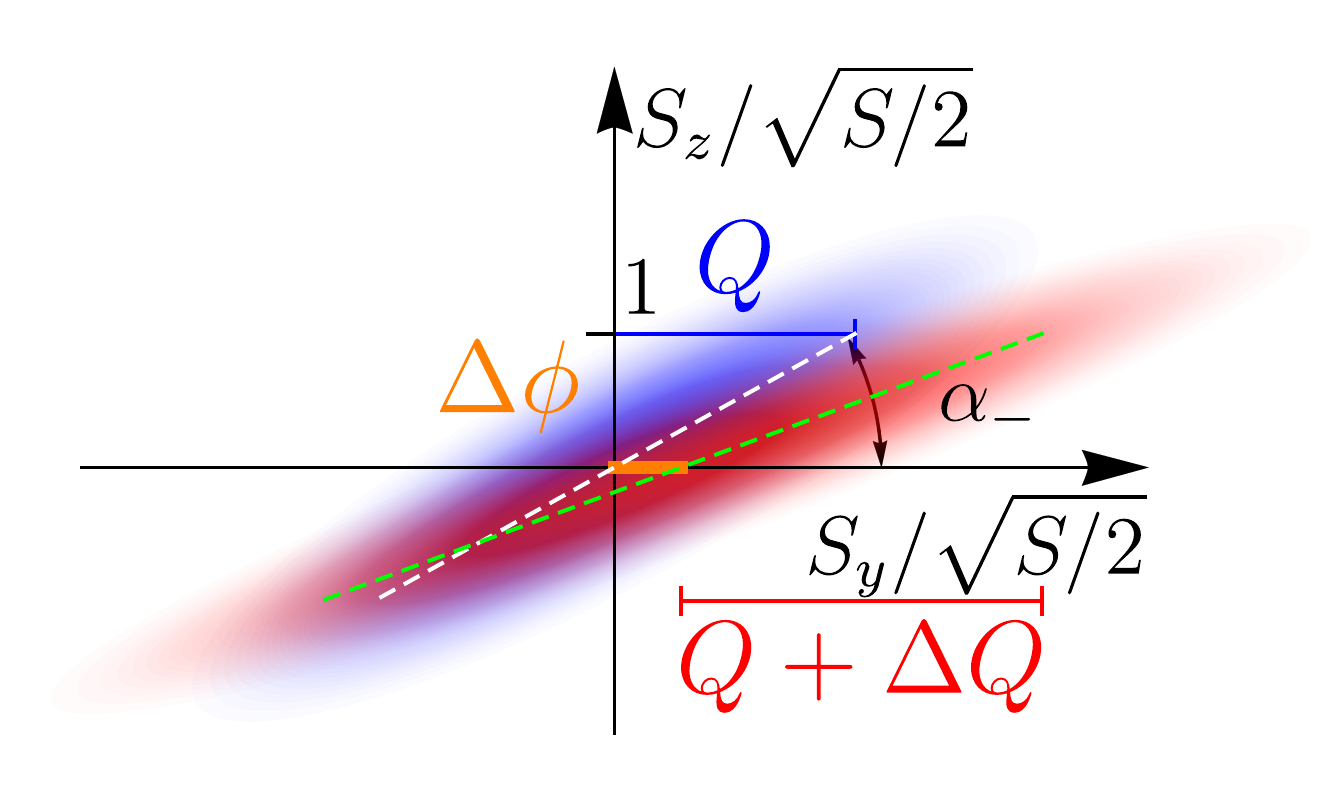}}
    \put(0.5,0.46){\includegraphics[width=.5\columnwidth]{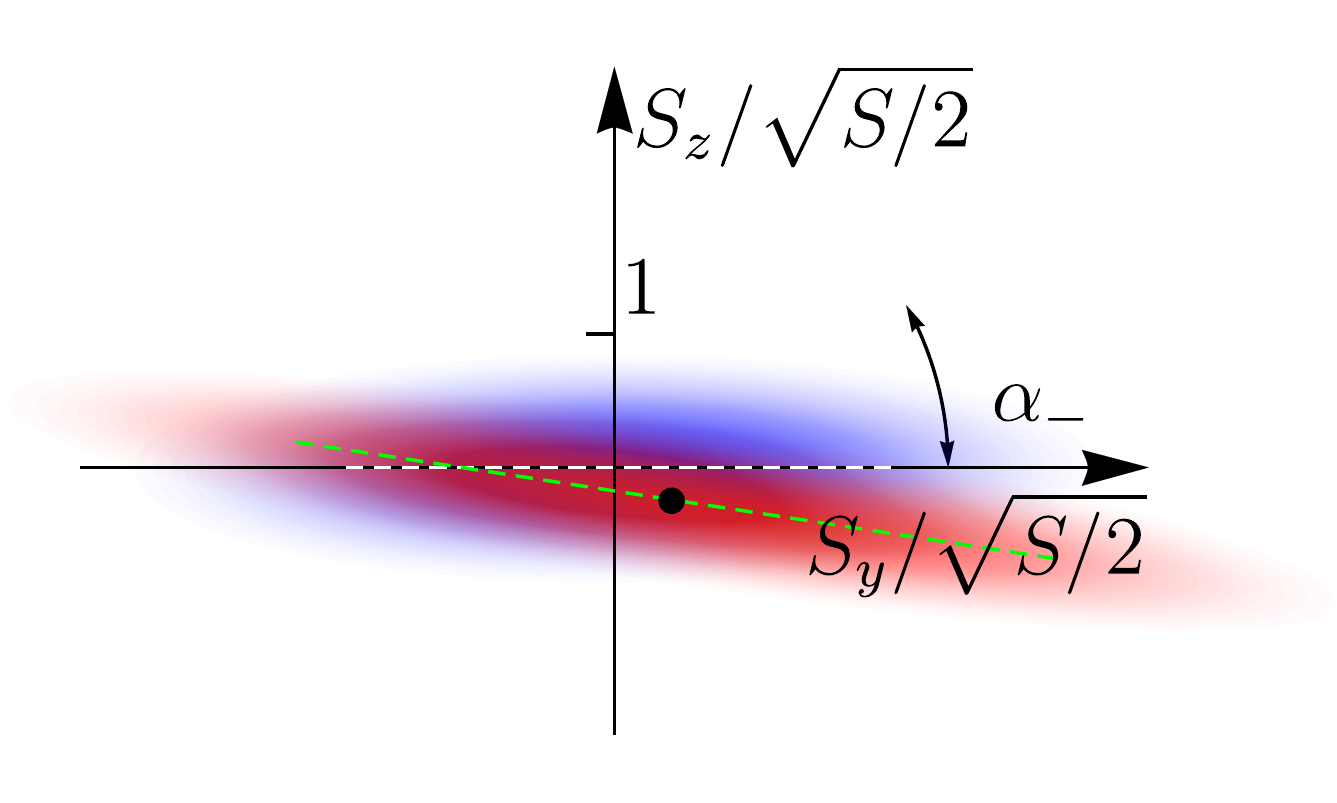}}
    \put(0.,0.43){{(b)}}
    \put(0.,0.0){\includegraphics[height=.42\columnwidth]{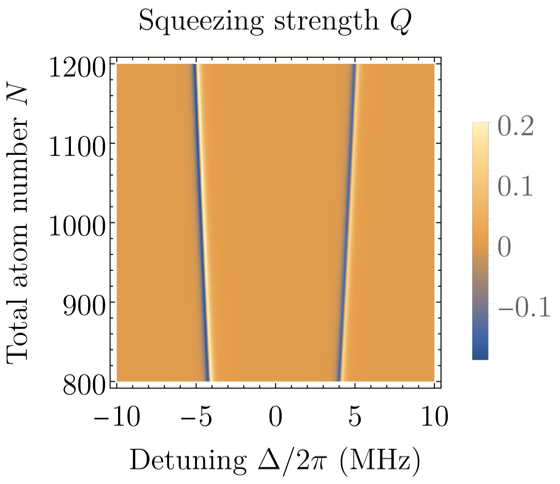}}
    \put(0.5,0.43){{(c)}}
    \put(0.5,0.0){\includegraphics[height=.42\columnwidth]{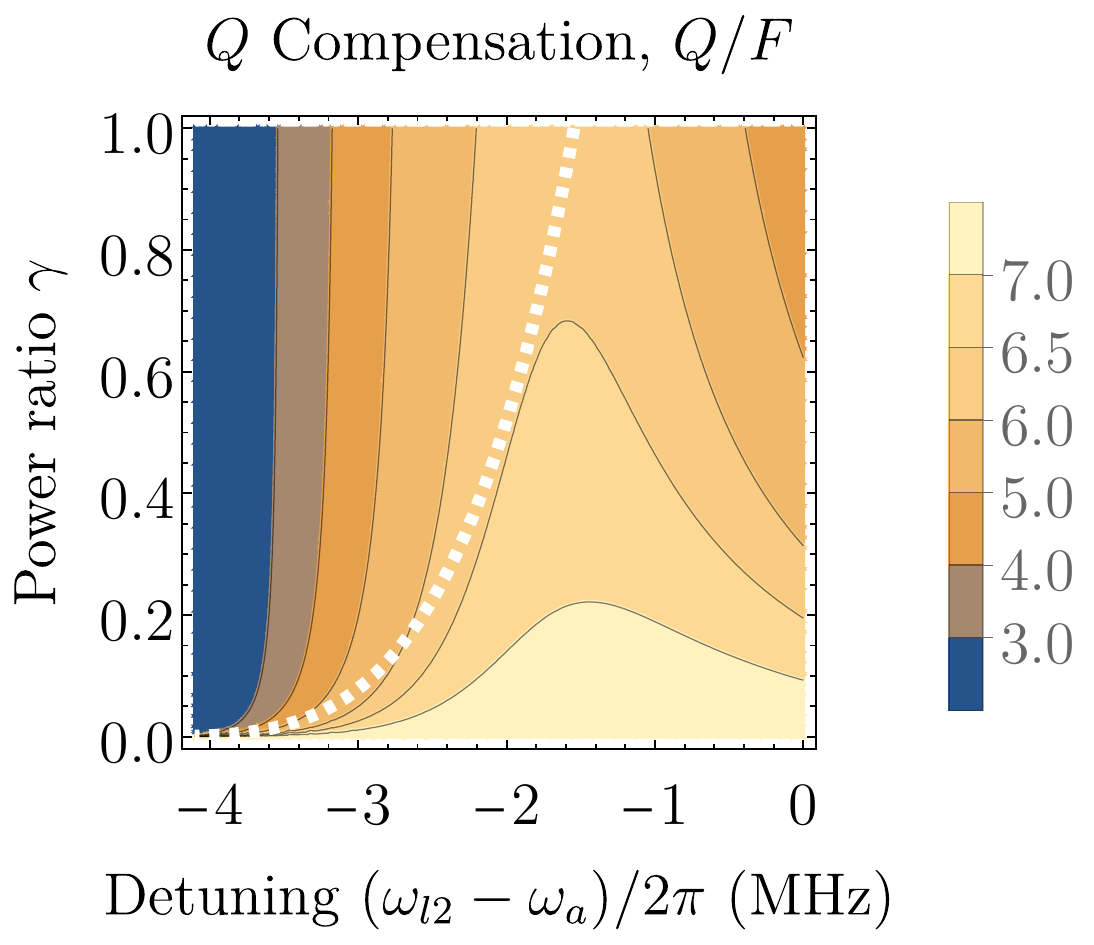}}
\end{picture}
\caption{(a) The quasiprobability for a squeezed state with/without final rotation for the ideal case (blue) and a situation with $Q$ and $\Delta\phi$ fluctuations (red). (b) Contour plot of squeezing strength $Q$ per incoming photon as a function of detuning $\Delta$ and total atom number $N$. In the calculation we use a four-level model with a cavity frequency $2\pi\times340\un{kHz}$ lower than atomic frequency. (c) Contour plot for the $Q/F$ ratio by fixing the detuning of the first squeezing pulse to be $7.333\un{MHz}$, and varying the detuning and power ratio of the second pulse. The white dashed line indicates the optimum parameters for automatic compensation against atom number fluctuations. The max ratio is $Q/F=6.47$ with power ratio $\gamma=0.52$ and red pulse detuning $-2\un{MHz}$. } \label{fig:QvsNdelta}
\end{figure}

We start by discussing the dependence of the squeezing strength $Q$ and phase shift $\Delta\phi$ on the total atom number $N$. At constant input photon number and $S_z=0$, variations in the total atom number $N$ result in a different transmitted photon number, and therefore different $Q$ and $\Delta\phi$ values. Such fluctuations, as illustrated in Fig.~\ref{fig:QvsNdelta}, can cause problems for the final measurement variance: the $Q$ fluctuations affect the orientation of the squeezing and the rotation angle that is needed to orient the maximally squeezed direction along the $z$-axis. This causes the antisqueezing to leak into the final measurement. The $\Delta\phi$ fluctuation are also deleterious in that they are converted into measured $S_z$ noise by the final $\pi/2$ rotation of the Ramsey sequence. In the scenario in \cite{Braverman2019}, where the imperfect $\pi$ pulse is at the level of $0.1\%$ and atom number $N=1000$, the induced variance by uncompensated phase shift gives $0.04$ SQL, i.e., $-14$ dB. Therefore it did not affect those measurements, however, for larger amounts of squeezing, this effect must be taken into account. 

To first order, the squeezing strength is linear in $N$, with the coefficient being either positive or negative, depending on the detuning of laser frequency to the atomic state $\Delta$, as in Fig. \ref{fig:QvsNdelta}. Therefore, by applying two pulses with positive and negative atomic detunings, this fluctuation can be compensated to first order. As in \eqref{eq:4levelQ}, the sign of $Q$ at a certain probing frequency is predominantly due to the detuning relative to the dressed atomic levels. Therefore, for a situation where the contributions to the squeezing $Q$ from two colors add constructively, while the first-order dependence on the atom number $N$ is opposite, the two pulses should be on different sides of the bare atomic resonance, but on the same side relative to the cavity transmission peaks (the dressed atomic resonance). Also, to avoid being close to the $\ket{\downarrow}\leftrightarrow\ket{e,\downarrow}$ transition which leads to a unfavorable large QFI, one needs a blue-detuned light pulse with frequency $\omega_{l1}$ tuned above the frequency of the higher-frequency Rabi peak and a red-detuned light pulse with frequency $\omega_{l2}$ tuned above the lower-frequency Rabi peak. At fixed $\omega_{l1}$, the second frequency $\omega_{l2}$ and power ratio $\gamma$ can be determined from
\begin{align}
    \frac{\mathrm{d}\hat{Q}_1^\mathrm{in}}{\mathrm{d}N}+\gamma\frac{\mathrm{d}\hat{Q}_2^\mathrm{in}}{\mathrm{d}N}=0, \quad
    \frac{\mathrm{d}\Delta\hat{\phi}_1^\mathrm{in}}{\mathrm{d}N}+\gamma\frac{\mathrm{d}\Delta\hat{\phi}_2^\mathrm{in}}{\mathrm{d}N}=0,
\end{align}
where $\hat{Q}_i^\mathrm{in}$ and $\Delta\hat{\phi}_i^\mathrm{in}$ indicate the $Q$ or $\Delta\phi$ value in the $i$-th pulse due to a single incoming photon. 

Besides the technical noise induced by the fluctuations in atom number, the QFI of the squeezing light also has a negative impact on unitarity. Since both $Q$ and $F$ are linear in the incoming photon number, we can choose the set of parameter $\{\omega_{l1},\omega_{l2},\gamma\}$ to maximize the $Q/F$ ratio while keeping the atom number fluctuation compensation discussed above. Since both $\omega_{l2}$ and $\gamma$ are determined by $\omega_{l1}$, we can calculate the $Q/F$ ratio for the combination of two pulses:
\[\frac{\hat{Q}_1^\mathrm{in}+\gamma\hat{Q}_2^\mathrm{in}}{\hat{F}_1^\mathrm{in}+\gamma\hat{F}_2^\mathrm{in}}. \]
If we select $\omega_{l1}=2\pi\times7.33\un{MHz}$, the best power ratio and frequency for the second pulse are $\gamma=0.52$ and $\omega_{l2}=-2\pi\times2.0\un{MHz}$, which provides a compensated and constructive effect of squeezing, as in Fig.~\ref{fig:QvsNdelta} (c). The detuning $\omega_{l1}=2\pi\times7.33\un{MHz}$ is chosen as a compromise between a larger $Q/F$ ratio and less contrast reduction. 

\section{Conclusions}\label{sec:Conclusion}

In conclusion, we have described a framework for understanding the two lowest-order effects of light on a collective spin state, i.e., the effective spin-spin interaction and the atomic-state measurement via measuring the light field that has been entangled with the collective atomic state. The interaction generates a phase shift of the collective spin state, a spin squeezing which is characterized by the shearing parameter $Q$, and a broadening of the atomic state due to the loss of information, which is characterized by the QFI $F$. We have given analytical expressions for all quantities. In addition to these coherent effects, we also consider the spontaneous emission from the atoms as a source of contrast loss. Under the assumption that the collective spin state can still be described by a Gaussian distribution, we have derived the Wineland parameter for the system as the attainable metrological gain, and studied its scaling with external parameters such as atom number $N$ and frequency detuning $\delta$. In particular, the light-induced evolution of the atomic state becomes more unitary with larger the detuning. Therefore it allows to optimize more fine-tuned phenomena~\cite{Ma2021,carrasco2022generating}. On the other hand, at sufficiently large atom number, the curvature of the generalized Bloch sphere effect limits the attainable metrological gain, i.e. the spin quasiprobability distribution is no longer a Gaussian distribution. 
In addition, we used this model to derive system parameters that we were used in experiments Ref.~\cite{Braverman2019,Colombo2021} to optimize the squeezing performance.  

We thank Monika Schleier-Smith and
Mikhail Lukin for valuable discussions. This work was
supported by NSF, DARPA, ONR, and the NSF Center for
Ultracold Atoms (CUA). S. C. and A. A acknowledge support as a
SNSF Early Postdoc.Mobility fellow. B. B. acknowledges
the support of the Banting Postdoctoral Fellowship.

\appendix

\section{Simulating a Cavity with Lossy Mirrors by a Lossless
Cavity with External Loss}
\label{sec:SimulatingLossyCavity}

\begin{figure}
    \centering
    \includegraphics[width=.8\columnwidth]{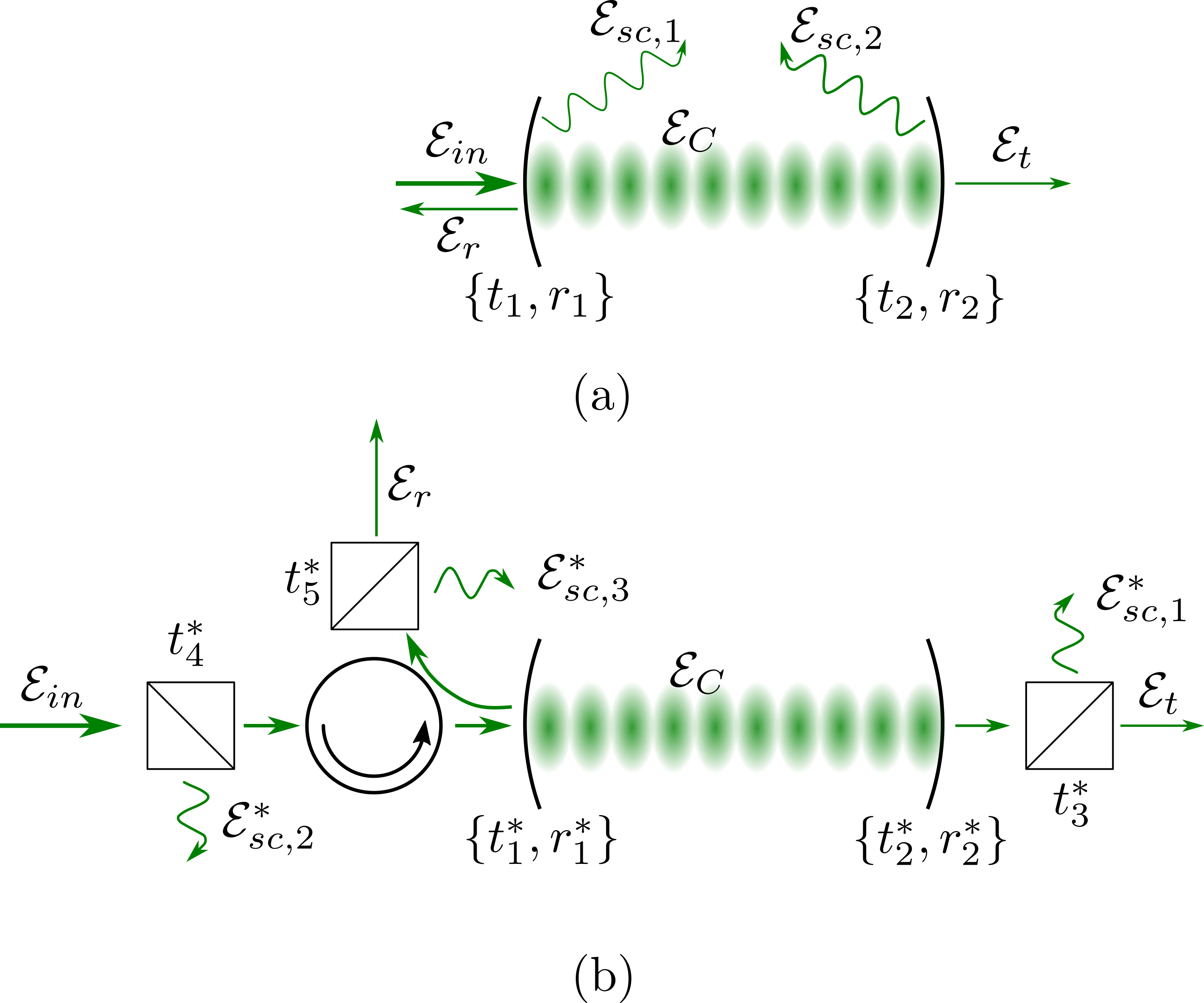}
    \caption{Analogy between (a) a realistic cavity with loss and (b) an idealized lossless cavity with external loss by using perfect beam splitters and optical circulators. }
    \label{fig:LossyAndLosslessCavity}
\end{figure}

In the main text, we consider the case of an idealized cavity with lossless mirrors, where $R_i + T_i = 1$ for $i=1,2$. This assumption simplifies many calculations of atom-light interactions. In reality, any cavity mirror has non-zero loss $L_i$ from a combination of absorption in the mirror coating material and scattering from imperfections on the mirror surface. Using $T_i = t_i^2$, $R_i = r_i^2$, we define the loss $L_i$ in mirror $j$ as
\begin{equation}\label{eq:MirrorLossDefinition}
L_i = 1 - T_i - R_i
\end{equation}
In fact, we can exactly reproduce the behaviour of a cavity with lossy mirrors, as shown in Figure \ref{fig:LossyAndLosslessCavity}(a), by a cavity with lossless mirrors and additional external loss, as shown in Figure \ref{fig:LossyAndLosslessCavity}(b).
To ensure the two systems are truly equivalent, we need to have the same values of $\mathcal{E}_c$, $\mathcal{E}_r$, and $\mathcal{E}_t$ for all cavity lengths $L$ and probing light wavenumber $k$. In addition, the scattered fields $\mathcal{E}^*_{sc,j}$ must be lost to the environment to ensure the correct decoherence rates for the system.

The resulting parameters for the optical elements in Fig.~\ref{fig:LossyAndLosslessCavity}(b) are: 
\begin{align}\label{eq:ConvertingLossyToLossLessCavityFinalResult}
t^*_{1} &= \left[1 + \frac{r^2_1}{t^2_1} \right]^{-1/2} \nonumber
\\
r^*_{2} &= \frac{r_1 r_2}{r^*_1} \nonumber
\\
t^*_{3} &= \frac{t_2}{t^*_2}
\\
t^*_{4} &= \frac{t_1}{t^*_1} \nonumber
\\
t^*_{5} &= \frac{r_1}{r^*_1 t^*_{4}}, \nonumber
\end{align}
where the lossless condition requires that $r^*_i = \sqrt{1 - \left(t^*\right)^2_i}$.

Note that in the case of a high-finesse cavity, we have $r_1 \approx 1$ and $t_1 \ll 1$, which allows us to simplify \eqref{eq:ConvertingLossyToLossLessCavityFinalResult} to 
\begin{align}\label{eq:ConvertingLossyToLossLessCavityFinalResult_Simplified}
t^*_{1} &\approx t_1 \nonumber
\\
r^*_{2} &\approx \frac{r_1 r_2}{r^*_1} \nonumber
\\
t^*_{3} &\approx \frac{t_2}{t^*_2}
\\
t^*_{4} &\approx 1 \nonumber
\\
t^*_{5} &\approx \frac{r_1}{r^*_1} \nonumber
\end{align}
The physical insight for these equations is the following: we preserve the transmissivity of the input coupling mirror ($T^*_1 \approx T_1$), transfer all the cavity loss into the transmissivity of the output coupling mirror ($T^*_2 \approx T_2 + L_1 + L_2$), and simulate the cavity loss through absorption after the light exits the cavity ($T^*_3 \approx \frac{T_2}{T_2 + L_1 + L_2}$). 
Table \ref{tab:AsymmetricandSymmetricCavity} gives the numerical values for the various relevant parameters for the experimental system that we consider in the main body of the paper \cite{Kawasaki2019,Braverman2019}.

\begin{table}[hbtp]
    \centering
    \begin{tabular}{cc}
    \toprule
        \multicolumn{2}{l}{\textbf{Lossy cavity}}\\
        \hline
        $T_1$ & $30\times 10^{-6}$ \\
        $L_1$ & $30\times 10^{-6}$ \\
        $T_2$ & $196\times 10^{-6}$ \\
        $L_2$ & $227.3\times 10^{-6}$ \\
        $\mathcal{F}$ & $13.0\times 10^3$ \\
    \botrule
    \end{tabular}
    \hspace{1cm}
    \begin{tabular}{cc}
    \toprule
        \multicolumn{2}{l}{\textbf{Lossless cavity}}\\
        \hline
        $T^*_1$ & $30\times 10^{-6}$ \\
        $T^*_2$ & $453.3\times 10^{-6}$ \\
        $T^*_3$ & $0.4324$ \\
        $T^*_4$ & $1$ \\
        $T^*_5$ & $1$ \\
    \botrule
    \end{tabular}
    \caption{Properties of the experimentally realized, lossy cavity \cite{Kawasaki2019,Braverman2019}, as well as the equivalent lossless cavity, with values of $T^*_i$ given with a relative accuracy of $10^{-4}$.}
    \label{tab:AsymmetricandSymmetricCavity}
\end{table}

\section{Contrast Loss due to Wrapping of the State around the Bloch Sphere} \label{sec:AppendixContrastLoss}

The contrast of a collective atomic state is commonly defined as the largest oscillation amplitude of the mean signal in a Rabi or Ramsey sequence as compared to a CSS. Even for a squeezed state generated by unitary evolution, the average collective spin vector is shortened and causing an effective contrast reduction
\begin{equation}\label{eq:BlochContrast}
    C_\mathrm{Bloch}(Q)=\frac{\langle S_x=S|e^{iS_z^2Q/N}S_xe^{-iS_z^2Q/N}|S_x=S\rangle}{S},
\end{equation}
where $\chi t= Q/ N$ is the action of the squeezing Hamiltonian on the initial state. In the limit of unitary squeezing, $F=0$, we rotate the coordinates to make the antisqueezing direction correspond to the $S_z$ direction, and the Wigner distribution is approximated as a Gaussian distribution
\begin{align}\begin{split}
    P(\theta)&=
    \frac{1}{\sqrt{2\pi}\xi_+\theta_\mathrm{SQL}}
    \exp\left(-\frac{\theta^2}{2\xi_+^2\theta_\mathrm{SQL}^2}\right),
\end{split}\end{align}
where $\theta\in(-\infty,\infty)$. 

\begin{figure}[!bp]\centering
\includegraphics[width=.7\columnwidth]{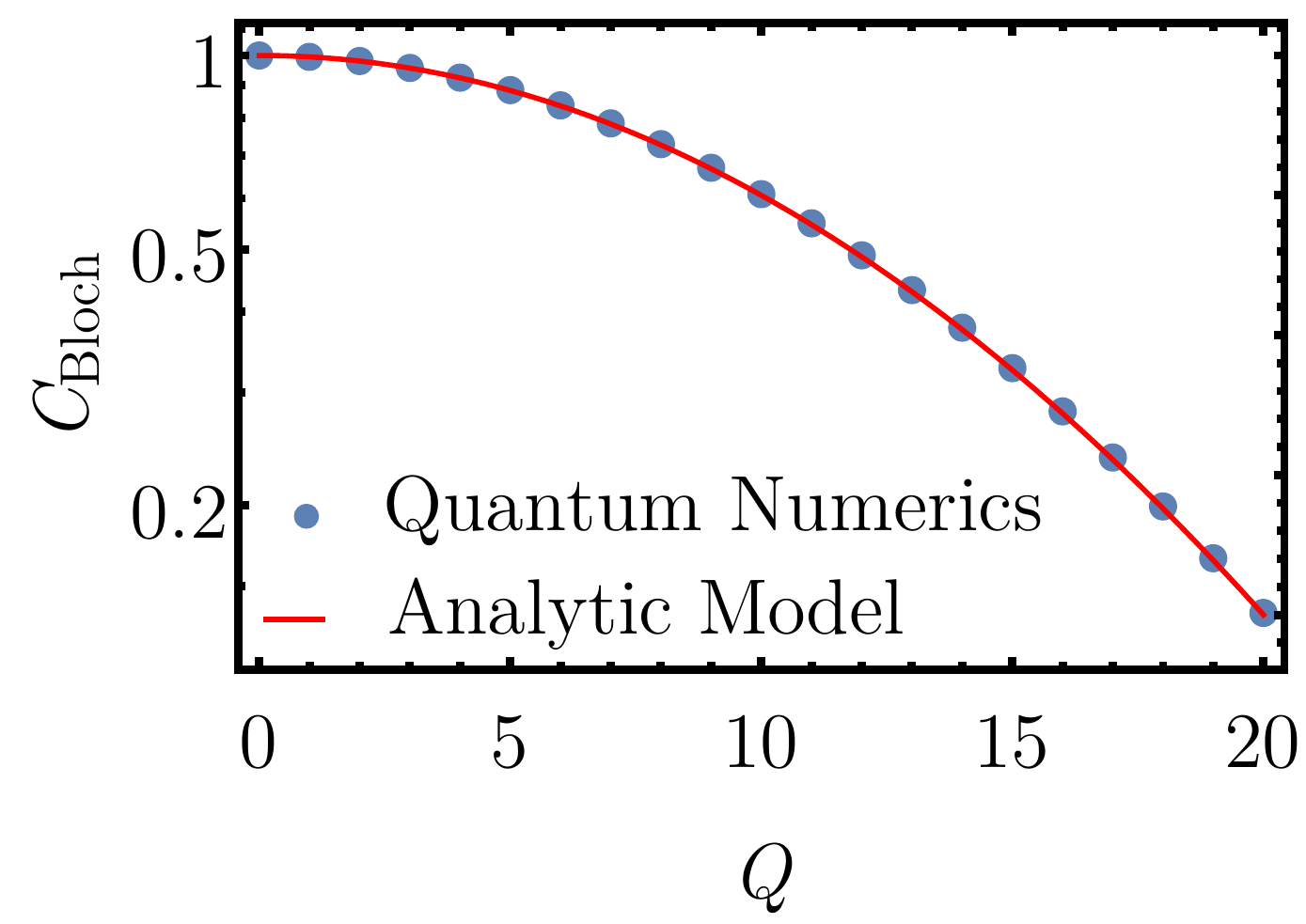}
\caption{Contrast loss (on a logarithmic scale) due to the finite size of the Bloch sphere  for $N=100$ atoms.}
\label{fig:BlochContrast}
\end{figure}

Then  Eq.~\eqref{eq:BlochContrast} can be evaluated by extending the integration to $\pm \infty$, yielding
\begin{align}\label{eq:ContrastLossBloch}
    C_\mathrm{Bloch}(Q)=\int\mathrm{d}\theta P(\theta)\cos\theta
    \to\exp\left(-\frac{Q^2}{2N}\right), 
\end{align}
Fig.~\ref{fig:BlochContrast} shows that the analytical formula \eqref{eq:ContrastLossBloch} describes the quantum behavior quite well even for $Q^2 \gtrsim N$, indicating that in this limit the spin distribution is still approximately a Gaussian distribution, but now wrapping around the whole Bloch sphere. 

\section{Derivation of Kitagawa-Ueda parameter from Shearing Strength and Quantum Fisher Information}\label{sec:Gaussian}

The effect of the shearing strength $Q$ and QFI $F$ on an initial CSS is presented in Fig.~\ref{fig:QFeffectonstate}(c, e). Mathematically, since the transformations at small $Q, F$ are all linear, the final Wigner function remains Gaussian. Writing all quantities in units of the SQL, the combination of $Q$, $F$, and another rotation by an angle $\alpha$ transforms the point $(S_y, S_z)^T$ on Bloch sphere to $(S_y', S_z')^T$ as given by
\begin{align}\begin{split}
    \left(\begin{matrix}
    S_y'\\ S_z'
    \end{matrix}\right) &= \left(\begin{matrix}
    \cos\alpha & \sin\alpha\\
    -\sin\alpha & \cos\alpha
    \end{matrix}\right)\left(\begin{matrix}
    \sqrt{1+F} & Q \\
    0 & 1
    \end{matrix}\right)\left(\begin{matrix}
    S_y\\ S_z
    \end{matrix}\right) \\
    &\equiv U\left(\begin{matrix}
    S_y\\ S_z
    \end{matrix}\right)
\end{split}\end{align}
The initial distribution of the CSS pointing along the $x$-axis in units of the SQL is
\[P(S_y,S_z)\sim\exp(-(S_y,S_z)(S_y,S_z)^T)\]
and after the transformation it is given by
\[P(S_y',S_z')\sim\exp(-(S_y',S_z')\Sigma^{-1}(S_y',S_z')^T).  \]

Making use of that $(S_y, S_z)^T=U^{-1}(S_y', S_z')^T$, the new covariance matrix is
\begin{eqnarray}
    \Sigma'&=&((U^{-1})^T U^{-1})^{-1}=UU^T
\end{eqnarray}
and therefore the variance on $S_z'$ is the second diagonal element of $\Sigma'$, 
\begin{equation}\label{eq:varSzfromCovariance}
    \mathrm{var}(S_z')=1-Q\sin2\alpha+(F+Q^2)\sin^2\alpha. 
\end{equation}

By changing $\alpha$, the minimum and maximum variance is obtained when $\alpha$ satisfies
\begin{equation}
    \frac{\mathrm{d}\mathrm{var}(S_z')}{\mathrm{d}\alpha}=-2Q\cos2\alpha+2(F+Q^2)\sin\alpha\cos\alpha = 0,
\end{equation}
which implies
\begin{equation}
    \tan\alpha=(\sqrt{4Q^2{+}(F{+}Q^2)^2}-(F{+}Q^2)){/}(2Q).
\end{equation}
The Kitagawa-Ueda parameter is given by
\begin{equation}\label{eq:WinelandFormula}
    \xi_\pm^2=\frac{1}{2} \left( 2+F+Q^2 \pm \sqrt{4Q^2 + (F+Q^2)^2}\right)
\end{equation}

\section{Fisher Information contained in a coherent state of light}\label{sec:FIderivation}

Suppose we have a coherent state of light which interacts with an arbitrary system, such that the state of light after the interaction remains a coherent state and is given by $\ket{\alpha(x)}$. What is the QFI of this quantum state in terms of the function $\alpha(x)$? In general, the QFI about the parameter $x$ is given by \cite{Braunstein1994,Demkowicz-Dobrzanski2015,Pezze2018}:
\begin{equation}
\label{eq:FisherInformationGeneralFormula}
F_Q\left[\ket{\alpha(x)}\right] = 4 \left( 
\braket{\partial_x\alpha(x)|\partial_x\alpha(x)} - \left|\braket{\partial_x\alpha(x)|\alpha(x)}
\right|^{2} \right)
\end{equation}
for coherent states parametrized by $\alpha(x)$, and where the state derivative is defined as
\begin{equation}
\ket{\partial_x\alpha(x)} \equiv \frac{\partial \ket{\alpha(x)}}{\partial x} = \lim\limits_{\epsilon\to 0} {\frac{\ket{\alpha(x + \epsilon)} - \ket{\alpha(x)}}{\epsilon}}. 
\end{equation}

To calculate $F_Q$, we begin with the well-known overlap relationship between coherent states:
\begin{equation} 
\label{eq:CoherentStateOverlapGeneral}
\braket{\beta|\alpha} = \exp \left[-\frac{1}{2} \left( |\beta|^2 + |\alpha|^2 - 2 \beta^* \alpha \right) \right]
\end{equation}
Substituting the states of interest gives
\begin{equation}
\label{eq:CoherentStateOverlapAlphaAlphaPlusEpsilon}
\begin{split}
    \braket{\alpha(x+\epsilon)|\alpha(x)} =& \exp \Big[-\frac{1}{2} \big( |\alpha(x+\epsilon)|^2 + |\alpha(x)|^2 \\
    &\phantom{\exp \Big[-\frac{1}{2} \big(} - 2 \alpha^*(x+\epsilon) \alpha(x) \big) \Big]
\end{split}
\end{equation}
We expand the exponential argument in \eqref{eq:CoherentStateOverlapAlphaAlphaPlusEpsilon} up to second order in $\epsilon$:
\begin{equation}
    \begin{split}
    & |\alpha(x+\epsilon)|^2 + |\alpha(x)|^2 - 2 \alpha^*(x+\epsilon) \alpha(x)  \\
    =& \epsilon \left(\alpha^* \dot{\alpha} - \dot{\alpha}^* \alpha \right) + 
    \frac{\epsilon^2}{2} \left(\alpha^* \ddot{\alpha} - \ddot{\alpha}^* \alpha + 2 \dot{\alpha}^* \dot{\alpha} \right)\\
    &+ \mathcal{O}(\epsilon^3), 
    \end{split}
\end{equation}
where $\dot{\alpha} = \mathrm{d}\alpha(x)/\mathrm{d}x$, and hence
\begin{eqnarray}
\label{eq:AlphaPlusEpsilonOverlap}
&& \braket{\alpha(x+\epsilon)|\alpha(x)} =
1 
- \frac{1}{2}  \left(\alpha^* \dot{\alpha} - \dot{\alpha}^* \alpha \right) \epsilon + \nonumber \\
&+& \frac{\epsilon^2}{2} \left[-\frac{1}{2} \left(\alpha^* \ddot{\alpha} - \ddot{\alpha}^* \alpha + 2 \dot{\alpha}^* \dot{\alpha} \right)  + \frac{1}{4} \left(\alpha^* \dot{\alpha} - \dot{\alpha}^* \alpha \right)^2  \right] \nonumber \\
&+& \mathcal{O}(\epsilon^3)
\end{eqnarray}
Note that to this order, the real part of $\braket{\alpha(x+\epsilon)|\alpha(x)}$ equals $\Re[\braket{\alpha(x+\epsilon)|\alpha(x)}] = 1 + \epsilon^2 [-\dot{\alpha}^* \dot{\alpha}/2 + (\alpha^* \dot{\alpha} - \dot{\alpha}^* \alpha )^2/8  ] + \mathcal{O}(\epsilon^3)$. The first term in bracket of \eqref{eq:FisherInformationGeneralFormula} therefore equals
\begin{eqnarray}
\label{eq:FisherInformationFirstTerm}
&& \lim\limits_{\epsilon\to 0} \left[\frac{\bra{\alpha(x + \epsilon)} - \bra{\alpha(x)}}{\epsilon}\right] \left[\frac{\ket{\alpha(x + \epsilon)} - \ket{\alpha(x)}}{\epsilon}\right] \nonumber \\
&=& \dot{\alpha}^* \dot{\alpha} - \frac{1}{4} \left(\alpha^* \dot{\alpha} - \dot{\alpha}^* \alpha \right)^2
\end{eqnarray}
while the second term equals
\begin{eqnarray}
\label{eq:FisherInformationSecondTerm}
&& \left| \lim\limits_{\epsilon\to 0} {\frac{\braket{\alpha(x + \epsilon)|\alpha(x)} - \braket{\alpha(x)|\alpha(x)}}{\epsilon}} \right|^2 \nonumber \\
&=& 
-\frac{1}{4} \left(\alpha^* \dot{\alpha} - \dot{\alpha}^* \alpha \right)^2
\end{eqnarray}

Combining \eqref{eq:FisherInformationFirstTerm} and \eqref{eq:FisherInformationSecondTerm}, we have the QFI for parameter-dependent coherent states:
\begin{equation}
\label{eq:FisherInformationForCoherentStates}
F_Q\left[\ket{\alpha(x)}\right] = 4 \left| \frac{\mathrm{d}\alpha(x)}{\mathrm{d}x} \right|^2
\end{equation}
The QFI only depends on the magnitude of the derivative of $\alpha$ with respect to $x$, which is consistent with the intuition that coherent states are symmetric around their mean values and behave the same way regardless of the direction in which they are translated. Furthermore, the QFI is independent of the value of $\alpha$ itself, because coherent states are invariant when translated in phase space. Note that the total derivative in \eqref{eq:FisherInformationForCoherentStates} becomes a partial derivative when $\alpha$ is a function of multiple parameters. The QFI \eqref{eq:FisherInformationForCoherentStates} implies a corresponding quantum Cram\'{e}r-Rao bound:
\begin{equation}
\label{eq:CramerRaoBoundForCoherentStates}
\left(\Delta x\right)^2 \geq  \frac{1}{4} \left| \frac{\mathrm{d}\alpha(x)}{\mathrm{d}x} \right|^{-2}
\end{equation}

We can further divide the total QFI \eqref{eq:FisherInformationForCoherentStates} into parts associated with varying amplitude $A(x)$ and varying phase $\phi(x)$, by defining
\begin{equation}
\alpha(x) = A(x) e^{i \phi(x)}
\end{equation}
with both amplitude $A(x)$ and phase $\phi(x)$ being real. Then,
\begin{equation}
\frac{\mathrm{d} \alpha(x)}{\mathrm{d}x} = \frac{\mathrm{d} A(x)}{\mathrm{d}x} e^{i \phi(x)} + i A(x) e^{i \phi(x)} \frac{\mathrm{d} \phi(x)}{\mathrm{d}x}
\end{equation}
and hence,
\begin{equation}
\label{eq:FisherInformationForCoherentStatesAmplitudeAndPhase}
F_Q\left[\ket{\alpha(x)}\right] = 4 \left(\frac{\mathrm{d} A(x)}{\mathrm{d}x}\right)^2 + 4 \left( A(x) \frac{\mathrm{d} \phi(x)}{\mathrm{d}x} \right)^2
\end{equation}
The amplitude component, $F_{Q,A} \left[\ket{\alpha(x)}\right] = 4  \left|\frac{\mathrm{d} A(x)}{\mathrm{d}x}\right|^2$, corresponds to the classical QFI associated with the photon number distribution, which is directly observable with a photon number-resolving detector. 

This can be seen by considering the probability of observing $n$ photons in the coherent state, which is given by a Poisson distribution:
\begin{equation}\label{eq:PoissonDistribution}
P_n(x) = \frac{\lambda^n(x) e^{-\lambda(x)}}{n!}
\end{equation}
where $\lambda(x)=A^2(x)$ is the mean photon number, i.e. $\sum_{n=0}^{\infty} n P_n(x) = \lambda(x)$.
Then, the classical QFI about the parameter $x$ contained in the photon number measurement is given by
\begin{eqnarray}\label{eq:PoissonFisherInformation}
&& F_C \left[\ket{\alpha(x)}\right] = -\sum_{n=0}^{\infty} \left(\frac{\mathrm{d}^2}{\mathrm{d}x^2} \log P_n(x) \right) P_n(x) \nonumber \\
&=&
-\sum_{n=0}^{\infty} \left(\frac{\mathrm{d}^2}{\mathrm{d}x^2} \left[n\log \lambda(x) - \lambda(x)\right] \right) P_n(x) \nonumber \\
&=&
\sum_{n=0}^{\infty} \left(\frac{1}{\lambda(x)} \left(\frac{\mathrm{d}\lambda(x)}{\mathrm{d}x}\right)^2 + \left(1 - \frac{n}{\lambda(x)}\right)\frac{\mathrm{d}^2 \lambda(x)}{\mathrm{d}x^2}  \right) P_n(x) \nonumber \\
&=&
\frac{1}{\lambda(x)} \left(\frac{\mathrm{d}\lambda(x)}{\mathrm{d} x}\right)^2 = 4  \left(\frac{\mathrm{d}A(x)}{\mathrm{d}x}\right)^2 \nonumber \\
&=& F_{Q,A} \left[\ket{\alpha(x)}\right]
\end{eqnarray}

The equality $F_C \left[\ket{\alpha(x)}\right] = F_{Q,A} \left[\ket{\alpha(x)}\right]$ implies that the quantum Cram\'{e}r-Rao bound \eqref{eq:CramerRaoBoundForCoherentStates} is saturated by an intensity measurement of the state if and only if the phase component of $F_Q$ is zero: $\frac{d \phi(x)}{d x} = 0$. This condition can always be satisfied by displacing the state $\ket{\alpha(x)}$ in phase space to $\beta(x) = \alpha(x) + \Delta$ so that $\frac{d \beta(x)}{dx} = \frac{d \alpha(x)}{dx}$ is parallel to $\beta(x)$. Physically, this is simply a homodyne measurement, where the phase of the reference beam is chosen so as to transform the signal of interest into a pure amplitude modulation, with no residual phase modulation. For a general $\alpha(x)$ and arbitrary $x$, this will require an $x$-dependent displacement $\Delta(x)$, which in turn requires some preexisting knowledge of $\alpha(x)$. Alternatively, half of the total QFI can be obtained by a heterodyne measurement, where modulating the direction of $\Delta$ periodically redistributes the QFI between amplitude and phase components of the state.

\section{Useful relations for the coupled atom-cavity system}\label{sec:CavityFieldAppendix}

Following Ref. \cite{Tanji-Suzuki2011}, we introduce a dimensionless ensemble-cavity coupling parameter $\beta = N \frac{k}{\pi w^2} \frac{\tilde{\alpha}}{\epsilon_0}$, where $k = 2\pi/\lambda$ is the wave number of the light, and $\tilde{\alpha}$ is the atomic polarizability. For a single two-level atom with transition frequency $\omega_a$ and natural linewidth $\Gamma$, the polarizability is given by \cite[Eq.~(1)]{Tanji-Suzuki2011}: 
$
\tilde{\alpha} = 6 \pi \epsilon_0 c^3 \frac{\Gamma / {\omega_a}^2}{{\omega_a}^2 - {\omega}^2 - i (\omega^3 / {\omega_a}^2) \Gamma}
$. The steady-state mode amplitudes of the cavity, transmitted and reflected fields are all proportional to $\mathcal{E}_\mathrm{in}$ and given by 
\begin{equation}\label{eq:CavityFieldFinalResult}
\mathcal{E}_c = \frac{i t_1}{1 - 4 i \beta - r_1 r_2 e^{2 i k L}} \mathcal{E}_\mathrm{in}. 
\end{equation}
The transmitted field amplitude is then
\begin{equation}
\mathcal{E}_t =  
- \frac{t_1 t_2 e^{i k L}}{1 - 4 i \beta - r_1 r_2 e^{2 i k L}} \mathcal{E}_\mathrm{in}. 
\end{equation}
Similarly, the reflected field amplitude is given by
\begin{equation}
\mathcal{E}_r = 
\left[ r_1-\frac{ t_1^2 r_2 e^{2 i k L} }{1 - 4 i \beta - r_1 r_2 e^{2 i k L}} \right] \mathcal{E}_\mathrm{in}. 
\end{equation}
Another relevant quantity is the power radiated by the atoms into free space. For a cavity with a large waist compared to the wavelength of the light, the cavity mode also subtends a small solid angle relative to the atom, and thus Purcell suppression \cite{Kleppner1981} does not come into play. In this case, the total power of radiation by the atoms into free space is approximately the same as when driven by the cavity field $\mathcal{E}_c$ in the absence of the cavity \cite[Eq.~(9)]{Tanji-Suzuki2011}
\begin{equation}\label{eq:CavityAtomsFreeSpaceRadiation}
P_{4\pi} = |\mathcal{E}_c|^2 \mathrm{Im}(4 \beta). 
\end{equation}

To further simplify the expressions, we assume that the probing beam frequency $\omega$ is close to a certain cavity resonance frequency $\omega_c$ which satisfies $e^{i\omega_c\cdot 2L/c}=1$, and both the cavity detuning $\delta = \omega - \omega_c$ and the atomic detuning $\Delta = \omega - \omega_a$ are much smaller than the free spectral range $\FSR = 2 \pi \times \frac{c}{2 L} $ of the bare cavity. This allows us to approximate the light propagation term in \eqref{eq:CavityFieldFinalResult} as $\exp \left( 2 i k L \right) \approx 1 + 2 i \frac{\delta}{c} L$. 
The latter is equivalent to the assumption that the optical cavity supports only a single field mode with frequency $\omega_c$ and decay constant $\kappa$, neglecting the influence of any other cavity modes.

The detuning from atomic resonance is also assumed to be much smaller than the atomic transition frequency $\Delta\ll\omega_a$, which corresponds to the condition of the rotating-wave approximation. Therefore
\begin{equation}
\tilde{\alpha} \approx 6 \pi \epsilon_0 \frac{c^3}{\omega_a^3} \frac{i}{1 - \frac{i \Delta}{\Gamma/2}}=6 \pi \epsilon_0 \frac{c^3}{\omega_a^3}\left(\mathcal{L}_d(\y)+i\mathcal{L}_a(\y)\right)
\end{equation}
where $\y{\equiv}\frac{\Delta}{\Gamma/2}$ is the normalized detuning of the probe laser from the atomic resonance, while $\mathcal{L}_d(\y){=}-\frac{\y}{1+\y^2}$ and $\mathcal{L}_a(\y){=}\frac{1}{1+\y^2}$
are the Lorentzian dispersive and absorptive lineshapes \cite{Tanji-Suzuki2011}.
Correspondingly, 
$\beta$ becomes 
\begin{equation}\label{eq:AtomicPolarizabilityBetaRWA}
\beta \approx N \frac{6 c ^2}{\omega_a^2 w^2} \frac{i}{1 - \frac{i \Delta}{\Gamma/2}}=N \eta_\mathrm{fs}
\left(\mathcal{L}_d(\y)+i\mathcal{L}_a(\y)\right), 
\end{equation}
where $\eta_\mathrm{fs}{=}6/k^2w^2$ is the free-space cooperativity which describes the interaction strength between a single atom and a free-space Gaussian mode \cite{Kimble1998}.

\section{Phenomenological description of multi-photon light field}\label{sec:AppendixMultiPhoton}

As in Ref.~\cite{Bishop2009}, the light field spectra deviate from \eqref{eq:CavityFieldFinalResultRWA} - \eqref{eq:ReflectedFieldFinalResult} with strong probing power. This effect corresponds to the saturation of 2-level atoms. However, the atomic ensemble is not equivalent to the 2-level atom. For simplicity, we consider the energy level structure used in Sec.~\ref{sec:ALI}, the collective angular momentum of $N$ 2-level atoms actually corresponds to an $(N_\uparrow+1)$-level system: $\{\ket{\downarrow\cdots\downarrow\uparrow\cdots\uparrow e\cdots e}_\mathrm{sym}\}$ with $N_\downarrow$ of $\ket{\downarrow}$, $n$ of $\ket{e}$, and $N_\uparrow-n$ of $\ket{\uparrow}$, where $n=0,1,\cdots N_\uparrow$ and symmetrized over all atoms. We refer to such a state $\ket{N_\uparrow,n}$ in later expressions for simplicity. 

We can truncate the Hilbert space by the total excitation number $m$, which equals to the sum of the atomic excitation number and the photon number. 
In the non-saturating regime, the excitation number satisfies $m\ll N_\uparrow$, and the sub-Hilbert space with $m$ excitation is spanned by $\{\ket{N_\uparrow,0}\otimes\ket{n_c=m},\cdots\ket{N_\uparrow,m}\otimes,\ket{n_c=0}\}$. The coupling strength between these levels are $\sqrt{(n_c+1)(N_\uparrow-n_c)}\sim\sqrt{n_c+1}\sqrt{N_\uparrow}$, and therefore can still be considered as a harmonic oscillator mode with $\sqrt{N_\uparrow}$ enhancement factor.

\section{Derivation of the effective Hamiltonian from the microscopic quantum model} \label{sec:QuantumHamiltonian}

The approach developed in Sec.~\ref{sec:ALI} is based on semiclassical cavity field, but the obtained effective Hamiltonian \eqref{eq:CavityExpansion} is equivalent to a microscopic model, as presented in Ref.~\cite{Davis2016}. Here we derive our Hamiltonian from a microscopic model of the atom-light interaction in the mean-field limit. We consider a three level atom as in Sec.~\ref{sec:ALI}, and an optical cavity mode (represented by its annihilation operator $\hat{b}$) interacting with the $\ket{\uparrow}$ state with coupling strength $g=\sqrt{\eta\kappa\Gamma}/2$~\cite{Tanji-Suzuki2011}. Also, we define the raising and lowering operators $\hat{a}^{(\dagger)}_{e,\uparrow}$ for the atomic states. The full Hamiltonian is 
\begin{align}
\begin{split}
    H=& \underbrace{\omega_c \hat{b}^\dagger \hat{b}}_{H_\mathrm{c}}+\underbrace{\omega_a\hat{a}_{e}^\dagger \hat{a}_{e}}_{H_\mathrm{a}}+\underbrace{g\left(\hat{b}\hat{a}_{e}^\dagger \hat{a}_{\uparrow}+h.c.\right)}_{H_\mathrm{int}}\\
    &+\underbrace{\sqrt{\dfrac{\kappa}{2}\dfrac{2T_1}{T_1+T_2}}\left[\beta^*\hat{b}^{-i\omega_lt}+\beta \hat{b}^\dagger e^{i\omega_lt}\right]}_{H_\mathrm{drive}},
\end{split}
\end{align}
and the atomic and cavity field decay are described by the Lindblad master equation:
\begin{align}
    \dot{\rho}=i[H, \rho]+\frac{1}{2} \sum_{j}\left(\left[L_{j} \rho, L_{j}^{\dagger}\right]+\left[L_{j}, \rho L_{j}^{\dagger}\right]\right),
\end{align}
and the corresponding Lindblad operators are
\[L_c=\sqrt{\kappa}\hat{b},L_a=\sqrt{\Gamma}\hat{a}_e\hat{a}_\uparrow^\dagger. \]

In the unsaturated regime, we can use a mean-field approximation so that the operator $a^{(\dagger)}_\uparrow$ can be replaced by $\sqrt{N_\uparrow}$, and the first line can be considered as a coupled harmonic oscillator problem. We can diagonalize the system into new modes described by annihilation operator $\hat{c}_+, \hat{c}_-$:
\begin{align}
    H_\mathrm{c}+H_\mathrm{a}+H_\mathrm{int}=\omega_+\hat{c}_+^\dagger\hat{c}_++\omega_-\hat{c}_-^\dagger\hat{c}_-,
\end{align}
where
\[\omega_{\pm}=\frac{\omega_a+\omega_c\pm\sqrt{4g^2N_\uparrow+(\omega_a-\omega_c)^2}}{2} \]
and
\[\begin{split}
    \hat{c}_{\pm}=&\frac{\omega_\pm-\omega_c}{\sqrt{(\omega_\pm-\omega_c)^2+g^2N_\uparrow}}\hat{a}_e + \frac{g\sqrt{N_\uparrow}}{\sqrt{(\omega_\pm-\omega_c)^2+g^2N_\uparrow}}\hat{b},
\end{split} \]
and the driving Hamiltonian $H_\mathrm{drive}$ can be written as
\[\begin{split}
    &H_\mathrm{drive}=\sqrt{\dfrac{\kappa}{2}\dfrac{2T_1}{T_1+T_2}}\frac{1}{\omega_+-\omega_-}\times\\
    \Bigg[&\beta^*e^{-i\omega_lt}\Big(\sqrt{(\omega_+-\omega_a)^2+g^2N_\uparrow}\hat{c}_+\\
    &\phantom{\beta^*e^{-i\omega_lt}\Big(}-\sqrt{(\omega_--\omega_a)^2+g^2N_\uparrow}\hat{c}_-\Big)\\
    &\beta e^{i\omega_lt}\Big(\sqrt{(\omega_+-\omega_a)^2+g^2N_\uparrow}\hat{c}^\dagger_+\\
    &\phantom{\beta e^{i\omega_lt}\Big(}-\sqrt{(\omega_--\omega_a)^2+g^2N_\uparrow}\hat{c}^\dagger_-\Big)\Bigg].
\end{split} \]

Lastly, the Lindblad operators can also be expressed in these new modes:
\[\begin{split}
    &L_c=\sqrt{\kappa}\hat{b}=\sqrt{\kappa}\frac{1}{\omega_+-\omega_-}\\
    &\times\left(\sqrt{(\omega_+-\omega_a)^2+g^2N_\uparrow}\hat{c}_+-\sqrt{(\omega_--\omega_a)^2+g^2N_\uparrow}\hat{c}_-\right),
    \\
    &L_a=\sqrt{N_\uparrow\Gamma} \hat{a}_e=\sqrt{N_\uparrow\Gamma}\frac{1}{\omega_+-\omega_-}\\
    &\times\left(\sqrt{(\omega_+-\omega_c)^2+g^2N_\uparrow}\hat{c}_+-\sqrt{(\omega_--\omega_c)^2+g^2N_\uparrow}\hat{c}_-\right).
\end{split}\]

Using the Reiter and S\o{}rensen method \cite{Reiter2012}, we can adiabatically eliminate the cavity field as well as the excited state by eliminating the $\pm$ modes, and therefore obtain the effective Hamiltonian in the $\hat{a}_\uparrow$ manifold. Further, we can assume that
\begin{itemize}
    \item $\kappa|\beta|^2\frac{2T_1}{T_1+T_2} \langle \hat{b}^{\dagger} \hat{b}\rangle \ll (\omega_l-\omega_{a,c})^2$
    \item $g^2\frac{2T_1}{T_1+T_2} \langle \hat{b}^{\dagger} \hat{b}\rangle \ll (\omega_l-\omega_{a,c})^2$
\end{itemize}
which simplifies the obtained Hamiltonian to
\begin{align}
    H_\mathrm{eff}=\Delta_\mathrm{eff}\hat{a}_\uparrow^\dagger\hat{a}_\uparrow,
\end{align}
where
\[
\begin{split}
    \Delta_\mathrm{eff}=&-|\beta|^2\frac{2T_1}{T_1+T_2}\frac{g^2}{\kappa\Gamma}(1+2(\omega_l-\omega_a)^2/\Gamma^2)^2\times\\
    &\big[((1+2(\omega_l-\omega_a)^2/\Gamma^2)^2+g^2N_\uparrow/\kappa\Gamma)^2\\
    &+(2(\omega_l-\omega_c)/\kappa\cdot(1+2(\omega_l-\omega_a)^2/\Gamma^2)^2)\\
    &-2(\omega_l-\omega_a)/\Gamma\cdot g^2N_\uparrow/\kappa\Gamma)^2\big]^{-1},
\end{split} \]
which is equivalent to Eq. \eqref{eq:CavityExpansion}. 

\newpage 
\onecolumngrid
\section{Table of symbols}
\begin{longtable}{c|c|l|c}
    \toprule
        Symbol  &   Definition  &   Meaning  & Experimental value   \\
        \hline
        $t_1$           &      $\phantom{\frac{1}{1}}$         &   Transmission amplitude of first cavity mirror & $5.5\times10^{-3}$ \\
        $r_1$           &      $\phantom{\frac{1}{1}}$         &   Reflection amplitude of first cavity mirror & $0.9999$ \\
        $t_2$           &     $\phantom{\frac{1}{1}}$          &   Transmission amplitude of second cavity mirror & $1.4\times10^{-2}$\\
        $r_2$           &      $\phantom{\frac{1}{1}}$         &   Reflection amplitude of second cavity mirror & $0.9992$ \\
        $\mathcal{F}$   &        $\frac{2\pi}{2-R_1-R_2}$       &   Cavity finesse & $1.2(1) \times 10^{4}$ \\
        $L$   &      $\phantom{\frac{1}{1}}$         &   Cavity length &  \\
        $\FSR$             &      $2\pi \times \frac{c}{2L}$         &   Cavity free spectral range & $2\pi \times 5970.04(4) \un{MHz}$ \\
        $\kappa$   &       $\frac{\FSR}{\mathcal{F}}$       &   Cavity linewidth & $2\pi\times520(10)\un{kHz}$ \\
        $\Gamma$   &      $\phantom{\frac{1}{1}}$        &   \Yb ground state to $^3P_1$ transition linewidth & $2\pi\times184\un{kHz}$ \\
        $\eta$   &     $\frac{24 \mathcal{F}}{\pi} \frac{1}{k^2 w^2} = \frac{4 g^2}{\kappa \Gamma}$    &   Single-atom cooperativity & \\
        $F$             &       $\phantom{\frac{1}{1}}$        &   Normalized Fisher information & \\
        $\zeta$             &      $\phantom{\frac{1}{1}}$         &   First-order phase shift rate & \\
        $\Delta \phi$             &   $\phantom{\frac{1}{1}}$            &   Linear AC Stark shift & \\
        $\chi$             &      $\phantom{\frac{1}{1}}$         &   One-axis twisting rate & \\
        $Q$             &      $\phantom{\frac{1}{1}}$         &   Normalized one-axis twisting strength & \\
        $\mathcal{E}_\mathrm{t},\mathcal{E}_\mathrm{r},\mathcal{E}_c$ &  $\phantom{\frac{1}{1}}$ & (Transmitted, reflected, intracavity) light mode amplitude &   \\
        $\mathcal{T}$ & $\left|\frac{\mathcal{E}_\mathrm{t}}{\mathcal{E}_\mathrm{in}}\right|^2$  &    Power transmission & \\
        $\mathcal{R}$ & $\left|\frac{\mathcal{E}_\mathrm{r}}{\mathcal{E}_\mathrm{in}}\right|^2$  &    Power reflection & \\
        $\omega$ &  $\phantom{\frac{1}{1}}$ &    Probe laser angular frequency & \\
        $\omega_c$ & $\phantom{\frac{1}{1}}$  &    Cavity resonance angular frequency & \\
        $\omega_a$ & $\phantom{\frac{1}{1}}$  &    Atomic resonance angular frequency & \\
        $\delta$ & $\omega-\omega_c$  &    Cavity detuning & $\phantom{\frac{1}{1}}$\\
        $\Delta$ & $\omega-\omega_a$   &    Atomic detuning & $\phantom{\frac{1}{1}}$\\
        $\x$ & $\frac{2\delta}{\kappa}$  &    Normalized cavity detuning & $\phantom{\dfrac{1}{1}}$ \\
        $\y$ & $\frac{2\Delta}{\Gamma}$  &    Normalized atomic detuning & $\phantom{\dfrac{1}{1}}$ \\
        $\Delta_z$ & $\phantom{\frac{1}{1}}$  &    Difference of two transitions in four-level model & $2\pi\times20\un{MHz}$ \\
        $b$ & $\frac{2\Delta_z}{\Gamma}$  &    Normalized atomic detuning of $\Delta_z$ $\phantom{\dfrac{1}{1}}$ & $230$ \\
    \botrule
    \caption{List of symbols used in this paper.}
    \label{tab:Symbols}
\end{longtable}
\twocolumngrid

\bibliographystyle{apsrev4-2}
\bibliography{UnitarySqueezingTheory}

\end{document}